\documentclass[%
 reprint,
superscriptaddress,
nofootinbib,
 amsmath,amssymb,
 aps,
]{revtex4-2}
\usepackage{graphicx}
\usepackage{dcolumn}
\usepackage{bm}
\usepackage[version=4]{mhchem} 
\usepackage{subfig}
\usepackage{float}
\usepackage{caption}
\usepackage{multirow}
\usepackage[colorlinks,
            linkcolor=red,
            anchorcolor=blue,
            citecolor=blue
           ]{hyperref}

\graphicspath{ {figure/} }
\begin{document}

\title{Reexamine the dark matter scenario accounting for the positron excess in a 
new cosmic ray propagation model}

\author{Xing-Jian Lv}
\email{lvxj@ihep.ac.cn}
 \affiliation{%
 Key Laboratory of Particle Astrophysics, Institute of High Energy Physics, Chinese Academy of Sciences, Beijing 100049, China}
\affiliation{
 University of Chinese Academy of Sciences, Beijing 100049, China 
}%
 \author{Xiao-Jun Bi}
 \email{bixj@ihep.ac.cn}
\affiliation{%
 Key Laboratory of Particle Astrophysics, Institute of High Energy Physics, Chinese Academy of Sciences, Beijing 100049, China}
\affiliation{
 University of Chinese Academy of Sciences, Beijing 100049, China 
}%
\author{Kun Fang}
\email{fangkun@ihep.ac.cn}
\affiliation{%
 Key Laboratory of Particle Astrophysics, Institute of High Energy Physics, Chinese Academy of Sciences, Beijing 100049, China}
 \author{Peng-Fei Yin}
\email{yinpf@ihep.ac.cn}
\affiliation{%
 Key Laboratory of Particle Astrophysics, Institute of High Energy Physics, Chinese Academy of Sciences, Beijing 100049, China}
\author{Meng-Jie Zhao}
\email{zhaomj@ihep.ac.cn}
 \affiliation{%
 Key Laboratory of Particle Astrophysics, Institute of High Energy Physics, Chinese Academy of Sciences, Beijing 100049, China}
\affiliation{
 University of Chinese Academy of Sciences, Beijing 100049, China 
}%



\date{\today}

\begin{abstract}
The positron excess in cosmic rays has stimulated a lot of interests in the last decade. The dark matter origin of the extra positrons has attracted great attention. However, the $\gamma$-ray search set very stringent constraints on the dark matter annihilation/decay rate, which leads to great disfavor of the dark matter scenario. In the work, we incorporate the recent progress in cosmic rays propagation and reexamine the dark matter scenario accounting for the positron excess. Recent observations indicate that cosmic rays propagation in the Milky Way may be not uniform and diffusion in the Galactic disk should be slower than that in the halo.
In the spatial-dependent propagation model, the positrons/electrons are more concentrated in the disk and lead to smaller dark matter annihilation/decay rate to account for the positron excess and also a smaller deficit in the background positron flux.
Especially for the $\mu^+\mu^-$ channel the positron spectrum fit the AMS-02 latest data perfectly and the annihilation rate satisfies all the present constraints from $\gamma$-ray and CMB observations.
\end{abstract}
\maketitle


\section{\label{sec:level1}INTRODUCTION}

As the observation of positron excess in cosmic rays (CR) by PAMELA~\cite{PAMELA:2008gwm} and later precise confirmation by AMS-02 \cite{AMS:2019rhg,AMS:2019iwo},
a multitude of studies have emerged, aiming to resolve its origin. In the literature, two primary conjectures have attracted significant attention: the involvement of dark matter (DM) annihilation or decay within the Galactic halo~\cite{Bergstrom:2008gr, Barger:2008su, Cirelli:2008pk, Yin:2008bs, Arkani-Hamed:2008hhe, Pospelov:2008jd, Cholis_2009,Hamaguchi_2009, Boudaud:2014dta, Lin:2014vja, Wang:2018pcc, Cheng:2016slx}, and the presence of nearby astrophysical sources~\cite{aharonianHighEnergyElectrons1995, Atoian:1995ux, Kobayashi:2003kp, Profumo:2008ms, Hooper:2008kg, Yuksel:2008rf, Malyshev:2009tw, Blasi:2009hv, Ioka_2010, DiMauro:2015jxa, Cholis:2022kio}. These interpretations have been thoroughly investigated. 

However, attempts to employ DM annihilation as an explanation for the observed excess have encountered challenges.
For instance, the annihilation/decay of DM into pairs of quarks or gauge bosons 
is ruled out due to the absence of corresponding excesses in the flux of cosmic ray antiprotons\cite{AMS:2016oqu}. For leptonic channels, the emission of high-energy photons in conjunction with charged leptons would generate discernible signals in systems with high DM densities and low baryon densities, such as dwarf galaxies\footnote{
While antiproton production does occur through internal bremsstrahlung for leptonic channels, the resulting quantity is comparatively minimal. Consequently, this leads to a weaker constraining power~\cite{caloreAMS02AntiprotonsDark2022} when compared with the more pronounced constraints imposed by gamma-ray observations.}. Consequently, the absence of such signals in the Fermi-LAT data strongly disfavor the DM-based explanations~\cite{Fermi-LAT:2016uux, Fermi-LAT:2012pls, Fermi-LAT:2015qzw, Liu:2016ngs}. Furthermore, the injection of energy resulting from DM annihilation/decay during recombination could impact the cosmic microwave background (CMB). Precise measurements carried out by Planck\cite{Planck:2015fie} have imposed stringent constraints on the properties of DM~\cite{Slatyer:2015jla, Slatyer:2016qyl}, which also conflict with the requirements to explain positron excess.
As a result, complicated DM models have been proposed to reconcile this apparent discrepancy. These attempts include proposals involving velocity-dependent annihilation cross sections, such as the Sommerfeld\cite{Hisano:2003ec,Hisano:2006nn,Feng:2009hw, Feng:2010zp,dasGalacticPositronExcess2020,Ding:2021zzg} and Breit-Wigner mechanisms\cite{Feldman:2008xs,Ibe:2008ye,Bi:2011qm,Bai:2017fav,Xiang:2017jou}, as well as local DM over-density\cite{Hektor:2013yga}.

The field of CR propagation has recently witnessed substantial progress owing to several observations. The identification of TeV halos around some middle-aged pulsars has unveiled diffusion coefficients in the vicinity of these pulsars that are more than two orders of magnitude lower than the galactic average~\cite{Fang:2022fof,HAWC:2017kbo,Hooper:2017gtd,LHAASO:2021crt}. Moreover, the spatial magnetic-energy spectrum within the Galaxy suggests that the intensity of magnetic field turbulence within the galactic disk surpasses that found in the halo~\cite{hanObservingInterstellarIntergalactic2017}, which implies a significant reduction in the diffusion coefficient within the Galactic disk\footnote{
The assertion in Ref.~\cite{hanObservingInterstellarIntergalactic2017} regarding stronger magnetic turbulence in the Galactic disk only pertains to larger scales, which are not directly translate to efficient cosmic ray scattering at smaller scales. Thus, the suggested reduction in the diffusion coefficient within the Galactic disk is speculative and requires further empirical investigation.}. Collectively, these findings indicate that the diffusion coefficient in the Galactic disk could be significantly smaller than that of the Galactic halo, thereby challenging the assumption of homogeneous diffusion embedded in conventional CR propagation models.

To address this issue, a spatial-dependent diffusion model, encompassing a slow-diffusion disk (SDD) proximate to the Galactic plane, has been proposed in a previous investigation~\cite{Zhao:2021yzf}. This model explains the observed spectral hardening of CRs at several hundreds of GeV energies, as reported by many experiments, including ATIC-2~\cite{Panov:2006kf, Panov:2009iih}, CREAM~\cite{Ahn:2010gv, Yoon:2017qjx}, PAMELA~\cite{PAMELA:2011mvy}, and AMS-02~\cite{AMS:2015azc, AMS:2015tnn}. Additionally, the model presents a plausible resolution for the relatively low magnitude of local CR anisotropy~\cite{Blasi:2011fm, Ahlers:2016rox} and addresses other related concerns. It is also found that the SDD model exhibited a higher prediction of secondary positrons compared to the conventional diffusive-reacceleration model due to the higher concentration of electrons/positrons in the disk. Therefore, it is pertinent to explore whether the excess can be solely attributed to DM annihilation within the framework of the SDD model.  

In this study, we undertake a quantitative analysis of the AMS-02 results within the context of the SDD model. To prevent potential biases resulting from the preselection of background parameters, a global fitting procedure is employed, simultaneously determining both the background and DM parameters. 
We adopt the Bayesian analysis based on a Markov chain Monte Carlo~\cite{gamermanMarkovChainMonte1997} sampling algorithm to constrain the model parameters.

 Additionally, this investigation employs a new electron/positron production cross-section model developed by Ref.~\cite{Orusa:2022pvp}, which is based on the latest collider data, in order to mitigate biases stemming from hadronic interactions. Furthermore, a charge-sign dependent solar modulation potential is incorporated, since particles of opposite charges explore distinct regions of the solar system~\cite{Potgieter:2013pdj}. To address systematic uncertainties among different detectors, only the most recent CR data provided by the AMS-02 collaboration is utilized~\cite{AMS:2021nhj,AMS:2019rhg,AMS:2019iwo}.

This paper is structured as follows. In Sec.~\ref{sec methology}, we provide a detailed description of the SDD propagation model and the employed methodology utilized to determine the propagation and source parameters, which serve as the foundation for calculating the background electron/positron flux. Additionally, we introduce our setups for DM annihilation/decay within this section. In Sec.~\ref{sec results}, we present the fitting results obtained under different DM setups, accompanied by comparisons with other DM indirect detection results. Finally, we summarize our findings and offer insightful discussions in Sec.~\ref{sec:conclusion}.

\section{Method}\label{sec methology}
\subsection{Description of the global fitting scheme}
The approach employed for the global data fit follows the methodology presented in our prior investigation of the AMS-02 positron fraction results~\cite{Yuan:2013eja,Lin:2014vja}. Initially, parameters associated with propagation in the SSD model are determined through the fitting to the secondary to primary ratios, and these parameters remain fixed throughout the study. Subsequently, the injection spectrum for protons and helium, which are crucial for calculating the secondary $e^{\pm}$ spectrum, is obtained by fitting against the latest AMS-02 proton and helium data~\cite{AMS:2021nhj}. These parameters are also held constant during the fitting process for the lepton data.

In the final step, we perform a fit to the latest AMS-02 lepton data, incorporating both the primary electrons and the electrons/positrons arising from DM annihilation/decay. Notably, we choose to utilize the positron flux $\Phi_{e^+}$, instead of the conventional positron fraction. 
Utilizing $\Phi_{e^+}$ offers a distinct advantage as it is independent of the energy dependence of electrons. The fit also takes into account the combined $e^+ + e^-$ spectrum. It is worth noting that the analysis excludes AMS-02 lepton data with energies below 7.5 GeV, as these measurements are substantially affected by solar modulation effects, as reported by the AMS-02 collaboration~\cite{AMS:2019rhg}.


In each fitting procedure, the Python package cobaya\footnote{\url{https://cobaya.readthedocs.io/en/latest/}}~\cite{Lewis:2002ah, Lewis:2013hha, Torrado:2020dgo} is utilized to implement the Markov Chain Monte Carlo technique, enabling the derivation of posterior probability distributions for the parameters based on observational data. Following Bayes' theorem, the posterior probability of a parameter set denoted as $\vec{\theta}$ with the given observational data is proportional to the product of the likelihood function $\mathcal{L}(\vec{\theta}) \propto\mathrm{exp}(-\chi^2(\vec{\theta})/2)$, which represents the model's fit to the data, and the prior probability $\mathcal{P}(\vec{\theta})$ of the model parameters prior to the current observations. 
The systematic errors in the AMS-02 data is added in quadrature with the statistical errors to get the total errors.
In this study, we adopt flat (constant) prior probabilities for all model parameters within specified ranges, some of which are logarithmic. Detailed information can be found in the provided tables.

\begin{table}[htbp]
\captionsetup{justification=raggedright}
\caption{\label{tab:best}The best-fit values and posterior 95\% range of all parameters in SDD model\footnote{
The prior range of the parameters can be found in Table II in Ref.~\cite{Zhao:2021yzf}}}
\begin{ruledtabular}
\begin{tabular}{ccc}
 Parameter&Best-fit values&posterior 95\% range\\ \hline
 $D_0(10^{28}cm^2s^{-1})$&3.379&[2.986,4.023]\\
 $\delta$&0.583&[0.557,0.608]\\
 $L$(kpc)&4.743&[4.323,5.625]\\
 $V_a$(km/s)&19.718&[17.130,21.706]\\
 $\eta$&-1.299&[-1.518,-1.099]\\
 $\xi$&1.153&[0.965,1.277]\\
 $h$(kpc)&0.468&[0.406,0.515]\\ \hline
 $\chi^2_{\rm{min}}/n_{\rm{dof}}$& 167.55/265 &-\\
\end{tabular}
\end{ruledtabular}
\end{table}

\subsection{CR propagation model}
Within the diffusive halo, the propagation of CRs can be mathematically described by the diffusion equation~\cite{Strong:2007nh}:
\begin{equation}
\label{eqn:prop}
\begin{aligned}
\frac{\partial \psi}{\partial t}= & Q(\mathbf{x}, p)+\nabla \cdot\left(D_{x x} \nabla \psi\right)+\frac{\partial}{\partial p}[p^2 D_{p p} \frac{\partial}{\partial p}(\frac{\psi}{p^2})] \\
& -\frac{\partial}{\partial p}\dot{p} \psi-\frac{\psi}{\tau_f}-\frac{\psi}{\tau_r}\;,
\end{aligned}
\end{equation}
where $Q(x, p)$ represents the CR source term, $\psi = \psi(x, p, t)$ denotes the CR density per unit momentum $p$ at position $\mathbf{x}$, $ \dot{p} \equiv dp/dt$ is the momentum loss rate, and the time scales $\tau_f$ and $\tau_r$ characterize fragmentation processes and radioactive decays, respectively.  In the framework of diffusive-reacceleration, the momentum space diffusion coefficient, $D_{p p}$ is related to the spatial diffusion coefficient $D_{x x}$ through the relation $D_{p p} D_{x x}=4 p^2 v_A^2/\left(3 \delta\left(4-\delta^2\right)(4-\delta)\right)$~\cite{seoStochasticReaccelerationCosmic1994,berezinskiiAstrophysicsCosmicRays1990a}, where $v_A$ is the Alfv\'en velocity. Notably, convection is not considered in this work, as previous studies have suggested that convection may not be necessary\cite{Yuan:2017ozr,Weinrich:2020ftb}. 

In the SDD model, the diffusion coefficient in the vicinity of the Galactic plane is suppressed. Consequently, the associated diffusion coefficient $D_{xx}$ is parameterized as follows:
\begin{subequations}
\begin{equation}
\label{eq:diff}
	D_{xx}(R,z)=aD_0\beta^\eta({\frac{R}{R_0}})^{b\delta}
\end{equation}
\begin{equation}
\label{eq:diff2}
	a=\begin{cases} \xi \,,  \quad& |z|\leq h \\
   1\,,  \quad& |z| > h\end{cases}\,
\end{equation}
\begin{equation}
\label{eq:diff3}
	b=\begin{cases} 0 \,,  \quad& |z|\leq h \\
   1\,,  \quad& |z| > h\end{cases}\;,
\end{equation}
\end{subequations}
where $\xi$ is a free parameter to be determined, $\beta=v/c$ is the particle velocity in natural units, and the factor $\beta^\eta$ describes the effect of the low-energy random-walk process. {
Here, $\eta\ne1$ is implemented in alignment with previous works~\cite{Yuan:2017ozr, Genolini:2019ewc}, serving as a phenomenological parameter employed to improve the model's fit.} The spatial variation of the diffusion coefficient is determined by the scale factors $a$ and $b$. $a$ alters the normalization at the reference rigidity of $R_0=4$~GV, and $b$ adjusts the slope index. The extent of the slow diffusion region is characterized by the parameter $h$. {
Note that the absence of rigidity-dependence for $D_{xx}$ in the disk is
data-driven rather than physically motivated, as elaborated in Appendix B.2 of Ref.~\cite{Zhao:2021yzf}.}

The accurate prediction of secondary $e^{\pm}$ relies heavily on the interstellar medium gas density and the treatment of energy losses. In this study, we employ the 2D default models implemented in GALPROP v56\footnote{Current version available at \url{https://galprop.stanford.edu/}}~\cite{Porter:2021tlr} to characterize the gas density and to account for the energy losses. The numerical solution takes account of the dominant losses, such as the synchrotron losses on the Galactic magnetic field and the inverse Compton losses on the interstellar radiation fields, for $e^{\pm}$ detected at energies exceeding approximately 10 GeV. Additionally, adiabatic, bremsstrahlung, and ionization losses, which impact the prediction at lower energies around a few GeV, are also taken into consideration. The interstellar radiation field model utilized in this study is the default GALPROP one~\cite{Porter:2005qx}.
Synchrotron energy losses are computed based on a regular magnetic field proposed in Ref.~\cite{Pshirkov:2011um}, along with a random component modeled according to Ref.~\cite{Sun:2007mx}.

We modify the default GALPROP code to enable the consideration of a spatially dependent diffusion coefficient. Following the methodology outlined in a previous investigation~\cite{Zhao:2021yzf}, the carbon flux, $\ce{^{10}Be}/\ce{^{9}Be}$ ratio, B/C ratio, and Be/B ratio serve as constraints for the parameters within the framework of the SDD model. The posterior distributions of all parameters are found to exhibit favorable behavior, successfully reproducing the nucleon fluxes and ratios. Table~\ref{tab:best} presents the posterior mean values and associated 95\% confidence intervals for the model parameters.

Before reaching Earth, local interstellar CRs undergo solar modulation effects within the heliosphere. Traditional approaches have relied on the force field approximation (FFA)~\cite{Gleeson:1968zza}, employing a single solar modulation potential $\phi$, to account for this phenomenon. However, this approximation assumes a spherical symmetry and overlooks the drift effect caused by the heliospheric magnetic field configuration. Recent studies employing realistic simulations and solving Parker's transport equation demonstrated that this drift effect induces charge-sign dependent behavior in CR spectra~\cite{Potgieter:2013pdj,Potgieter:2014pka,Cholis:2015gna}. Consequently, employing the FFA with a single modulation potential $\phi$ proves insufficient in accurately describing all CR particles. In this study, we incorporate the FFA to account for solar modulation effects while considering two modulation potentials $\phi_{e^+}$ and $\phi_{e^-}$  for positrons and electrons, respectively.

\subsection{CR injection sources}
The detected CR $e^{\pm}$ particles consist of three distinct components: the primary electrons originating from supernova remnants, the  secondary electrons and positrons arising from spallation processes of primary nuclei within the interstellar medium, and the $e^{\pm}$ pairs generated by exotic sources like DM annihilation or decay. The combined impact of the primary and secondary components is regarded as the background. In this section, we provide a comprehensive overview of the injection CR $e^{\pm}$ spectra pertaining to both the background and DM annihilation/decay sources.

\subsubsection{The \texorpdfstring{$e^{\pm}$}{e} background spectrum}
The distribution of regular CR sources is expected to align with the radial profile of supernova remnants around the Galactic disk, which can be described as follows:
\begin{equation}
f(r, z)=\left(\frac{r}{r_{\odot}}\right)^a \exp \left(-b \cdot \frac{r-r_{\odot}}{r_{\odot}}\right) \exp \left(-\frac{|z|}{z_s}\right)\;,
\end{equation}
where $r_{\odot}= 8.5$ kpc represents the distance between the Sun and the Galactic center, and $z_s \approx 0.2$ kpc denotes the characteristic height of the Galactic disk. In accordance with Ref.~\cite{Trotta:2010mx}, we adopt the parameters $a = 1.25$ and $b = 3.56$, which are adjusted to match the observed $\gamma$-ray gradient. 
Regarding the energy dependence of the source term, the shock acceleration theory~\cite{Achterberg:2001rx} predicts that the injection spectra of primary CRs follow a power-law relation in rigidity. {
Additionally, we introduce a low-energy break $R_{\mathrm{br}}$,  serving as a phenomenological parameter to account for the observed low-energy spectral bumps observed in all nuclei.}
\begin{equation}
q^i(R) \propto\left\{\begin{array}{cc}
\left(R / R^i_{\mathrm{br}}\right)^{-\nu^i_0} , & R\leq R^i_{\mathrm{br}} \\
\left(R / R^i_{\mathrm{br}}\right)^{-\nu^i_1} , & R>R^i_{\mathrm{br}}\;\qquad ,
\end{array}\right.
\end{equation}
where $i$ denotes the species of the nuclei.
The spectral indices below and above the break are denoted as $\nu^i_0$ and $\nu^i_1$, respectively. 
\begin{table}[htbp]
\captionsetup{justification=raggedright}
\caption{\label{tab:inj}The prior range, best-fit values and posterior 95\% range of the proton and helium injection spectra.}
\begin{ruledtabular}
\begin{tabular}{lll}
 Parameter&Prior range&posterior 95\% range\\ \hline
 $\nu_0^{p}$&[1.4,2.8]&$2.17^{+0.02}_{-0.02}   $\\
 $\nu_1^{p}$&[1.8,3.0]&$2.428^{+0.004}_{-0.005}$\\
 $R^p_{\mathrm{br}}$(GV)&[5.0,25.0]&$13.7^{+1.1}_{-1.1}        $\\
 $A_p$\footnote{The normalization of post-propagated proton flux at 100 GeV}&[2.6,5.4]&$4.135^{+0.018}_{-0.019}   $\\
 $\nu_1^{\mathrm{He}}$\footnote{$\nu_0^{\mathrm{He}}$ and $R^{\mathrm{He}}_{\mathrm{br}}$ are fixed at $2.0$ and $3.1$ GV, respectively.}&[1.5,3.0]&$2.377^{+0.005}_{-0.005}$\\
 Abund. He\footnote{Source abundance of the helium, when one fixes the abundance of the proton {\color{red}to} $1.06 \times 10^6$ at 100 GeV/n.}&[8.6,11.8]&$9.88^{+0.06}_{-0.05}   $\\
 $\phi_{\mathrm{nuc.}}$(MV)&[0.1,0.9]&$0.709^{+0.029}_{-0.028}   $\\ 
\end{tabular}
\end{ruledtabular}
\end{table}

\begin{figure}[htbp]
\includegraphics[width=0.48\textwidth]{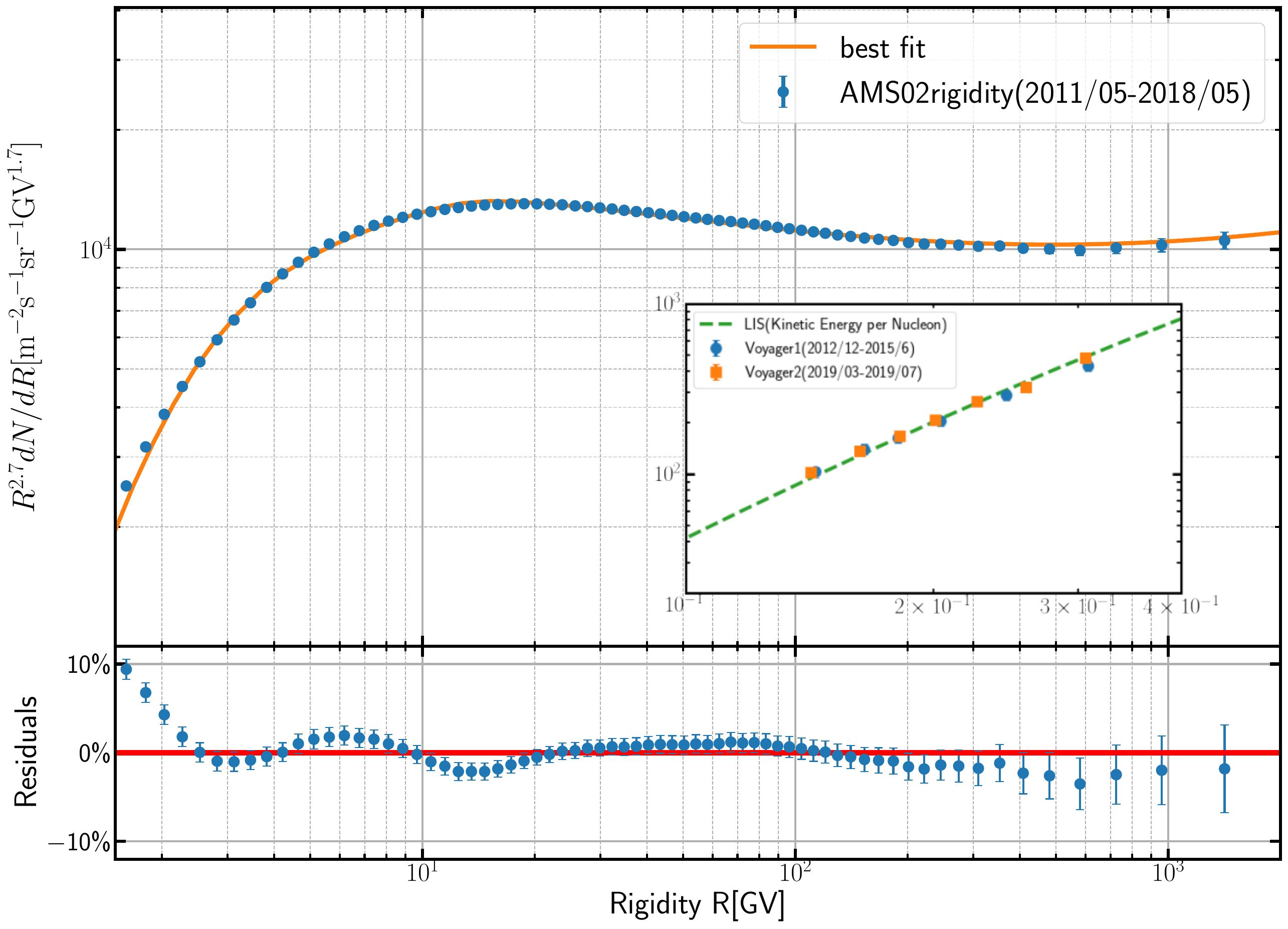}\\
\includegraphics[width=0.48\textwidth]{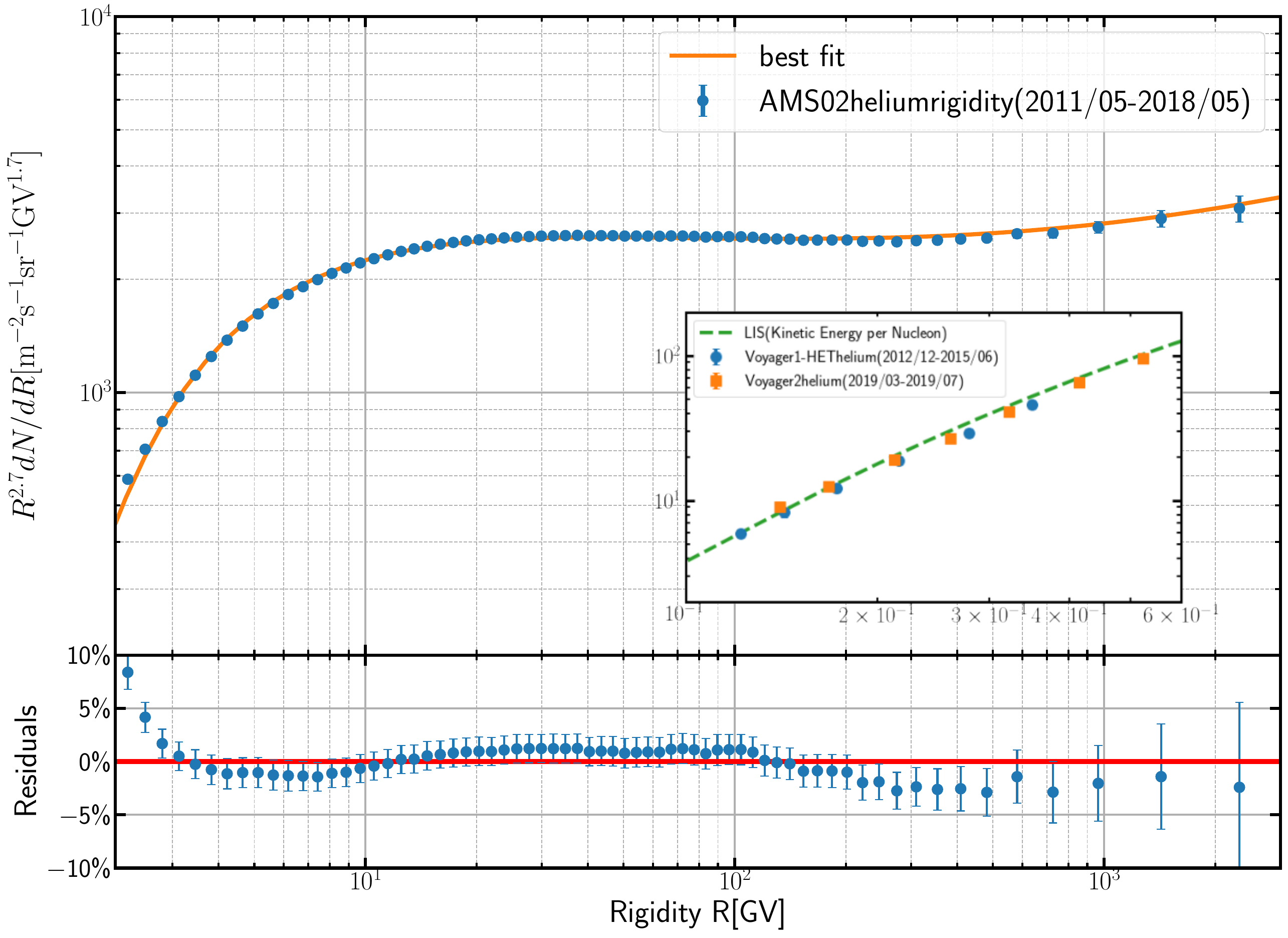}\\
\captionsetup{justification=raggedright}
\caption{
The fluxes of the proton (Top) and helium (Bottom) after solar modulation for the parameters shown in Table~\ref{tab:inj} and the corresponding residuals, compared with the latest data from AMS-02~\cite{AMS:2021nhj}. The inset is the local interstellar spectrum in units of kinetic energy per nucleon compared with Voyager~\cite{stoneVoyagerObservesLowEnergy2013} data. \label{fig:proton}}
\end{figure}

Upon adopting the best-fit propagation parameters outlined in Table~\ref{tab:best}, we further constrain the injection parameters based on the proton and helium flux data. The resulting injection parameters can be found in Table~\ref{tab:inj}. In Fig.~\ref{fig:proton}, 
we present a comparison between the best-fitting spectrum and the corresponding observational data. Notably, our calculated proton and helium fluxes, both before and after solar modulation, exhibit excellent agreement with observations across the entire energy range.

The determination of secondary electrons and positrons is a straightforward process given the known injection spectrum and propagation parameters. In this study, we adopt a parameterization for the production cross section of secondary leptons as presented in Ref.~\cite{Orusa:2022pvp}, incorporating the latest collider data from experiments such as NA49~\cite{NA49:2005qor,NA49:2009wth} and NA61~\cite{NA61SHINE:2017fne}. We also introduce a renormalization parameter $c_{e^{\pm}}$ to account for uncertainties arising from factors including the $e^{\pm}$ production cross section, enhancement factor from heavier nuclei, and uncertainties in propagation {\color{red} as ~\cite{Yuan:2013eja,Lin:2014vja,Cheng:2016slx,Wang:2018pcc,Ding:2021zzg}}.
It is also important to note that these uncertainties may not be accurately captured by a constant factor, as they likely possess an energy-dependent nature. The utilization of this constant factor $c_{e^{\pm}}$ is merely an approximation employed for the data fitting purpose.

Regarding the primary electron injection spectrum, we assume a broken power-law relation in rigidity, featuring a low-energy break suggested by synchrotron observations~\cite{Strong:2011wd,Bringmann:2011py,Orlando:2013ysa}. We fix the position of the break at 5 GV and the spectral index below the break at 1.5, 
since we do not include data points below 7.5 GV.

In summary, the free parameters governing the background electron and positron spectra are as follows:
\begin{equation*}
    \boldsymbol\theta = \{A_{e},\:\nu_1,\:\phi_{e^-},\:\phi_{e^+},\:c_{e^{\pm}}\}\;,
\end{equation*}
where $A_{e}$ is the post-propagated normalization flux of primary $e^-$ at $25$ GeV, and $\nu_1$ stands for the spectral index above the spectral break. The solar modulation potentials for the electrons and positrons are denoted by $\phi_{e^-}$ and $\phi_{e^+}$, respectively, and $c_{e^{\pm}}$ represents the rescaling factor for secondary $e^{\pm}$.

\begin{figure*}[htbp]
    \begin{center}
        \subfloat{\includegraphics[width=0.45\textwidth]{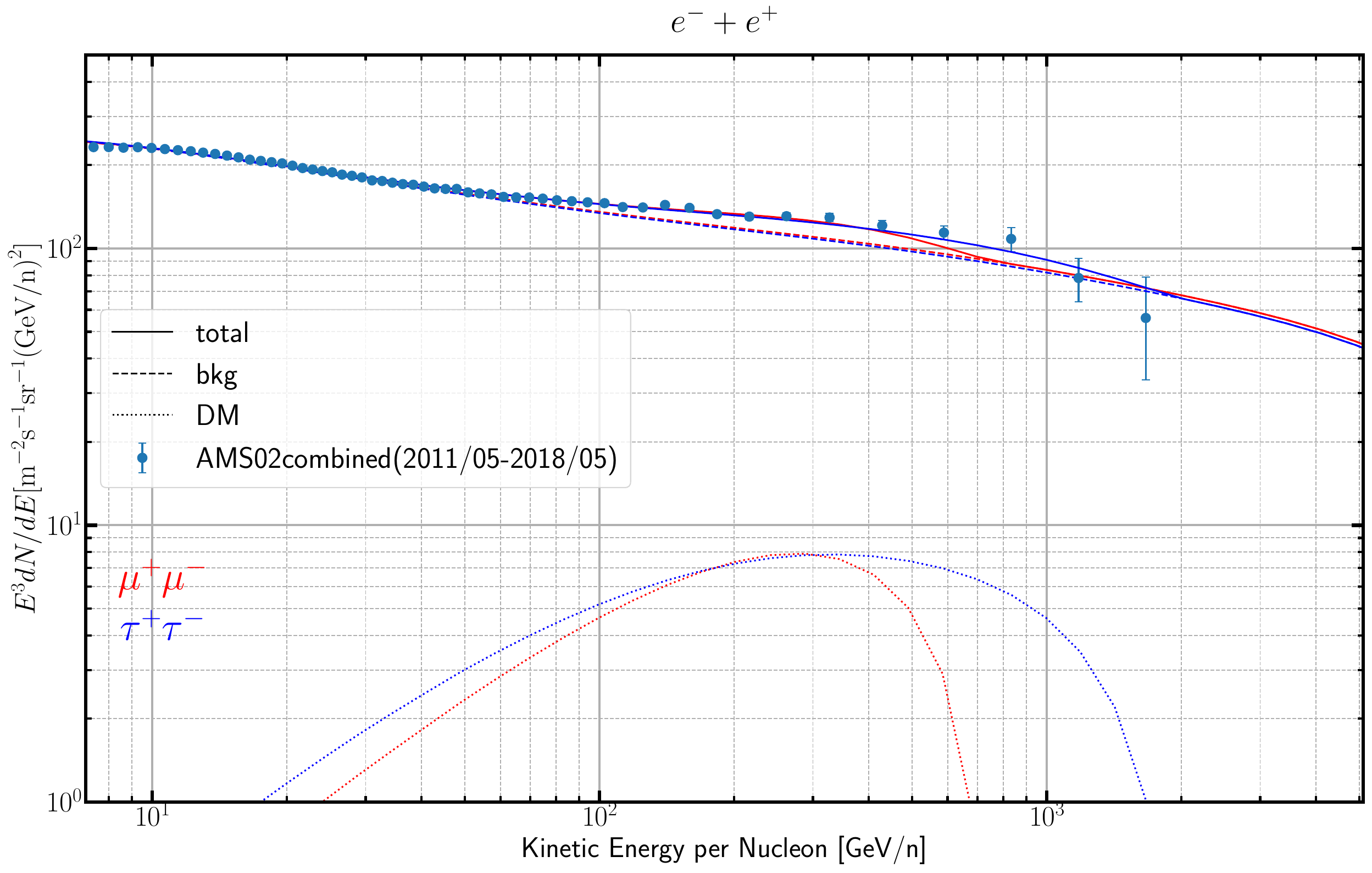}} \hskip 0.03\textwidth
        \subfloat{\includegraphics[width=0.45\textwidth]{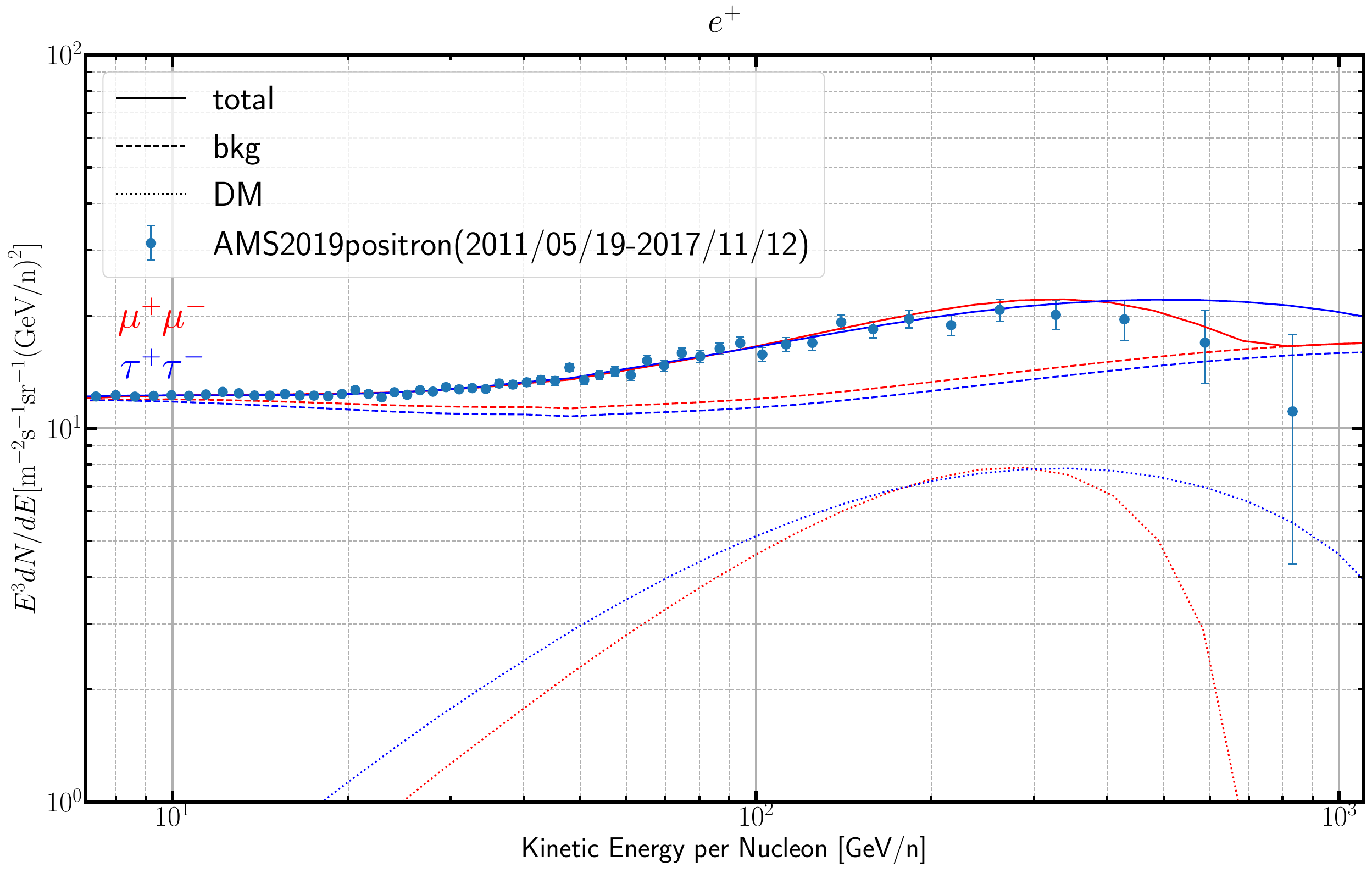}}
    \end{center}
    \captionsetup{justification=raggedright}
    \caption{The expected spectra of the best-fit results for the case of DM annihilation. The left panel shows the total $e^{\pm}$ spectra compared with the AMS-02~\cite{AMS:2019iwo} data,
    while the right panel shows the $e^+$ spectrum alone with AMS-02~\cite{AMS:2019rhg} results. 
    The dashed, dotted, and solid lines represent the backgrounds, DM contributions, and total results, respectively.}\label{fig:ann.}
\end{figure*}
\subsubsection{\texorpdfstring{$e^{\pm}$}{e} from DM annihilations}
\begin{table*}[htbp]
\captionsetup{justification=raggedright}
\caption{The prior ranges, best-fit values, mean values, and posterior 95\% range of the model parameters for DM annihilation. The number of data points for $e^+$ and $e^+ + e^-$ are 54 and 56, respectively\label{table:ann.}.}
\begin{tabular}{lcccccc}
\hline
\multirow{2}{*}{}               & \multirow{2}{*}{Prior Range} & \multicolumn{2}{c}{$\tau^+\tau^-$}                           &  & \multicolumn{2}{c}{$\mu^+\mu^-$}                             \\ \cline{3-4} \cline{6-7} 
                                &                              & Best                         & Mean                          &  & Best                         & Mean                          \\ \hline
log($A_e$\footnote{Post-propagated normalization flux of $e^-$ at 25 GeV in unit 
$\mathrm{cm}^{-2} \mathrm{s}^{-1} \mathrm{sr}^{-1} \mathrm{MeV}^{-1}$})                      & {[}-10.5,-7.5{]}             & -8.947                      & $-8.946^{+0.009}_{-0.009}$ &  & -8.946                      & $-8.945^{+0.009}_{-0.009}$ \\
$\nu_1$                         & {[}1.5,4.0{]}                & 2.83                        & $2.83^{+0.02}_{-0.02}   $  &  & 2.83                       & $2.83^{+0.02}_{-0.02}   $  \\
$\phi_{e^-}$/GV                 & {[}0.1,1.8{]}                & 0.466                        & $0.476^{+0.071}_{-0.069}   $  &  & 0.478                        & $0.480^{+0.069}_{-0.067}   $  \\
$\phi_{e^+}$/GV                 & {[}0.1,1.8{]}                & 0.639                        & $0.636^{+0.080}_{-0.083}   $  &  & 0.724                        & $0.728^{+0.056}_{-0.055}   $  \\
$c_{e^{\pm}}$                   & {[}0.25,4.0{]}               & 1.71                        & $1.71^{+0.08}_{-0.08}   $  &  & 1.80                        & $1.81^{+0.05}_{-0.05}   $  \\ \hline
log($m_{\mathrm{DM}}$/GeV)      & {[}1.0,5.5{]}                & 3.30                         & $3.31^{+0.13}_{-0.11}      $  &  & 2.87                        & $2.88^{+0.09}_{-0.08}   $  \\
log($\langle\sigma v\rangle\footnote{In unit $\mathrm{cm}^3\mathrm{s}^{-1}$}$) & {[}-28.0,-22.0{]}            & -23.04                       & $-23.02^{+0.17}_{-0.15}    $  &  & -23.96                       & $-23.95^{+0.14}_{-0.13}    $  \\ \hline
$\chi^2_{e^+}$                  & -                            & 35.0                     & -                             &  & 37.0                    & -                             \\
$\chi^2_{e^{\pm}}$              & -                            & 50.2                     & -                             &  & 50.1 & -                             \\
$\chi^2_{\mathrm{tot.}}$/DoF    & -                            & 85.2/103 & -                             &  & 87.1/103 & -                             \\ \hline
\end{tabular}
\end{table*}
The extensive DM halo surrounding the Milky Way provides a distinctive opportunity to explore the potential non-gravitational interactions between the DM and standard model particles~\cite{profumoIntroductionParticleDark2017}.  If such interactions exist, they could give rise to the production of CRs, presenting an unconventional CR source. Specifically, for CR electrons and positrons, the source term arising from DM annihilation/decay can be expressed as follows:
\begin{equation}
\begin{split}
Q^{\mathrm{ann.}}_{\mathrm{DM}}(\vec{r}, E)&=\frac{1}{2} \left(\frac{\rho_{\mathrm{DM}}(\vec{r})}{m_{\mathrm{DM}}}\right)^2\langle\sigma v\rangle \sum_k B_k \frac{\mathrm{d} N_{e^{ \pm}}^k}{\mathrm{d} E}\;,\\
Q^{\mathrm{dec.}}_{\mathrm{DM}}(\vec{r}, E)&= \left(\frac{\rho_{\mathrm{DM}}(\vec{r})}{m_{\mathrm{DM}}}\right)\frac{1}{\tau} \sum_k B_k \frac{\mathrm{d} N_{e^{ \pm}}^k}{\mathrm{d} E} \;,
\end{split}
\end{equation}
where the factor 1/2 corresponds to the DM particle being scalar or Majorana fermion, $m_{\mathrm{DM}}$ denotes the mass of the DM particle, $\langle\sigma v\rangle$ represents the thermally averaged DM annihilation cross section in the case of DM annihilation, and $\tau$ stands for the DM lifetime in the case of DM decay. The $e^{\pm}$ production spectrum per annihilation/decay to final state $k$ with the branching ratio $B_k$, obtained from the PPPC 4 DM ID~\cite{Cirelli:2010xx}, is denoted by $\mathrm{d} N_{e^{ \pm}}^k/\mathrm{d} E$. The DM density profile in the Milky Way  $\rho_{\mathrm{DM}}(r)$ is assumed to follow the Navarro-Frenk-White density profile~\cite{Navarro:1996gj}:
\begin{equation}
\rho(\vec{r})=\frac{\rho_s}{\left(r / r_s\right)\left(1+r / r_s\right)^2}\;,
\end{equation}
where $r_s = 20$ kpc and $\rho_s = 0.35$ GeV have been selected, resulting in a local DM density of 0.4 GeV $\text{cm}^{-3}$. This choice of parameters is in agreement with the latest constraints derived from the Galactic rotation curve~\cite{Karukes:2019jxv, Benito:2019ngh}. Alternative density profiles, such as the Einasto~\cite{Graham:2005xx,Navarro:2008kc} or Burkert~\cite{Burkert:1995yz, Salucci:2000ps} profiles, are not taken into account, as they yield similar $e^{\pm}$ spectrum at Earth.

The propagation of the DM induced $e^{\pm}$ is simulated using GALPROP, utilizing the same configuration as the background $e^{\pm}$, ensuring a unified treatment. The free parameters for DM annihilation include the DM particle mass, $m_{\mathrm{DM}}$ and the thermally
averaged DM annihilation cross section $\langle\sigma v\rangle$.  Conversely, in the case of decay, the parameters of interest are the mass, $m_{\mathrm{DM}}$, and the lifetime $\tau$.

\section{RESULTS \& DISCUSSION}\label{sec results}
\subsection{DM annihilation}
\begin{figure*}
    \begin{center}
        \subfloat{\includegraphics[width=0.45\textwidth]{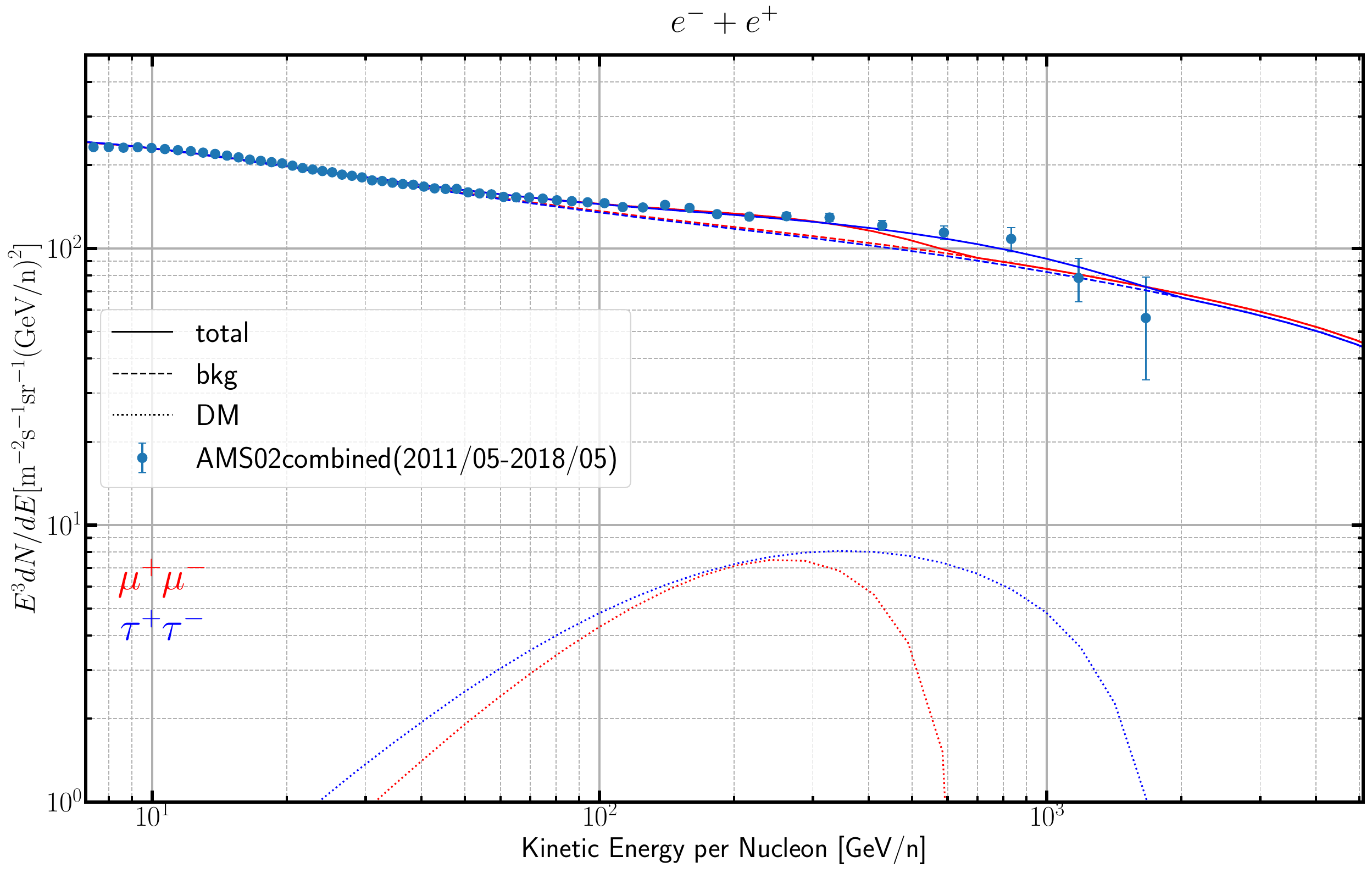}} \hskip 0.03\textwidth
        \subfloat{\includegraphics[width=0.45\textwidth]{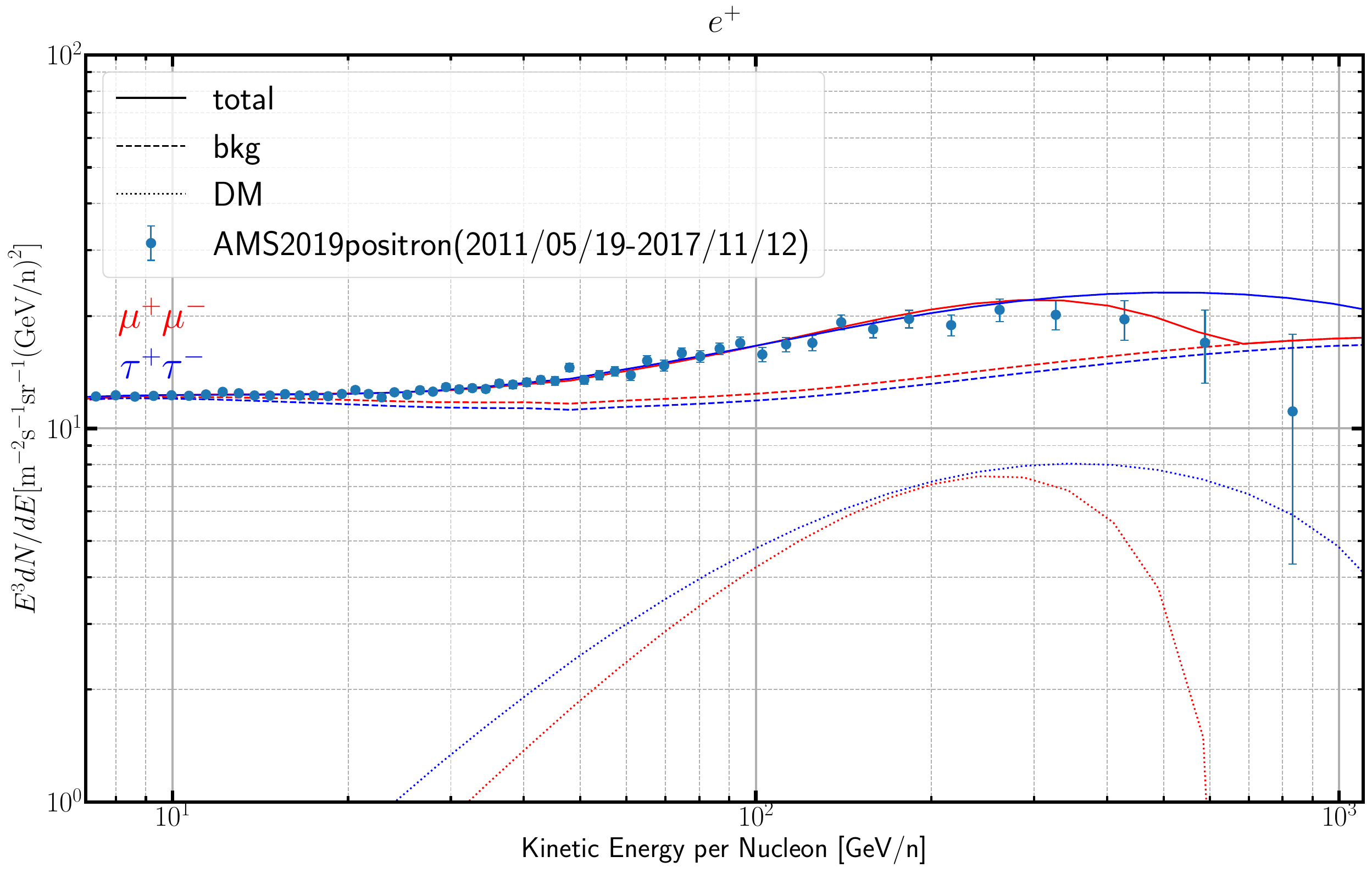}}
    \end{center}
    \captionsetup{justification=raggedright}
    \caption{The same as Fig.~\ref{fig:ann.}, but for DM decay.}\label{fig:dec.}
\end{figure*}

\begin{table*}[]
\captionsetup{justification=raggedright}
\caption{The prior ranges, best-fit values, mean values, and posterior 95\% range of the model parameters for DM decay. The number of data points for $e^+$ and $e^+ + e^-$ are 54 and 56, respectively\label{tab:dec.}.}
\begin{tabular}{lcccccc}
\hline
\multirow{2}{*}{}               & \multirow{2}{*}{Prior Range} & \multicolumn{2}{c}{$\tau^+\tau^-$}                           &  & \multicolumn{2}{c}{$\mu^+\mu^-$}                             \\ \cline{3-4} \cline{6-7} 
                                &                              & Best                         & Mean                          &  & Best                         & Mean                          \\ \hline
log($A_e$\footnote{Post-propagated normalization flux of $e^-$ at 25 GeV in unit 
$\mathrm{cm}^{-2} \mathrm{s}^{-1} \mathrm{sr}^{-1} \mathrm{MeV}^{-1}$})                      & {[}-10.5,-7.5{]}             & -8.943                      & $-8.945^{+0.009}_{-0.009}$ &  & -8.944                      & $-8.945^{+0.009}_{-0.009}$ \\
$\nu_1$                         & {[}1.5,4.0{]}                & 2.84                        & $2.84^{+0.02}_{-0.02}   $  &  & 2.84                        & $2.83^{+0.02}_{-0.02}   $  \\
$\phi_{e^-}$/GV                 & {[}0.1,1.8{]}                & 0.500                        & $0.481^{+0.072}_{-0.068}   $  &  & 0.494                        & $0.482^{+0.071}_{-0.068}   $  \\
$\phi_{e^+}$/GV                 & {[}0.1,1.8{]}                & 0.692                       & $0.687^{+0.072}_{-0.071}   $  &  & 0.765                        & $0.779^{+0.056}_{-0.055}   $  \\
$c_{e^{\pm}}$                   & {[}0.25,4.0{]}               & 1.79                        & $1.78^{+0.07}_{-0.07}   $  &  & 1.87                        & $1.88^{+0.05}_{-0.05}   $  \\ \hline
log($m_{\mathrm{DM}}$/GeV)      & {[}1.0,5.5{]}                & 3.58                         & $3.55^{+0.12}_{-0.11}      $  &  & 3.13                        & $3.13^{+0.08}_{-0.08}   $  \\
log($\tau$/s) & {[}20.0,30.0{]}            & 26.69                       & $26.70^{+0.06}_{-0.06}  $  &  & 27.20                       & $27.22^{+0.06}_{-0.06}  $  \\ \hline
$\chi^2_{e^+}$                  & -                            & 40.1                     & -                             &  & 40.2                     & -                             \\
$\chi^2_{e^{\pm}}$              & -                            & 44.3                     & -                             &  & 49.9 & -                             \\
$\chi^2_{\mathrm{tot.}}$/DoF    & -                            & 84.4/103 & -                             &  & 90.1/103 & -                             \\ \hline
\end{tabular}
\end{table*}

We summarize the fitting results in Table~\ref{table:ann.} and show the corresponding spectra compared with data in Fig.~\ref{fig:ann.}. 
We find that both the $\mu^+\mu^-$ and $\tau^+\tau^-$ annihilation channels yield reasonable fits to the AMS-02 data, with the reduced $\chi^2$ values smaller than one\footnote{
Note that the very low $\chi^2$ values could be attributed to the absence of the correlations in systematic errors. While the AMS-02 collaboration does not provide these correlations, there are some theoretical attempts to construct them~\cite{Derome:2019jfs,Cuoco:2019kuu,Heisig:2020nse}. However, their validity remains unconfirmed.}. 
The $\tau^+\tau^-$ channel provides slightly better agreement with the total $e^{\pm}$ spectrum, exhibiting a more gradual decline at high energies compared to $\mu^+\mu^-$ and resulting in improved concordance with three data points in the $428.5-832.3$ GeV/n range. 
Since the high-energy end of the electron spectrum suffers severe energy losses, making it more easily influenced by nearby sources~\cite{Atoian:1995ux, Kobayashi:2003kp, Profumo:2008ms, Hooper:2008kg, Yuksel:2008rf, Malyshev:2009tw, Blasi:2009hv, DiMauro:2015jxa, Fang:2016wid,Fang:2017tvj}, it is acceptable that the $\mu^+\mu^-$ channel does not well reproduce these high-energy data points.

Regarding the positron spectrum, although both channels yield comparable results in terms of $\chi^2$ statistics, the $\tau^+\tau^-$ channel appears to generate an excess of positrons at the high-energy end because it exhibits a more gradual decline in its spectrum at high energies. It is important to note that this finding contradicts previous studies conducted with the standard propagation model using older AMS-02 data from the 2011-2015 period~\cite{AMS:2014bun}, where the $\tau^+\tau^-$ channel was strongly favored over the $\mu^+\mu^-$ channel. This discrepancy can be attributed to two factors. Firstly, the new AMS-02 data (2011-2017)\cite{AMS:2019rhg} extends to higher energies compared to the previous data, and for the first time, a spectral cut is observed. Secondly, the SDD model assumes that CRs propagating in the slow disk could contribute a harder component at high energies, which give rise to the positron flux compared to the standard propagation model, thereby compensating for the hard spectrum of the $\mu^+\mu^-$ channel.

It is important to note that due to the enhanced production of secondary $e^{\pm}$ in the SDD model, the rescaling factors required to reproduce the data for secondary $e^{\pm}$ are significantly smaller compared to previous works~\cite{Lin:2014vja,Wang:2018pcc} utilizing the standard diffusion-reacceleration model. Specifically, the rescaling factors for secondary $e^{\pm}$ in the SDD model are found to be 1.694 and 1.783 for the $\tau^+\tau^-$ and $\mu^+\mu^-$ channels respectively, in contrast to previous studies where values of $c_{e^{\pm}}$ around 3 are typically employed. Another noteworthy consequence of the SDD model's capability to generate a greater number of secondary $e^{\pm}$ is that the contribution of positrons from the additional source never surpasses that of the secondary component. This finding stands in contrast to the results obtained using the standard propagation model, where the contribution of positrons from the additional source dominates over the secondary component above tens of GeV.

As for the solar modulation potential, we obtain reasonable results within the range of $0.5-0.7$ GV, consistent with the values obtained from the fitting to nuclei data. This represents a significant improvement compared to the results obtained using the standard propagation model, where the modulation potentials are typically larger than 1 GV, {
which is at odds with the potentials reconstructed from the CR data and neutron monitor observations in the AMS-02 periods~\cite{Ghelfi_2017,Zhu:2018jbk,Koldobskiy:2023prp}.}
An interesting observation regarding the modulation potential is that the potential for positively charged particles $\phi_{e^+}$, is larger than that for negatively charged particles $\phi_{e^-}$, by approximately $0.1$ GV. This finding is in agreement with the case of the CR proton and antiproton~\cite{Lin:2019ljc}. 

\subsection{DM decay}
\begin{figure*}
    \begin{center}
        \subfloat{\includegraphics[width=0.45\textwidth]{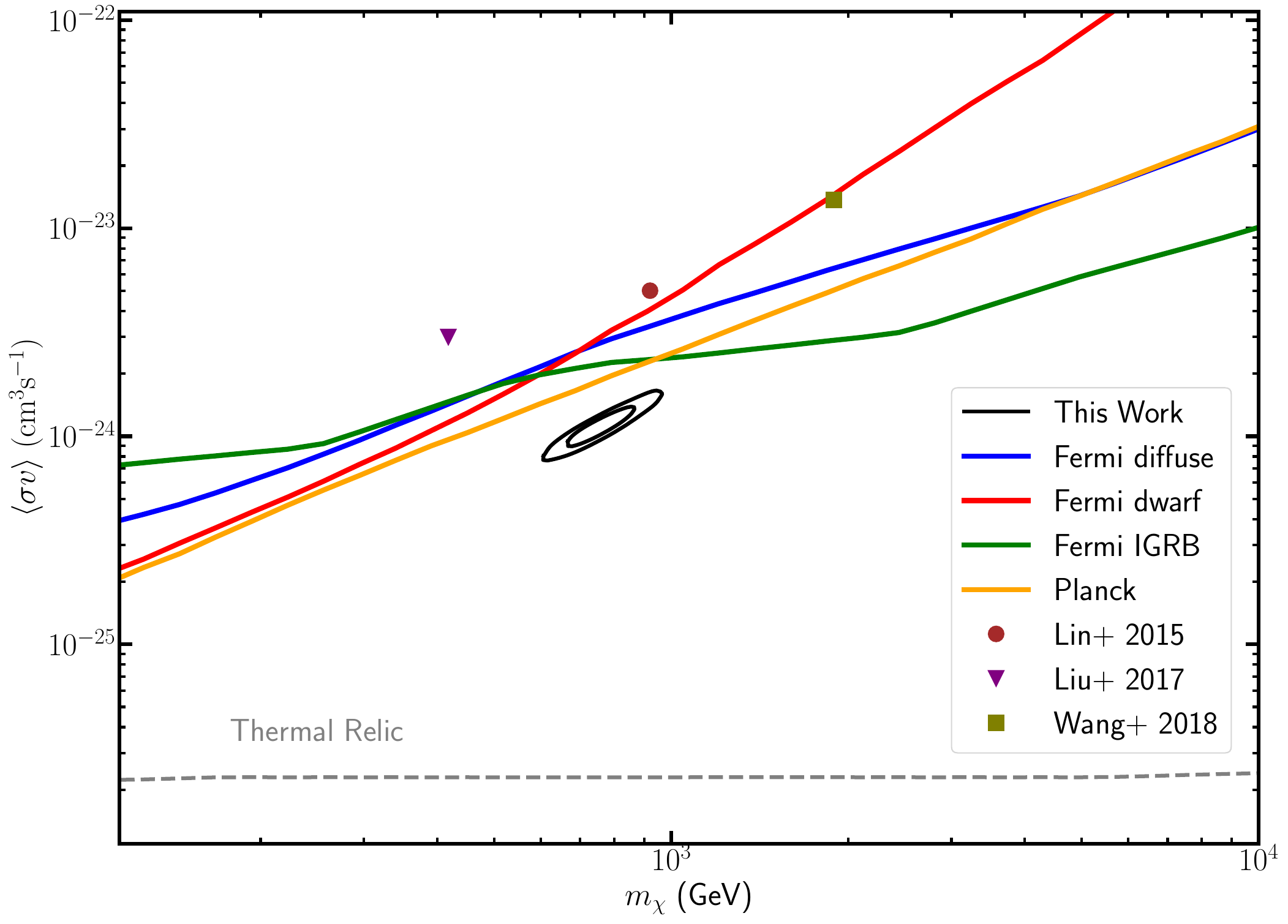}} \hskip 0.03\textwidth
        \subfloat{\includegraphics[width=0.45\textwidth]{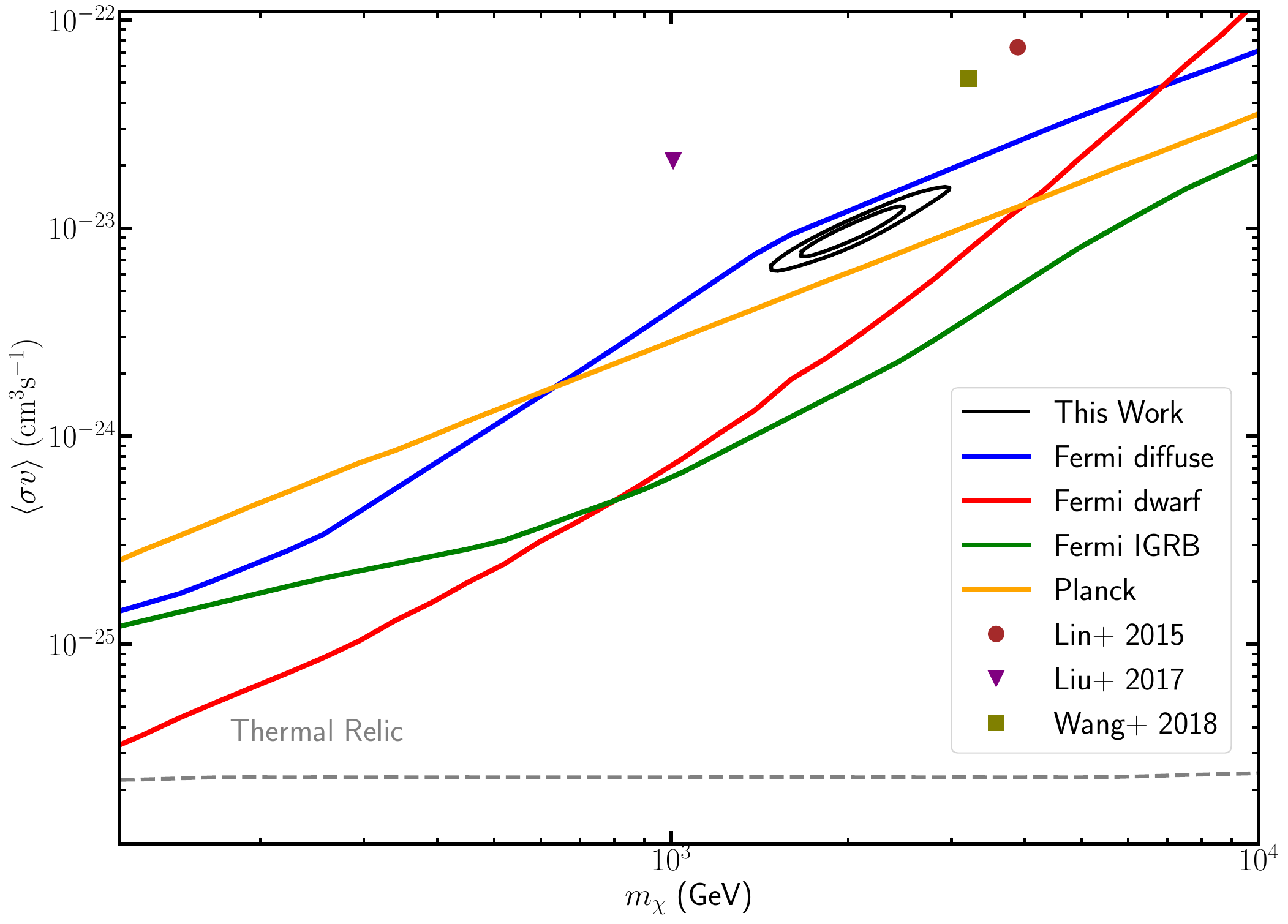}}
    \end{center}
\end{figure*}
\begin{figure*}
    \begin{center}
        \subfloat{\includegraphics[width=0.45\textwidth]{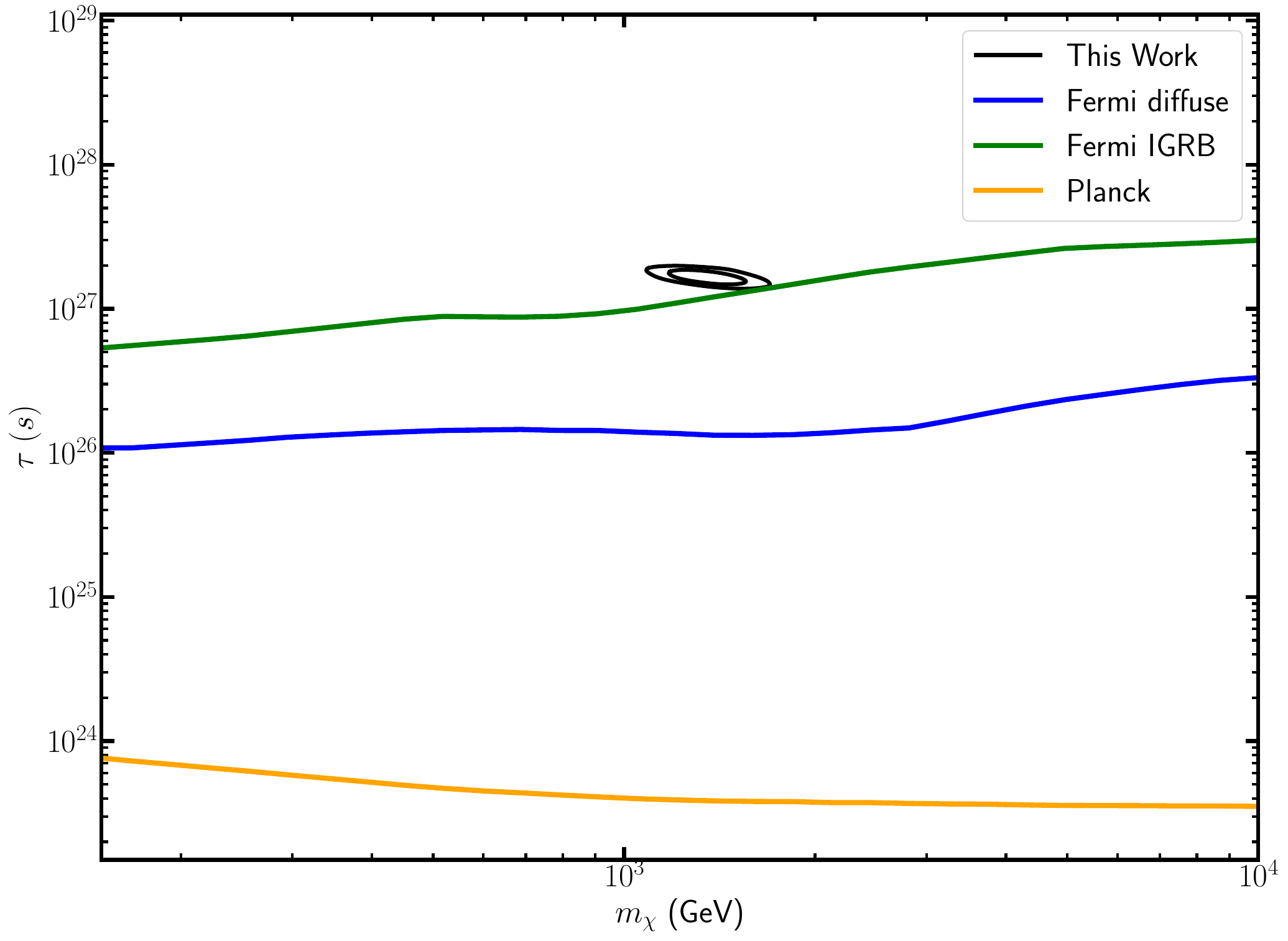}} \hskip 0.03\textwidth
        \subfloat{\includegraphics[width=0.45\textwidth]{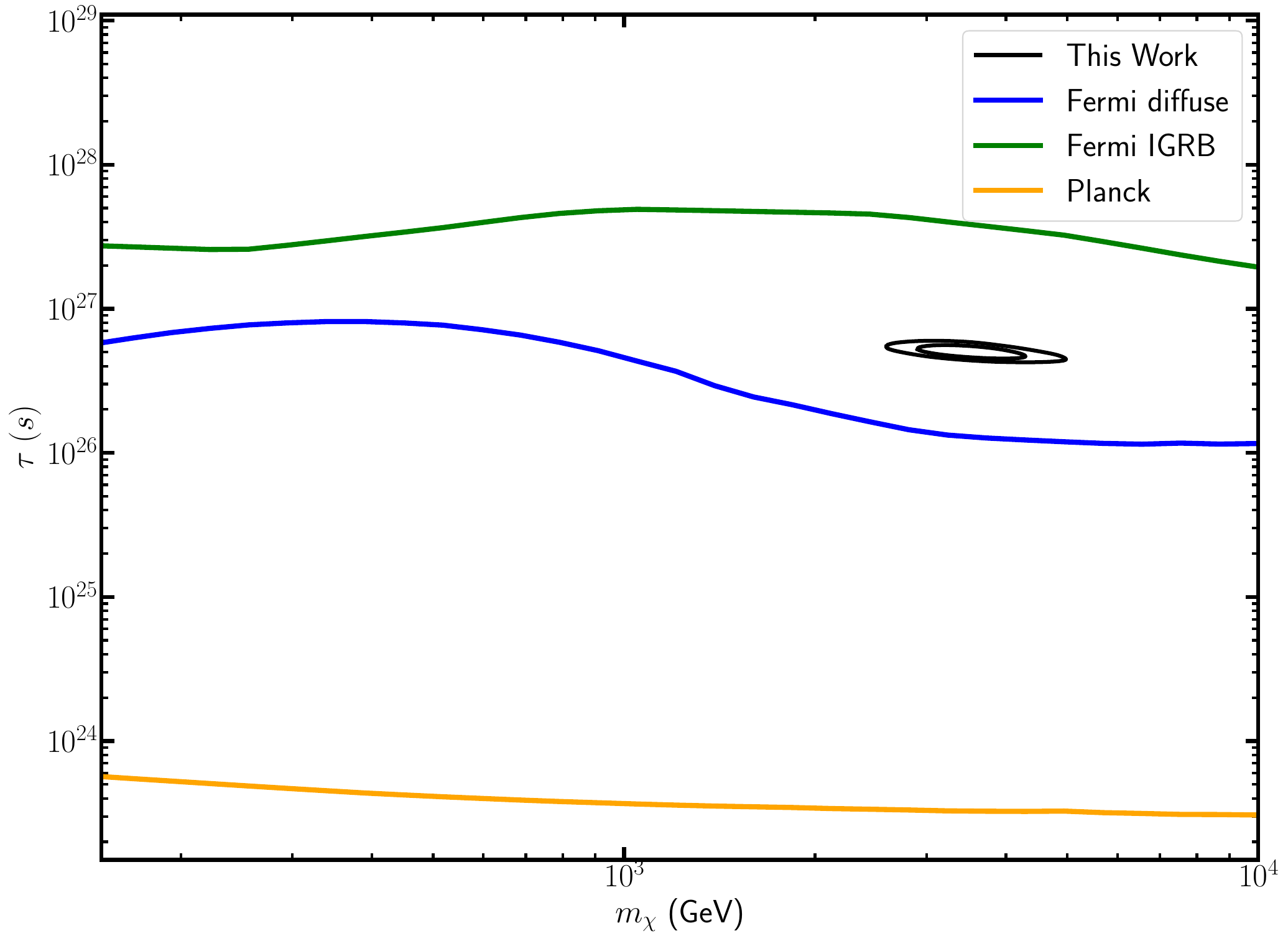}}
    \end{center}
    \captionsetup{justification=raggedright}
    \caption{ 
    $1\sigma$ and $2\sigma$ confidence regions in the $m_{\mathrm{DM}}-\langle\sigma v\rangle$ plane, together with the exclusion lines from the Fermi observations of dwarf galaxies~\cite{Fermi-LAT:2016uux}, diffuse gamma-rays in the Milky Way halo~\cite{Fermi-LAT:2012pls}, IGRB~\cite{Liu:2016ngs}, and the Planck CMB observations~\cite{Slatyer:2015jla}. The left and right panels show the results for the $\mu^+\mu^-$ and $\tau^+\tau^-$ channels, respectively. The upper and lower panels correspond to DM annihilation and decay, respectively. The fitting results to the AMS-02 observations from some previous analyses~\cite{Lin:2014vja,Liu:2016ngs,Wang:2018pcc} are shown as \textit{colored points}. Additionally, the thermal relic cross section from Ref.~\cite{Steigman:2012nb} is illustrated as a dashed gray line in the upper panels.}\label{fig:dm}
\end{figure*}
Although DM particles are generally assumed to be stable, the potential for DM decay cannot be completely disregarded, particularly if the decay process unfolds over a timescale exceeding the age of the universe~\cite{Ibarra:2013cra}. If such decay occurs, the resulting products within the Milky Way halo could potentially account for the observed excess of positrons~\cite{Nardi:2008ix,Yin:2008bs,Meade:2009iu}.

In this section, we perform fits to the AMS-02 total $e^{\pm}$ spectrum and $e^+$ spectrum under the assumption of DM decay. A summary of the fitting outcomes is presented in Table~\ref{tab:dec.}, and visual comparison with the data is depicted in Fig.~\ref{fig:dec.}. Similar to the case of DM annihilation, both the $\mu^+\mu^-$ and $\tau^+\tau^-$ decay channels yield satisfactory fits to the AMS-02 data, as indicated by reduced $\chi^2$ values smaller than one. Furthermore, the resulting parameters are well constrained and demonstrate reasonable values similar to those obtained in the annihilation scenario. 

Specifically, the rescaling factors for secondary $e^{\pm}$ in both the two decay channels are approximately 1.8, highlighting the production of leptons in the high-energy region facilitated by the SDD model. Furthermore, the solar modulation potentials for the electron and positron are determined to be approximately 0.7 and 0.5 GV, respectively, emphasizing their distinct impacts on the CR flux. Importantly, it should be noted that the contribution from the additional source never surpasses that of the secondary origins, affirming the dominance of the secondary component under the SDD model. 

The $\tau^+\tau^-$ channel gives a better fit to the total $e^{\pm}$ spectrum, because the resulting spectrum from decay is slightly harder than that from annihilation, and this is favored by the AMS-02 data. It is worth noting, however, that caution is warranted in interpreting this improvement in the $\chi^2$ value as a definitive physical significance. As previously discussed, the presence of nearby electron sources can significantly influence the high-energy end of the electron spectrum. Therefore, while the $\tau^+\tau^-$ channel yields the best statistical agreement with the data, further investigations and considerations are required to verify the true physical implications of this result.

Regarding the positron spectrum, although the $\tau^+\tau^-$ channel provides the best fit in terms of the reduced $\chi^2$ value, it is evident that it also overestimates the positron flux at the highest energies accessible to AMS-02, similar to the $\tau^+\tau^-$ annihilation scenario. Therefore, with the ongoing enhancement in the precision of positron data, it is possible that the $\tau^+\tau^-$ channel may be less favored in future investigations.

\subsection{Comparison with other constraints}
The DM annihilation/decay hypothesis as an explanation for the positron excess is subject to strong constraints imposed by various other observations, including the CMB~\cite{Slatyer:2015jla} and $\gamma$-ray measurements, which encompass observations such as dwarf galaxy gamma-rays~\cite{Fermi-LAT:2016uux}, diffuse gamma-rays in the Milky Way halo~\cite{Fermi-LAT:2012pls}, and the isotropic gamma-ray background (IGRB)~\cite{Fermi-LAT:2015qzw, Liu:2016ngs}. In the case of DM annihilation, attempts to reconcile the tension between these observations have involved the introduction of complex velocity-dependent cross sections, such as the 
Breit-Wigner mechanism~\cite{Feldman:2008xs,Ibe:2008ye,Bi:2011qm,Bai:2017fav,Xiang:2017jou}. However, these models are somewhat \emph{ad hoc} in nature and are unable to circumvent the constraints imposed by IGRB observations. On the other hand, the constraints from $\gamma$-ray observations are even more stringent for DM decay scenarios, rendering them exceedingly difficult, if not impossible, to evade. 

Given that the SDD model predicts an increased abundance of high-energy secondary $e^{\pm}$, the required cross section/decay rate for DM to account for the positron excess is smaller than that predicted by conventional models. Consequently, our models have the potential to elude the constraints imposed by various observations. To illustrate this, we present the exclusion limits derived from Fermi observations of dwarf galaxies~\cite{Fermi-LAT:2016uux}, diffuse gamma-rays in the Milky Way halo~\cite{Fermi-LAT:2012pls}, IGRB~\cite{Liu:2016ngs}, and the Planck CMB observations~\cite{Slatyer:2015jla}. In addition, we depict the 68\% and 95\% confidence regions for the DM mass and the thermally averaged annihilation cross section/lifetime of the DM particle in Fig.\ref{fig:dm}. For comparison, we also include DM properties from previous works~\cite{Boudaud:2014dta, DiMauro:2015jxa,Lin:2014vja,Liu:2016ngs,Wang:2018pcc} utilizing the standard propagation model to fit the lepton data.

The inspection of the top panels of Fig.~\ref{fig:dm} reveals that for the $\mu^+\mu^-$ channel, the required DM annihilation cross section in the SDD model is consistent with all the imposed constraints. Conversely, the standard models' cross sections are excluded by nearly all the constraints. However, when considering the $\tau^+\tau^-$ channel, despite the SDD model's cross section being smaller than that of the conventional models, it remains excluded by the majority of observational constraints. This is because that the $\tau^+\tau^-$ final state produces a larger amount of gamma-ray emissions compared to the $\mu^+\mu^-$ channel, resulting in stronger constraints. Additionally, the rescaling factor for the $\tau^+\tau^-$ channel is smaller than that of the $\mu^+\mu^-$ channel, necessitating a larger cross section to compensate for the lower rescaling factor to achieve agreement with the data.

{
It should be noted that while the required annihilation cross section aligns with the upper bounds set by $\gamma$-ray and CMB data, specifically in the $\mu^+\mu^-$
channel in the SDD model, it nonetheless exceeds the standard cross section of approximately $3\times 10^{-26} \mathrm{cm}^3\mathrm{s}^{-1}$ for the correct thermal relic. This discrepancy necessitates the introduction of supplementary mechanisms, such as velocity-dependent cross-sections, to account for the observed DM relic density.}

Similar trends are observed in the case of DM decay. The $\mu^+\mu^-$ channel appears to be compatible with the available data, while the $\tau^+\tau^-$ channel is not. These findings reinforce the notion that the $\mu^+\mu^-$ channel, whether in the context of annihilation or decay, exhibits more promising agreement with the data compared to the $\tau^+\tau^-$ channel.

\section{SUMMARY\label{sec:conclusion}}
In this study, we present a quantitative study of the CR electron and positron fluxes observed by the AMS-02 experiment, employing the SDD model. Our analysis incorporates the latest advancements in secondary electron/positron production cross sections and accounts for the  charge-sign dependent solar modulation potentials. {
Notably, the SDD model outperforms conventional propagation models in several key aspects due to the enhanced secondary $e^{\pm}$ at energies above 10 GeV.}

{
The direct consequence of the SDD model's ability to predict an increased population of secondary $e^{\pm}$ is that it mitigates the issue of positron/electron excess compared to conventional propagation models.} This reduction in excess is exemplified by the rescaling factor, denoted as $c_{e^{\pm}}$, which is nearly halved when compared to conventional models (reducing from approximately 3 to around 1.7). This substantial improvement is noteworthy since assuming the production uncertainties greater than 200\% may not be convincing.

{
Due to the reduced rescaling factor, the solar modulation potential required by the SDD model also aligns more closely with the modulation potential observed for cosmic nuclei.} In contrast, conventional models often necessitate a large modulation potential to compensate for the significant rescaling factor $c_{e^{\pm}}$.
Furthermore, we find that, unlike the conventional propagation models where primary sources dominate the positron spectrum at high energies, the secondary component overwhelmingly contributes to the positron fluxes across all energy ranges within the SDD model. 

All considered scenarios demonstrate a satisfactory fit to the AMS-02 data, as evidenced by reduced $\chi^2$ values below unity. In the case of DM annihilating into the $\mu^+\mu^-$ final states, the corresponding DM mass required is approximately 790 GeV, accompanied by a thermally averaged cross section of around $1.26\times 10^{-24}\mathrm{cm}^3\mathrm{s}^{-1}$. Importantly, these parameters remain consistent with the constraints from Fermi and Planck.

For DM decay into the $\mu^+\mu^-$ final states, the fitting analysis yields a DM mass estimate of roughly 1.4 TeV, coupled with a lifetime of approximately $1.58\times 10^{27}$ s. Notably, this scenario also conforms to available constraints from other observations.

Conversely, both the annihilation and decay channels associated with the $\tau^+\tau^-$ final states are excluded by independent constraints derived from $\gamma$-ray and CMB observations. These stringent constraints provide compelling evidence against the $\tau^+\tau^-$ channel as a viable explanation for the observed CR electron and positron excess.


The agreement of DM annihilation/decay into the $\mu^+\mu^-$ final states with the AMS-02 data, coupled with their compliance with the relevant CMB and gamma-ray constraints, underscores the potential as a plausible explanation for the observed phenomena. To pinpoint the precise mechanism behind the excess of positrons, the acquisition of additional data becomes imperative, either through the extension of measurements to higher energy ranges\cite{Fang:2017nww} or by reducing statistical and systematic errors to enable a more thorough analysis of spectral fluctuations\cite{Cholis:2022kio}.

\acknowledgments
This work is supported by the National Natural Science Foundation of China under
the grants No. 12175248, No. 12105292, and No. 2022YFA1604802.

\bibliography{apssamp}

\begin{thebibliography}{116}%
\makeatletter
\providecommand \@ifxundefined [1]{%
 \@ifx{#1\undefined}
}%
\providecommand \@ifnum [1]{%
 \ifnum #1\expandafter \@firstoftwo
 \else \expandafter \@secondoftwo
 \fi
}%
\providecommand \@ifx [1]{%
 \ifx #1\expandafter \@firstoftwo
 \else \expandafter \@secondoftwo
 \fi
}%
\providecommand \natexlab [1]{#1}%
\providecommand \enquote  [1]{``#1''}%
\providecommand \bibnamefont  [1]{#1}%
\providecommand \bibfnamefont [1]{#1}%
\providecommand \citenamefont [1]{#1}%
\providecommand \href@noop [0]{\@secondoftwo}%
\providecommand \href [0]{\begingroup \@sanitize@url \@href}%
\providecommand \@href[1]{\@@startlink{#1}\@@href}%
\providecommand \@@href[1]{\endgroup#1\@@endlink}%
\providecommand \@sanitize@url [0]{\catcode `\\12\catcode `\$12\catcode
  `\&12\catcode `\#12\catcode `\^12\catcode `\_12\catcode `\%12\relax}%
\providecommand \@@startlink[1]{}%
\providecommand \@@endlink[0]{}%
\providecommand \url  [0]{\begingroup\@sanitize@url \@url }%
\providecommand \@url [1]{\endgroup\@href {#1}{\urlprefix }}%
\providecommand \urlprefix  [0]{URL }%
\providecommand \Eprint [0]{\href }%
\providecommand \doibase [0]{https://doi.org/}%
\providecommand \selectlanguage [0]{\@gobble}%
\providecommand \bibinfo  [0]{\@secondoftwo}%
\providecommand \bibfield  [0]{\@secondoftwo}%
\providecommand \translation [1]{[#1]}%
\providecommand \BibitemOpen [0]{}%
\providecommand \bibitemStop [0]{}%
\providecommand \bibitemNoStop [0]{.\EOS\space}%
\providecommand \EOS [0]{\spacefactor3000\relax}%
\providecommand \BibitemShut  [1]{\csname bibitem#1\endcsname}%
\let\auto@bib@innerbib\@empty
\bibitem [{\citenamefont {Adriani}\ \emph {et~al.}(2009)\citenamefont {Adriani}
  \emph {et~al.}}]{PAMELA:2008gwm}%
  \BibitemOpen
  \bibfield  {author} {\bibinfo {author} {\bibfnamefont {O.}~\bibnamefont
  {Adriani}} \emph {et~al.} (\bibinfo {collaboration} {PAMELA}),\ }\href
  {https://doi.org/10.1038/nature07942} {\bibfield  {journal} {\bibinfo
  {journal} {Nature}\ }\textbf {\bibinfo {volume} {458}},\ \bibinfo {pages}
  {607} (\bibinfo {year} {2009})},\ \Eprint {https://arxiv.org/abs/0810.4995}
  {arxiv:0810.4995 [astro-ph]} \BibitemShut {NoStop}%
\bibitem [{\citenamefont {Aguilar}\ \emph
  {et~al.}(2019{\natexlab{a}})\citenamefont {Aguilar} \emph
  {et~al.}}]{AMS:2019rhg}%
  \BibitemOpen
  \bibfield  {author} {\bibinfo {author} {\bibfnamefont {M.}~\bibnamefont
  {Aguilar}} \emph {et~al.} (\bibinfo {collaboration} {AMS}),\ }\href
  {https://doi.org/10.1103/PhysRevLett.122.041102} {\bibfield  {journal}
  {\bibinfo  {journal} {Physical Review Letters}\ }\textbf {\bibinfo {volume}
  {122}},\ \bibinfo {pages} {041102} (\bibinfo {year}
  {2019}{\natexlab{a}})}\BibitemShut {NoStop}%
\bibitem [{\citenamefont {Aguilar}\ \emph
  {et~al.}(2019{\natexlab{b}})\citenamefont {Aguilar} \emph
  {et~al.}}]{AMS:2019iwo}%
  \BibitemOpen
  \bibfield  {author} {\bibinfo {author} {\bibfnamefont {M.}~\bibnamefont
  {Aguilar}} \emph {et~al.} (\bibinfo {collaboration} {AMS}),\ }\href
  {https://doi.org/10.1103/PhysRevLett.122.101101} {\bibfield  {journal}
  {\bibinfo  {journal} {Physical Review Letters}\ }\textbf {\bibinfo {volume}
  {122}},\ \bibinfo {pages} {101101} (\bibinfo {year}
  {2019}{\natexlab{b}})}\BibitemShut {NoStop}%
\bibitem [{\citenamefont {Bergstrom}\ \emph {et~al.}(2008)\citenamefont
  {Bergstrom}, \citenamefont {Bringmann},\ and\ \citenamefont
  {Edsjo}}]{Bergstrom:2008gr}%
  \BibitemOpen
  \bibfield  {author} {\bibinfo {author} {\bibfnamefont {L.}~\bibnamefont
  {Bergstrom}}, \bibinfo {author} {\bibfnamefont {T.}~\bibnamefont
  {Bringmann}},\ and\ \bibinfo {author} {\bibfnamefont {J.}~\bibnamefont
  {Edsjo}},\ }\href {https://doi.org/10.1103/PhysRevD.78.103520} {\bibfield
  {journal} {\bibinfo  {journal} {Physical Review D}\ }\textbf {\bibinfo
  {volume} {78}},\ \bibinfo {pages} {103520} (\bibinfo {year} {2008})},\
  \Eprint {https://arxiv.org/abs/0808.3725} {arxiv:0808.3725 [astro-ph]}
  \BibitemShut {NoStop}%
\bibitem [{\citenamefont {Barger}\ \emph {et~al.}(2009)\citenamefont {Barger},
  \citenamefont {Keung}, \citenamefont {Marfatia},\ and\ \citenamefont
  {Shaughnessy}}]{Barger:2008su}%
  \BibitemOpen
  \bibfield  {author} {\bibinfo {author} {\bibfnamefont {V.}~\bibnamefont
  {Barger}}, \bibinfo {author} {\bibfnamefont {W.-Y.}\ \bibnamefont {Keung}},
  \bibinfo {author} {\bibfnamefont {D.}~\bibnamefont {Marfatia}},\ and\
  \bibinfo {author} {\bibfnamefont {G.}~\bibnamefont {Shaughnessy}},\ }\href
  {https://doi.org/10.1016/j.physletb.2009.01.016} {\bibfield  {journal}
  {\bibinfo  {journal} {Physics Letters B}\ }\textbf {\bibinfo {volume}
  {672}},\ \bibinfo {pages} {141} (\bibinfo {year} {2009})},\ \Eprint
  {https://arxiv.org/abs/0809.0162} {arxiv:0809.0162 [hep-ph]} \BibitemShut
  {NoStop}%
\bibitem [{\citenamefont {Cirelli}\ \emph {et~al.}(2013)\citenamefont
  {Cirelli}, \citenamefont {Kadastik}, \citenamefont {Raidal},\ and\
  \citenamefont {Strumia}}]{Cirelli:2008pk}%
  \BibitemOpen
  \bibfield  {author} {\bibinfo {author} {\bibfnamefont {M.}~\bibnamefont
  {Cirelli}}, \bibinfo {author} {\bibfnamefont {M.}~\bibnamefont {Kadastik}},
  \bibinfo {author} {\bibfnamefont {M.}~\bibnamefont {Raidal}},\ and\ \bibinfo
  {author} {\bibfnamefont {A.}~\bibnamefont {Strumia}},\ }\href
  {https://doi.org/10.1016/j.nuclphysb.2008.11.031} {\bibfield  {journal}
  {\bibinfo  {journal} {Nuclear Physics B}\ }\textbf {\bibinfo {volume}
  {813}},\ \bibinfo {pages} {1} (\bibinfo {year} {2013})},\ \Eprint
  {https://arxiv.org/abs/0809.2409} {arxiv:0809.2409 [hep-ph]} \BibitemShut
  {NoStop}%
\bibitem [{\citenamefont {Yin}\ \emph {et~al.}(2009)\citenamefont {Yin},
  \citenamefont {Yuan}, \citenamefont {Liu}, \citenamefont {Zhang},
  \citenamefont {Bi}, \citenamefont {Zhu},\ and\ \citenamefont
  {Zhang}}]{Yin:2008bs}%
  \BibitemOpen
  \bibfield  {author} {\bibinfo {author} {\bibfnamefont {P.-f.}\ \bibnamefont
  {Yin}}, \bibinfo {author} {\bibfnamefont {Q.}~\bibnamefont {Yuan}}, \bibinfo
  {author} {\bibfnamefont {J.}~\bibnamefont {Liu}}, \bibinfo {author}
  {\bibfnamefont {J.}~\bibnamefont {Zhang}}, \bibinfo {author} {\bibfnamefont
  {X.-j.}\ \bibnamefont {Bi}}, \bibinfo {author} {\bibfnamefont {S.-h.}\
  \bibnamefont {Zhu}},\ and\ \bibinfo {author} {\bibfnamefont {X.}~\bibnamefont
  {Zhang}},\ }\href {https://doi.org/10.1103/PhysRevD.79.023512} {\bibfield
  {journal} {\bibinfo  {journal} {Physical Review D}\ }\textbf {\bibinfo
  {volume} {79}},\ \bibinfo {pages} {023512} (\bibinfo {year} {2009})},\
  \Eprint {https://arxiv.org/abs/0811.0176} {arxiv:0811.0176 [hep-ph]}
  \BibitemShut {NoStop}%
\bibitem [{\citenamefont {{Arkani-Hamed}}\ \emph {et~al.}(2009)\citenamefont
  {{Arkani-Hamed}}, \citenamefont {Finkbeiner}, \citenamefont {Slatyer},\ and\
  \citenamefont {Weiner}}]{Arkani-Hamed:2008hhe}%
  \BibitemOpen
  \bibfield  {author} {\bibinfo {author} {\bibfnamefont {N.}~\bibnamefont
  {{Arkani-Hamed}}}, \bibinfo {author} {\bibfnamefont {D.~P.}\ \bibnamefont
  {Finkbeiner}}, \bibinfo {author} {\bibfnamefont {T.~R.}\ \bibnamefont
  {Slatyer}},\ and\ \bibinfo {author} {\bibfnamefont {N.}~\bibnamefont
  {Weiner}},\ }\href {https://doi.org/10.1103/PhysRevD.79.015014} {\bibfield
  {journal} {\bibinfo  {journal} {Physical Review D}\ }\textbf {\bibinfo
  {volume} {79}},\ \bibinfo {pages} {015014} (\bibinfo {year} {2009})},\
  \Eprint {https://arxiv.org/abs/0810.0713} {arxiv:0810.0713 [hep-ph]}
  \BibitemShut {NoStop}%
\bibitem [{\citenamefont {Pospelov}\ and\ \citenamefont
  {Ritz}(2009)}]{Pospelov:2008jd}%
  \BibitemOpen
  \bibfield  {author} {\bibinfo {author} {\bibfnamefont {M.}~\bibnamefont
  {Pospelov}}\ and\ \bibinfo {author} {\bibfnamefont {A.}~\bibnamefont
  {Ritz}},\ }\href {https://doi.org/10.1016/j.physletb.2008.12.012} {\bibfield
  {journal} {\bibinfo  {journal} {Physics Letters B}\ }\textbf {\bibinfo
  {volume} {671}},\ \bibinfo {pages} {391} (\bibinfo {year} {2009})},\ \Eprint
  {https://arxiv.org/abs/0810.1502} {arxiv:0810.1502 [hep-ph]} \BibitemShut
  {NoStop}%
\bibitem [{\citenamefont {Cholis}\ \emph {et~al.}(2009)\citenamefont {Cholis},
  \citenamefont {Goodenough}, \citenamefont {Hooper}, \citenamefont {Simet},\
  and\ \citenamefont {Weiner}}]{Cholis_2009}%
  \BibitemOpen
  \bibfield  {author} {\bibinfo {author} {\bibfnamefont {I.}~\bibnamefont
  {Cholis}}, \bibinfo {author} {\bibfnamefont {L.}~\bibnamefont {Goodenough}},
  \bibinfo {author} {\bibfnamefont {D.}~\bibnamefont {Hooper}}, \bibinfo
  {author} {\bibfnamefont {M.}~\bibnamefont {Simet}},\ and\ \bibinfo {author}
  {\bibfnamefont {N.}~\bibnamefont {Weiner}},\ }\bibfield  {journal} {\bibinfo
  {journal} {Physical Review D}\ }\textbf {\bibinfo {volume} {80}},\ \href
  {https://doi.org/10.1103/physrevd.80.123511} {10.1103/physrevd.80.123511}
  (\bibinfo {year} {2009})\BibitemShut {NoStop}%
\bibitem [{\citenamefont {Hamaguchi}\ \emph {et~al.}(2009)\citenamefont
  {Hamaguchi}, \citenamefont {Nakamura}, \citenamefont {Shirai},\ and\
  \citenamefont {Yanagida}}]{Hamaguchi_2009}%
  \BibitemOpen
  \bibfield  {author} {\bibinfo {author} {\bibfnamefont {K.}~\bibnamefont
  {Hamaguchi}}, \bibinfo {author} {\bibfnamefont {E.}~\bibnamefont {Nakamura}},
  \bibinfo {author} {\bibfnamefont {S.}~\bibnamefont {Shirai}},\ and\ \bibinfo
  {author} {\bibfnamefont {T.}~\bibnamefont {Yanagida}},\ }\href
  {https://doi.org/10.1016/j.physletb.2009.03.025} {\bibfield  {journal}
  {\bibinfo  {journal} {Physics Letters B}\ }\textbf {\bibinfo {volume}
  {674}},\ \bibinfo {pages} {299} (\bibinfo {year} {2009})}\BibitemShut
  {NoStop}%
\bibitem [{\citenamefont {Boudaud}\ \emph {et~al.}(2015)\citenamefont {Boudaud}
  \emph {et~al.}}]{Boudaud:2014dta}%
  \BibitemOpen
  \bibfield  {author} {\bibinfo {author} {\bibfnamefont {M.}~\bibnamefont
  {Boudaud}} \emph {et~al.},\ }\href
  {https://doi.org/10.1051/0004-6361/201425197} {\bibfield  {journal} {\bibinfo
   {journal} {Astron. Astrophys.}\ }\textbf {\bibinfo {volume} {575}},\
  \bibinfo {pages} {A67} (\bibinfo {year} {2015})},\ \Eprint
  {https://arxiv.org/abs/1410.3799} {arXiv:1410.3799 [astro-ph.HE]}
  \BibitemShut {NoStop}%
\bibitem [{\citenamefont {Lin}\ \emph {et~al.}(2015)\citenamefont {Lin},
  \citenamefont {Yuan},\ and\ \citenamefont {Bi}}]{Lin:2014vja}%
  \BibitemOpen
  \bibfield  {author} {\bibinfo {author} {\bibfnamefont {S.-J.}\ \bibnamefont
  {Lin}}, \bibinfo {author} {\bibfnamefont {Q.}~\bibnamefont {Yuan}},\ and\
  \bibinfo {author} {\bibfnamefont {X.-J.}\ \bibnamefont {Bi}},\ }\href
  {https://doi.org/10.1103/PhysRevD.91.063508} {\bibfield  {journal} {\bibinfo
  {journal} {Physical Review D}\ }\textbf {\bibinfo {volume} {91}},\ \bibinfo
  {pages} {063508} (\bibinfo {year} {2015})},\ \Eprint
  {https://arxiv.org/abs/1409.6248} {arxiv:1409.6248 [astro-ph.HE]}
  \BibitemShut {NoStop}%
\bibitem [{\citenamefont {Wang}\ \emph {et~al.}(2018)\citenamefont {Wang},
  \citenamefont {Bi}, \citenamefont {Lin},\ and\ \citenamefont
  {Yin}}]{Wang:2018pcc}%
  \BibitemOpen
  \bibfield  {author} {\bibinfo {author} {\bibfnamefont {B.}~\bibnamefont
  {Wang}}, \bibinfo {author} {\bibfnamefont {X.}~\bibnamefont {Bi}}, \bibinfo
  {author} {\bibfnamefont {S.}~\bibnamefont {Lin}},\ and\ \bibinfo {author}
  {\bibfnamefont {P.}~\bibnamefont {Yin}},\ }\href
  {https://doi.org/10.1007/s11433-018-9244-y} {\bibfield  {journal} {\bibinfo
  {journal} {Science China Physics, Mechanics, and Astronomy}\ }\textbf
  {\bibinfo {volume} {61}},\ \bibinfo {pages} {101004} (\bibinfo {year}
  {2018})}\BibitemShut {NoStop}%
\bibitem [{\citenamefont {Cheng}\ \emph {et~al.}(2017)\citenamefont {Cheng},
  \citenamefont {Huang}, \citenamefont {Huang}, \citenamefont {Low},
  \citenamefont {Tsai},\ and\ \citenamefont {Yuan}}]{Cheng:2016slx}%
  \BibitemOpen
  \bibfield  {author} {\bibinfo {author} {\bibfnamefont {H.-C.}\ \bibnamefont
  {Cheng}}, \bibinfo {author} {\bibfnamefont {W.-C.}\ \bibnamefont {Huang}},
  \bibinfo {author} {\bibfnamefont {X.}~\bibnamefont {Huang}}, \bibinfo
  {author} {\bibfnamefont {I.}~\bibnamefont {Low}}, \bibinfo {author}
  {\bibfnamefont {Y.-L.~S.}\ \bibnamefont {Tsai}},\ and\ \bibinfo {author}
  {\bibfnamefont {Q.}~\bibnamefont {Yuan}},\ }\href
  {https://doi.org/10.1088/1475-7516/2017/03/041} {\bibfield  {journal}
  {\bibinfo  {journal} {Journal of Cosmology and Astroparticle Physics}\
  }\textbf {\bibinfo {volume} {2017}}\bibinfo  {number} { (03)},\ \bibinfo
  {pages} {041}}\BibitemShut {NoStop}%
\bibitem [{\citenamefont {Aharonian}\ \emph {et~al.}(1995)\citenamefont
  {Aharonian}, \citenamefont {Atoyan},\ and\ \citenamefont
  {Voelk}}]{aharonianHighEnergyElectrons1995}%
  \BibitemOpen
\bibfield  {number} {  }\bibfield  {author} {\bibinfo {author} {\bibfnamefont
  {F.~A.}\ \bibnamefont {Aharonian}}, \bibinfo {author} {\bibfnamefont {A.~M.}\
  \bibnamefont {Atoyan}},\ and\ \bibinfo {author} {\bibfnamefont {H.~J.}\
  \bibnamefont {Voelk}},\ }\href@noop {} {\bibfield  {journal} {\bibinfo
  {journal} {Astronomy and Astrophysics}\ }\textbf {\bibinfo {volume} {294}},\
  \bibinfo {pages} {L41} (\bibinfo {year} {1995})}\BibitemShut {NoStop}%
\bibitem [{\citenamefont {Atoian}\ \emph {et~al.}(1995)\citenamefont {Atoian},
  \citenamefont {Aharonian},\ and\ \citenamefont {Volk}}]{Atoian:1995ux}%
  \BibitemOpen
  \bibfield  {author} {\bibinfo {author} {\bibfnamefont {A.~M.}\ \bibnamefont
  {Atoian}}, \bibinfo {author} {\bibfnamefont {F.~A.}\ \bibnamefont
  {Aharonian}},\ and\ \bibinfo {author} {\bibfnamefont {H.~J.}\ \bibnamefont
  {Volk}},\ }\href {https://doi.org/10.1103/PhysRevD.52.3265} {\bibfield
  {journal} {\bibinfo  {journal} {Phys. Rev. D}\ }\textbf {\bibinfo {volume}
  {52}},\ \bibinfo {pages} {3265} (\bibinfo {year} {1995})}\BibitemShut
  {NoStop}%
\bibitem [{\citenamefont {Kobayashi}\ \emph {et~al.}(2004)\citenamefont
  {Kobayashi}, \citenamefont {Komori}, \citenamefont {Yoshida},\ and\
  \citenamefont {Nishimura}}]{Kobayashi:2003kp}%
  \BibitemOpen
  \bibfield  {author} {\bibinfo {author} {\bibfnamefont {T.}~\bibnamefont
  {Kobayashi}}, \bibinfo {author} {\bibfnamefont {Y.}~\bibnamefont {Komori}},
  \bibinfo {author} {\bibfnamefont {K.}~\bibnamefont {Yoshida}},\ and\ \bibinfo
  {author} {\bibfnamefont {J.}~\bibnamefont {Nishimura}},\ }\href
  {https://doi.org/10.1086/380431} {\bibfield  {journal} {\bibinfo  {journal}
  {Astrophys. J.}\ }\textbf {\bibinfo {volume} {601}},\ \bibinfo {pages} {340}
  (\bibinfo {year} {2004})},\ \Eprint {https://arxiv.org/abs/astro-ph/0308470}
  {arxiv:astro-ph/0308470} \BibitemShut {NoStop}%
\bibitem [{\citenamefont {Profumo}(2011)}]{Profumo:2008ms}%
  \BibitemOpen
  \bibfield  {author} {\bibinfo {author} {\bibfnamefont {S.}~\bibnamefont
  {Profumo}},\ }\href {https://doi.org/10.2478/s11534-011-0099-z} {\bibfield
  {journal} {\bibinfo  {journal} {Open Physics}\ }\textbf {\bibinfo {volume}
  {10}},\ \bibinfo {pages} {1} (\bibinfo {year} {2011})},\ \Eprint
  {https://arxiv.org/abs/0812.4457} {arxiv:0812.4457 [astro-ph]} \BibitemShut
  {NoStop}%
\bibitem [{\citenamefont {Hooper}\ \emph {et~al.}(2009)\citenamefont {Hooper},
  \citenamefont {Blasi},\ and\ \citenamefont {Serpico}}]{Hooper:2008kg}%
  \BibitemOpen
  \bibfield  {author} {\bibinfo {author} {\bibfnamefont {D.}~\bibnamefont
  {Hooper}}, \bibinfo {author} {\bibfnamefont {P.}~\bibnamefont {Blasi}},\ and\
  \bibinfo {author} {\bibfnamefont {P.~D.}\ \bibnamefont {Serpico}},\ }\href
  {https://doi.org/10.1088/1475-7516/2009/01/025} {\bibfield  {journal}
  {\bibinfo  {journal} {Journal of Cosmology and Astroparticle Physics}\
  }\textbf {\bibinfo {volume} {01}}\bibfield  {number} {\bibinfo  {number} {
  (01)},\ \bibinfo {pages} {025}},\ }\Eprint {https://arxiv.org/abs/0810.1527}
  {arxiv:0810.1527 [astro-ph]} \BibitemShut {NoStop}%
\bibitem [{\citenamefont {Yuksel}\ \emph {et~al.}(2009)\citenamefont {Yuksel},
  \citenamefont {Kistler},\ and\ \citenamefont {Stanev}}]{Yuksel:2008rf}%
  \BibitemOpen
  \bibfield  {author} {\bibinfo {author} {\bibfnamefont {H.}~\bibnamefont
  {Yuksel}}, \bibinfo {author} {\bibfnamefont {M.~D.}\ \bibnamefont
  {Kistler}},\ and\ \bibinfo {author} {\bibfnamefont {T.}~\bibnamefont
  {Stanev}},\ }\href {https://doi.org/10.1103/PhysRevLett.103.051101}
  {\bibfield  {journal} {\bibinfo  {journal} {Physical Review Letters}\
  }\textbf {\bibinfo {volume} {103}},\ \bibinfo {pages} {051101} (\bibinfo
  {year} {2009})},\ \Eprint {https://arxiv.org/abs/0810.2784} {arxiv:0810.2784
  [astro-ph]} \BibitemShut {NoStop}%
\bibitem [{\citenamefont {Malyshev}\ \emph {et~al.}(2009)\citenamefont
  {Malyshev}, \citenamefont {Cholis},\ and\ \citenamefont
  {Gelfand}}]{Malyshev:2009tw}%
  \BibitemOpen
  \bibfield  {author} {\bibinfo {author} {\bibfnamefont {D.}~\bibnamefont
  {Malyshev}}, \bibinfo {author} {\bibfnamefont {I.}~\bibnamefont {Cholis}},\
  and\ \bibinfo {author} {\bibfnamefont {J.}~\bibnamefont {Gelfand}},\ }\href
  {https://doi.org/10.1103/PhysRevD.80.063005} {\bibfield  {journal} {\bibinfo
  {journal} {Physical Review D}\ }\textbf {\bibinfo {volume} {80}},\ \bibinfo
  {pages} {063005} (\bibinfo {year} {2009})},\ \Eprint
  {https://arxiv.org/abs/0903.1310} {arxiv:0903.1310 [astro-ph.HE]}
  \BibitemShut {NoStop}%
\bibitem [{\citenamefont {Blasi}(2009)}]{Blasi:2009hv}%
  \BibitemOpen
  \bibfield  {author} {\bibinfo {author} {\bibfnamefont {P.}~\bibnamefont
  {Blasi}},\ }\href {https://doi.org/10.1103/PhysRevLett.103.051104} {\bibfield
   {journal} {\bibinfo  {journal} {Physical Review Letters}\ }\textbf {\bibinfo
  {volume} {103}},\ \bibinfo {pages} {051104} (\bibinfo {year} {2009})},\
  \Eprint {https://arxiv.org/abs/0903.2794} {arxiv:0903.2794 [astro-ph.HE]}
  \BibitemShut {NoStop}%
\bibitem [{\citenamefont {Ioka}(2010)}]{Ioka_2010}%
  \BibitemOpen
  \bibfield  {author} {\bibinfo {author} {\bibfnamefont {K.}~\bibnamefont
  {Ioka}},\ }\href {https://doi.org/10.1143/ptp.123.743} {\bibfield  {journal}
  {\bibinfo  {journal} {Progress of Theoretical Physics}\ }\textbf {\bibinfo
  {volume} {123}},\ \bibinfo {pages} {743} (\bibinfo {year}
  {2010})}\BibitemShut {NoStop}%
\bibitem [{\citenamefont {Di~Mauro}\ \emph {et~al.}(2016)\citenamefont
  {Di~Mauro}, \citenamefont {Donato}, \citenamefont {Fornengo},\ and\
  \citenamefont {Vittino}}]{DiMauro:2015jxa}%
  \BibitemOpen
  \bibfield  {author} {\bibinfo {author} {\bibfnamefont {M.}~\bibnamefont
  {Di~Mauro}}, \bibinfo {author} {\bibfnamefont {F.}~\bibnamefont {Donato}},
  \bibinfo {author} {\bibfnamefont {N.}~\bibnamefont {Fornengo}},\ and\
  \bibinfo {author} {\bibfnamefont {A.}~\bibnamefont {Vittino}},\ }\href
  {https://doi.org/10.1088/1475-7516/2016/05/031} {\bibfield  {journal}
  {\bibinfo  {journal} {Journal of Cosmology and Astroparticle Physics}\
  }\textbf {\bibinfo {volume} {05}}\bibfield  {number} {\bibinfo  {number} {
  (05)},\ \bibinfo {pages} {031}},\ }\Eprint {https://arxiv.org/abs/1507.07001}
  {arxiv:1507.07001 [astro-ph.HE]} \BibitemShut {NoStop}%
\bibitem [{\citenamefont {Cholis}\ and\ \citenamefont
  {Hoover}(2023)}]{Cholis:2022kio}%
  \BibitemOpen
  \bibfield  {author} {\bibinfo {author} {\bibfnamefont {I.}~\bibnamefont
  {Cholis}}\ and\ \bibinfo {author} {\bibfnamefont {T.}~\bibnamefont
  {Hoover}},\ }\href {https://doi.org/10.1103/PhysRevD.107.063003} {\bibfield
  {journal} {\bibinfo  {journal} {Phys. Rev. D}\ }\textbf {\bibinfo {volume}
  {107}},\ \bibinfo {pages} {063003} (\bibinfo {year} {2023})},\ \Eprint
  {https://arxiv.org/abs/2211.15709} {arxiv:2211.15709 [astro-ph.HE]}
  \BibitemShut {NoStop}%
\bibitem [{\citenamefont {Aguilar}\ \emph {et~al.}(2016)\citenamefont
  {Aguilar}, \citenamefont {Ali~Cavasonza}, \citenamefont {Alpat} \emph
  {et~al.}}]{AMS:2016oqu}%
  \BibitemOpen
  \bibfield  {author} {\bibinfo {author} {\bibfnamefont {M.}~\bibnamefont
  {Aguilar}}, \bibinfo {author} {\bibfnamefont {L.}~\bibnamefont
  {Ali~Cavasonza}}, \bibinfo {author} {\bibfnamefont {B.}~\bibnamefont
  {Alpat}}, \emph {et~al.} (\bibinfo {collaboration} {AMS}),\ }\href
  {https://doi.org/10.1103/PhysRevLett.117.091103} {\bibfield  {journal}
  {\bibinfo  {journal} {Physical Review Letters}\ }\textbf {\bibinfo {volume}
  {117}},\ \bibinfo {pages} {091103} (\bibinfo {year} {2016})}\BibitemShut
  {NoStop}%
\bibitem [{\citenamefont {Calore}\ \emph {et~al.}(2022)\citenamefont {Calore},
  \citenamefont {Cirelli}, \citenamefont {Derome}, \citenamefont {Genolini},
  \citenamefont {Maurin}, \citenamefont {Salati},\ and\ \citenamefont
  {Serpico}}]{caloreAMS02AntiprotonsDark2022}%
  \BibitemOpen
  \bibfield  {author} {\bibinfo {author} {\bibfnamefont {F.}~\bibnamefont
  {Calore}}, \bibinfo {author} {\bibfnamefont {M.}~\bibnamefont {Cirelli}},
  \bibinfo {author} {\bibfnamefont {L.}~\bibnamefont {Derome}}, \bibinfo
  {author} {\bibfnamefont {Y.}~\bibnamefont {Genolini}}, \bibinfo {author}
  {\bibfnamefont {D.}~\bibnamefont {Maurin}}, \bibinfo {author} {\bibfnamefont
  {P.}~\bibnamefont {Salati}},\ and\ \bibinfo {author} {\bibfnamefont {P.~D.}\
  \bibnamefont {Serpico}},\ }\href
  {https://doi.org/10.21468/SciPostPhys.12.5.163} {\bibfield  {journal}
  {\bibinfo  {journal} {SciPost Phys.}\ }\textbf {\bibinfo {volume} {12}},\
  \bibinfo {pages} {163} (\bibinfo {year} {2022})}\BibitemShut {NoStop}%
\bibitem [{\citenamefont {Albert}\ \emph {et~al.}(2017)\citenamefont {Albert}
  \emph {et~al.}}]{Fermi-LAT:2016uux}%
  \BibitemOpen
  \bibfield  {author} {\bibinfo {author} {\bibfnamefont {A.}~\bibnamefont
  {Albert}} \emph {et~al.} (\bibinfo {collaboration} {Fermi-LAT, DES}),\ }\href
  {https://doi.org/10.3847/1538-4357/834/2/110} {\bibfield  {journal} {\bibinfo
   {journal} {The Astrophysical Journal}\ }\textbf {\bibinfo {volume} {834}},\
  \bibinfo {pages} {110} (\bibinfo {year} {2017})},\ \Eprint
  {https://arxiv.org/abs/1611.03184} {arxiv:1611.03184 [astro-ph.HE]}
  \BibitemShut {NoStop}%
\bibitem [{\citenamefont {Ackermann}\ \emph {et~al.}(2012)\citenamefont
  {Ackermann} \emph {et~al.}}]{Fermi-LAT:2012pls}%
  \BibitemOpen
  \bibfield  {author} {\bibinfo {author} {\bibfnamefont {M.}~\bibnamefont
  {Ackermann}} \emph {et~al.} (\bibinfo {collaboration} {Fermi-LAT}),\ }\href
  {https://doi.org/10.1088/0004-637X/761/2/91} {\bibfield  {journal} {\bibinfo
  {journal} {The Astrophysical Journal}\ }\textbf {\bibinfo {volume} {761}},\
  \bibinfo {pages} {91} (\bibinfo {year} {2012})},\ \Eprint
  {https://arxiv.org/abs/1205.6474} {arxiv:1205.6474 [astro-ph.CO]}
  \BibitemShut {NoStop}%
\bibitem [{\citenamefont {Ackermann}\ \emph {et~al.}(2015)\citenamefont
  {Ackermann}, \citenamefont {Ajello}, \citenamefont {Albert} \emph
  {et~al.}}]{Fermi-LAT:2015qzw}%
  \BibitemOpen
  \bibfield  {author} {\bibinfo {author} {\bibfnamefont {M.}~\bibnamefont
  {Ackermann}}, \bibinfo {author} {\bibfnamefont {M.}~\bibnamefont {Ajello}},
  \bibinfo {author} {\bibfnamefont {A.}~\bibnamefont {Albert}}, \emph {et~al.}
  (\bibinfo {collaboration} {Fermi-LAT}),\ }\href
  {https://doi.org/10.1088/1475-7516/2015/09/008} {\bibfield  {journal}
  {\bibinfo  {journal} {Journal of Cosmology and Astroparticle Physics}\
  }\textbf {\bibinfo {volume} {09}}\bibfield  {number} {\bibinfo  {number} {
  (09)},\ \bibinfo {pages} {008}},\ }\Eprint {https://arxiv.org/abs/1501.05464}
  {arxiv:1501.05464 [astro-ph.CO]} \BibitemShut {NoStop}%
\bibitem [{\citenamefont {Liu}\ \emph {et~al.}(2017)\citenamefont {Liu},
  \citenamefont {Bi}, \citenamefont {Lin},\ and\ \citenamefont
  {Yin}}]{Liu:2016ngs}%
  \BibitemOpen
  \bibfield  {author} {\bibinfo {author} {\bibfnamefont {W.}~\bibnamefont
  {Liu}}, \bibinfo {author} {\bibfnamefont {X.-J.}\ \bibnamefont {Bi}},
  \bibinfo {author} {\bibfnamefont {S.-J.}\ \bibnamefont {Lin}},\ and\ \bibinfo
  {author} {\bibfnamefont {P.-F.}\ \bibnamefont {Yin}},\ }\href
  {https://doi.org/10.1088/1674-1137/41/4/045104} {\bibfield  {journal}
  {\bibinfo  {journal} {Chinese Physics C}\ }\textbf {\bibinfo {volume} {41}},\
  \bibinfo {pages} {045104} (\bibinfo {year} {2017})},\ \Eprint
  {https://arxiv.org/abs/1602.01012} {arxiv:1602.01012 [astro-ph.CO]}
  \BibitemShut {NoStop}%
\bibitem [{\citenamefont {Ade}\ \emph {et~al.}(2016)\citenamefont {Ade} \emph
  {et~al.}}]{Planck:2015fie}%
  \BibitemOpen
  \bibfield  {author} {\bibinfo {author} {\bibfnamefont {P.}~\bibnamefont
  {Ade}} \emph {et~al.} (\bibinfo {collaboration} {Planck}),\ }\href
  {https://doi.org/10.1051/0004-6361/201525830} {\bibfield  {journal} {\bibinfo
   {journal} {Astronomy \& Astrophysics}\ }\textbf {\bibinfo {volume} {594}},\
  \bibinfo {pages} {A13} (\bibinfo {year} {2016})},\ \Eprint
  {https://arxiv.org/abs/1502.01589} {arxiv:1502.01589 [astro-ph.CO]}
  \BibitemShut {NoStop}%
\bibitem [{\citenamefont {Slatyer}(2016)}]{Slatyer:2015jla}%
  \BibitemOpen
  \bibfield  {author} {\bibinfo {author} {\bibfnamefont {T.~R.}\ \bibnamefont
  {Slatyer}},\ }\href {https://doi.org/10.1103/PhysRevD.93.023527} {\bibfield
  {journal} {\bibinfo  {journal} {Physical Review D}\ }\textbf {\bibinfo
  {volume} {93}},\ \bibinfo {pages} {023527} (\bibinfo {year} {2016})},\
  \Eprint {https://arxiv.org/abs/1506.03811} {arxiv:1506.03811 [hep-ph]}
  \BibitemShut {NoStop}%
\bibitem [{\citenamefont {Slatyer}\ and\ \citenamefont
  {Wu}(2017)}]{Slatyer:2016qyl}%
  \BibitemOpen
  \bibfield  {author} {\bibinfo {author} {\bibfnamefont {T.~R.}\ \bibnamefont
  {Slatyer}}\ and\ \bibinfo {author} {\bibfnamefont {C.-L.}\ \bibnamefont
  {Wu}},\ }\href {https://doi.org/10.1103/PhysRevD.95.023010} {\bibfield
  {journal} {\bibinfo  {journal} {Physical Review D}\ }\textbf {\bibinfo
  {volume} {95}},\ \bibinfo {pages} {023010} (\bibinfo {year} {2017})},\
  \Eprint {https://arxiv.org/abs/1610.06933} {arxiv:1610.06933 [astro-ph.CO]}
  \BibitemShut {NoStop}%
\bibitem [{\citenamefont {Hisano}\ \emph {et~al.}(2004)\citenamefont {Hisano},
  \citenamefont {Matsumoto},\ and\ \citenamefont {Nojiri}}]{Hisano:2003ec}%
  \BibitemOpen
  \bibfield  {author} {\bibinfo {author} {\bibfnamefont {J.}~\bibnamefont
  {Hisano}}, \bibinfo {author} {\bibfnamefont {S.}~\bibnamefont {Matsumoto}},\
  and\ \bibinfo {author} {\bibfnamefont {M.~M.}\ \bibnamefont {Nojiri}},\
  }\href {https://doi.org/10.1103/PhysRevLett.92.031303} {\bibfield  {journal}
  {\bibinfo  {journal} {Physical Review Letters}\ }\textbf {\bibinfo {volume}
  {92}},\ \bibinfo {pages} {031303} (\bibinfo {year} {2004})},\ \Eprint
  {https://arxiv.org/abs/hep-ph/0307216} {arxiv:hep-ph/0307216} \BibitemShut
  {NoStop}%
\bibitem [{\citenamefont {Hisano}\ \emph {et~al.}(2007)\citenamefont {Hisano},
  \citenamefont {Matsumoto}, \citenamefont {Nagai}, \citenamefont {Saito},\
  and\ \citenamefont {Senami}}]{Hisano:2006nn}%
  \BibitemOpen
  \bibfield  {author} {\bibinfo {author} {\bibfnamefont {J.}~\bibnamefont
  {Hisano}}, \bibinfo {author} {\bibfnamefont {S.}~\bibnamefont {Matsumoto}},
  \bibinfo {author} {\bibfnamefont {M.}~\bibnamefont {Nagai}}, \bibinfo
  {author} {\bibfnamefont {O.}~\bibnamefont {Saito}},\ and\ \bibinfo {author}
  {\bibfnamefont {M.}~\bibnamefont {Senami}},\ }\href
  {https://doi.org/10.1016/j.physletb.2007.01.012} {\bibfield  {journal}
  {\bibinfo  {journal} {Physics Letters B}\ }\textbf {\bibinfo {volume}
  {646}},\ \bibinfo {pages} {34} (\bibinfo {year} {2007})},\ \Eprint
  {https://arxiv.org/abs/hep-ph/0610249} {arxiv:hep-ph/0610249} \BibitemShut
  {NoStop}%
\bibitem [{\citenamefont {Feng}\ \emph
  {et~al.}(2010{\natexlab{a}})\citenamefont {Feng}, \citenamefont
  {Kaplinghat},\ and\ \citenamefont {Yu}}]{Feng:2009hw}%
  \BibitemOpen
  \bibfield  {author} {\bibinfo {author} {\bibfnamefont {J.~L.}\ \bibnamefont
  {Feng}}, \bibinfo {author} {\bibfnamefont {M.}~\bibnamefont {Kaplinghat}},\
  and\ \bibinfo {author} {\bibfnamefont {H.-B.}\ \bibnamefont {Yu}},\ }\href
  {https://doi.org/10.1103/PhysRevLett.104.151301} {\bibfield  {journal}
  {\bibinfo  {journal} {Physical Review Letters}\ }\textbf {\bibinfo {volume}
  {104}},\ \bibinfo {pages} {151301} (\bibinfo {year} {2010}{\natexlab{a}})},\
  \Eprint {https://arxiv.org/abs/0911.0422} {arxiv:0911.0422 [hep-ph]}
  \BibitemShut {NoStop}%
\bibitem [{\citenamefont {Feng}\ \emph
  {et~al.}(2010{\natexlab{b}})\citenamefont {Feng}, \citenamefont
  {Kaplinghat},\ and\ \citenamefont {Yu}}]{Feng:2010zp}%
  \BibitemOpen
  \bibfield  {author} {\bibinfo {author} {\bibfnamefont {J.~L.}\ \bibnamefont
  {Feng}}, \bibinfo {author} {\bibfnamefont {M.}~\bibnamefont {Kaplinghat}},\
  and\ \bibinfo {author} {\bibfnamefont {H.-B.}\ \bibnamefont {Yu}},\ }\href
  {https://doi.org/10.1103/PhysRevD.82.083525} {\bibfield  {journal} {\bibinfo
  {journal} {Physical Review D}\ }\textbf {\bibinfo {volume} {82}},\ \bibinfo
  {pages} {083525} (\bibinfo {year} {2010}{\natexlab{b}})},\ \Eprint
  {https://arxiv.org/abs/1005.4678} {arxiv:1005.4678 [hep-ph]} \BibitemShut
  {NoStop}%
\bibitem [{\citenamefont {Das}\ \emph {et~al.}(2020)\citenamefont {Das},
  \citenamefont {Dasgupta},\ and\ \citenamefont
  {Ray}}]{dasGalacticPositronExcess2020}%
  \BibitemOpen
  \bibfield  {author} {\bibinfo {author} {\bibfnamefont {A.}~\bibnamefont
  {Das}}, \bibinfo {author} {\bibfnamefont {B.}~\bibnamefont {Dasgupta}},\ and\
  \bibinfo {author} {\bibfnamefont {A.}~\bibnamefont {Ray}},\ }\href
  {https://doi.org/10.1103/PhysRevD.101.063014} {\bibfield  {journal} {\bibinfo
   {journal} {Physical Review D}\ }\textbf {\bibinfo {volume} {101}},\ \bibinfo
  {pages} {063014} (\bibinfo {year} {2020})},\ \Eprint
  {https://arxiv.org/abs/1911.03488} {arxiv:1911.03488 [astro-ph,
  physics:hep-ph]} \BibitemShut {NoStop}%
\bibitem [{\citenamefont {Ding}\ \emph {et~al.}(2021)\citenamefont {Ding},
  \citenamefont {Ku}, \citenamefont {Wei},\ and\ \citenamefont
  {Zhou}}]{Ding:2021zzg}%
  \BibitemOpen
  \bibfield  {author} {\bibinfo {author} {\bibfnamefont {Y.-C.}\ \bibnamefont
  {Ding}}, \bibinfo {author} {\bibfnamefont {Y.-L.}\ \bibnamefont {Ku}},
  \bibinfo {author} {\bibfnamefont {C.-C.}\ \bibnamefont {Wei}},\ and\ \bibinfo
  {author} {\bibfnamefont {Y.-F.}\ \bibnamefont {Zhou}},\ }\href
  {https://doi.org/10.1088/1475-7516/2021/09/005} {\bibfield  {journal}
  {\bibinfo  {journal} {Journal of Cosmology and Astroparticle Physics}\
  }\textbf {\bibinfo {volume} {09}}\bibfield  {number} {\bibinfo  {number} {
  (09)},\ \bibinfo {pages} {005}},\ }\Eprint {https://arxiv.org/abs/2104.14881}
  {arxiv:2104.14881 [hep-ph]} \BibitemShut {NoStop}%
\bibitem [{\citenamefont {Feldman}\ \emph {et~al.}(2009)\citenamefont
  {Feldman}, \citenamefont {Liu},\ and\ \citenamefont {Nath}}]{Feldman:2008xs}%
  \BibitemOpen
  \bibfield  {author} {\bibinfo {author} {\bibfnamefont {D.}~\bibnamefont
  {Feldman}}, \bibinfo {author} {\bibfnamefont {Z.}~\bibnamefont {Liu}},\ and\
  \bibinfo {author} {\bibfnamefont {P.}~\bibnamefont {Nath}},\ }\href
  {https://doi.org/10.1103/PhysRevD.79.063509} {\bibfield  {journal} {\bibinfo
  {journal} {Physical Review D}\ }\textbf {\bibinfo {volume} {79}},\ \bibinfo
  {pages} {063509} (\bibinfo {year} {2009})},\ \Eprint
  {https://arxiv.org/abs/0810.5762} {arxiv:0810.5762 [hep-ph]} \BibitemShut
  {NoStop}%
\bibitem [{\citenamefont {Ibe}\ \emph {et~al.}(2009)\citenamefont {Ibe},
  \citenamefont {Murayama},\ and\ \citenamefont {Yanagida}}]{Ibe:2008ye}%
  \BibitemOpen
  \bibfield  {author} {\bibinfo {author} {\bibfnamefont {M.}~\bibnamefont
  {Ibe}}, \bibinfo {author} {\bibfnamefont {H.}~\bibnamefont {Murayama}},\ and\
  \bibinfo {author} {\bibfnamefont {T.~T.}\ \bibnamefont {Yanagida}},\ }\href
  {https://doi.org/10.1103/PhysRevD.79.095009} {\bibfield  {journal} {\bibinfo
  {journal} {Physical Review D}\ }\textbf {\bibinfo {volume} {79}},\ \bibinfo
  {pages} {095009} (\bibinfo {year} {2009})},\ \Eprint
  {https://arxiv.org/abs/0812.0072} {arxiv:0812.0072 [hep-ph]} \BibitemShut
  {NoStop}%
\bibitem [{\citenamefont {Bi}\ \emph {et~al.}(2012)\citenamefont {Bi},
  \citenamefont {Yin},\ and\ \citenamefont {Yuan}}]{Bi:2011qm}%
  \BibitemOpen
  \bibfield  {author} {\bibinfo {author} {\bibfnamefont {X.-J.}\ \bibnamefont
  {Bi}}, \bibinfo {author} {\bibfnamefont {P.-F.}\ \bibnamefont {Yin}},\ and\
  \bibinfo {author} {\bibfnamefont {Q.}~\bibnamefont {Yuan}},\ }\href
  {https://doi.org/10.1103/PhysRevD.85.043526} {\bibfield  {journal} {\bibinfo
  {journal} {Physical Review D}\ }\textbf {\bibinfo {volume} {85}},\ \bibinfo
  {pages} {043526} (\bibinfo {year} {2012})},\ \Eprint
  {https://arxiv.org/abs/1106.6027} {arxiv:1106.6027 [hep-ph]} \BibitemShut
  {NoStop}%
\bibitem [{\citenamefont {Bai}\ \emph {et~al.}(2018)\citenamefont {Bai},
  \citenamefont {Berger},\ and\ \citenamefont {Lu}}]{Bai:2017fav}%
  \BibitemOpen
  \bibfield  {author} {\bibinfo {author} {\bibfnamefont {Y.}~\bibnamefont
  {Bai}}, \bibinfo {author} {\bibfnamefont {J.}~\bibnamefont {Berger}},\ and\
  \bibinfo {author} {\bibfnamefont {S.}~\bibnamefont {Lu}},\ }\href
  {https://doi.org/10.1103/PhysRevD.97.115012} {\bibfield  {journal} {\bibinfo
  {journal} {Physical Review D}\ }\textbf {\bibinfo {volume} {97}},\ \bibinfo
  {pages} {115012} (\bibinfo {year} {2018})},\ \Eprint
  {https://arxiv.org/abs/1706.09974} {arxiv:1706.09974 [hep-ph]} \BibitemShut
  {NoStop}%
\bibitem [{\citenamefont {Xiang}\ \emph {et~al.}(2017)\citenamefont {Xiang},
  \citenamefont {Bi}, \citenamefont {Lin},\ and\ \citenamefont
  {Yin}}]{Xiang:2017jou}%
  \BibitemOpen
  \bibfield  {author} {\bibinfo {author} {\bibfnamefont {Q.-F.}\ \bibnamefont
  {Xiang}}, \bibinfo {author} {\bibfnamefont {X.-J.}\ \bibnamefont {Bi}},
  \bibinfo {author} {\bibfnamefont {S.-J.}\ \bibnamefont {Lin}},\ and\ \bibinfo
  {author} {\bibfnamefont {P.-F.}\ \bibnamefont {Yin}},\ }\href
  {https://doi.org/10.1016/j.physletb.2017.09.003} {\bibfield  {journal}
  {\bibinfo  {journal} {Physics Letters B}\ }\textbf {\bibinfo {volume}
  {773}},\ \bibinfo {pages} {448} (\bibinfo {year} {2017})},\ \Eprint
  {https://arxiv.org/abs/1707.09313} {arxiv:1707.09313 [astro-ph.HE]}
  \BibitemShut {NoStop}%
\bibitem [{\citenamefont {Hektor}\ \emph {et~al.}(2014)\citenamefont {Hektor},
  \citenamefont {Raidal}, \citenamefont {Strumia},\ and\ \citenamefont
  {Tempel}}]{Hektor:2013yga}%
  \BibitemOpen
  \bibfield  {author} {\bibinfo {author} {\bibfnamefont {A.}~\bibnamefont
  {Hektor}}, \bibinfo {author} {\bibfnamefont {M.}~\bibnamefont {Raidal}},
  \bibinfo {author} {\bibfnamefont {A.}~\bibnamefont {Strumia}},\ and\ \bibinfo
  {author} {\bibfnamefont {E.}~\bibnamefont {Tempel}},\ }\href
  {https://doi.org/10.1016/j.physletb.2013.11.017} {\bibfield  {journal}
  {\bibinfo  {journal} {Physics Letters B}\ }\textbf {\bibinfo {volume}
  {728}},\ \bibinfo {pages} {58} (\bibinfo {year} {2014})},\ \Eprint
  {https://arxiv.org/abs/1307.2561} {arxiv:1307.2561 [hep-ph]} \BibitemShut
  {NoStop}%
\bibitem [{\citenamefont {Fang}(2022)}]{Fang:2022fof}%
  \BibitemOpen
  \bibfield  {author} {\bibinfo {author} {\bibfnamefont {K.}~\bibnamefont
  {Fang}},\ }\href {https://doi.org/10.3389/fspas.2022.1022100} {\bibfield
  {journal} {\bibinfo  {journal} {Front. Astron. Space Sci.}\ }\textbf
  {\bibinfo {volume} {9}},\ \bibinfo {pages} {1022100} (\bibinfo {year}
  {2022})},\ \Eprint {https://arxiv.org/abs/2209.13294} {arxiv:2209.13294
  [astro-ph.HE]} \BibitemShut {NoStop}%
\bibitem [{\citenamefont {Abeysekara}\ \emph {et~al.}(2017)\citenamefont
  {Abeysekara} \emph {et~al.}}]{HAWC:2017kbo}%
  \BibitemOpen
  \bibfield  {author} {\bibinfo {author} {\bibfnamefont {A.~U.}\ \bibnamefont
  {Abeysekara}} \emph {et~al.} (\bibinfo {collaboration} {HAWC}),\ }\href
  {https://doi.org/10.1126/science.aan4880} {\bibfield  {journal} {\bibinfo
  {journal} {Science}\ }\textbf {\bibinfo {volume} {358}},\ \bibinfo {pages}
  {911} (\bibinfo {year} {2017})},\ \Eprint {https://arxiv.org/abs/1711.06223}
  {arxiv:1711.06223 [astro-ph.HE]} \BibitemShut {NoStop}%
\bibitem [{\citenamefont {Hooper}\ \emph {et~al.}(2017)\citenamefont {Hooper},
  \citenamefont {Cholis}, \citenamefont {Linden},\ and\ \citenamefont
  {Fang}}]{Hooper:2017gtd}%
  \BibitemOpen
  \bibfield  {author} {\bibinfo {author} {\bibfnamefont {D.}~\bibnamefont
  {Hooper}}, \bibinfo {author} {\bibfnamefont {I.}~\bibnamefont {Cholis}},
  \bibinfo {author} {\bibfnamefont {T.}~\bibnamefont {Linden}},\ and\ \bibinfo
  {author} {\bibfnamefont {K.}~\bibnamefont {Fang}},\ }\href
  {https://doi.org/10.1103/PhysRevD.96.103013} {\bibfield  {journal} {\bibinfo
  {journal} {Physical Review D}\ }\textbf {\bibinfo {volume} {96}},\ \bibinfo
  {pages} {103013} (\bibinfo {year} {2017})},\ \Eprint
  {https://arxiv.org/abs/1702.08436} {arxiv:1702.08436 [astro-ph.HE]}
  \BibitemShut {NoStop}%
\bibitem [{\citenamefont {Aharonian}\ \emph {et~al.}(2021)\citenamefont
  {Aharonian} \emph {et~al.}}]{LHAASO:2021crt}%
  \BibitemOpen
  \bibfield  {author} {\bibinfo {author} {\bibnamefont {Aharonian}} \emph
  {et~al.} (\bibinfo {collaboration} {LHAASO}),\ }\href
  {https://doi.org/10.1103/PhysRevLett.126.241103} {\bibfield  {journal}
  {\bibinfo  {journal} {Physical Review Letters}\ }\textbf {\bibinfo {volume}
  {126}},\ \bibinfo {pages} {241103} (\bibinfo {year} {2021})},\ \Eprint
  {https://arxiv.org/abs/2106.09396} {arxiv:2106.09396 [astro-ph.HE]}
  \BibitemShut {NoStop}%
\bibitem [{\citenamefont
  {Han}(2017)}]{hanObservingInterstellarIntergalactic2017}%
  \BibitemOpen
  \bibfield  {author} {\bibinfo {author} {\bibfnamefont {J.}~\bibnamefont
  {Han}},\ }\href {https://doi.org/10.1146/annurev-astro-091916-055221}
  {\bibfield  {journal} {\bibinfo  {journal} {Annual Review of Astronomy and
  Astrophysics}\ }\textbf {\bibinfo {volume} {55}},\ \bibinfo {pages} {111}
  (\bibinfo {year} {2017})}\BibitemShut {NoStop}%
\bibitem [{\citenamefont {Zhao}\ \emph {et~al.}(2021)\citenamefont {Zhao},
  \citenamefont {Fang},\ and\ \citenamefont {Bi}}]{Zhao:2021yzf}%
  \BibitemOpen
  \bibfield  {author} {\bibinfo {author} {\bibfnamefont {M.-J.}\ \bibnamefont
  {Zhao}}, \bibinfo {author} {\bibfnamefont {K.}~\bibnamefont {Fang}},\ and\
  \bibinfo {author} {\bibfnamefont {X.-J.}\ \bibnamefont {Bi}},\ }\href
  {https://doi.org/10.1103/PhysRevD.104.123001} {\bibfield  {journal} {\bibinfo
   {journal} {Physical Review D}\ }\textbf {\bibinfo {volume} {104}},\ \bibinfo
  {pages} {123001} (\bibinfo {year} {2021})},\ \Eprint
  {https://arxiv.org/abs/2109.04112} {arxiv:2109.04112 [astro-ph.HE]}
  \BibitemShut {NoStop}%
\bibitem [{\citenamefont {Panov}\ \emph {et~al.}(2007)\citenamefont {Panov}
  \emph {et~al.}}]{Panov:2006kf}%
  \BibitemOpen
  \bibfield  {author} {\bibinfo {author} {\bibfnamefont {A.~D.}\ \bibnamefont
  {Panov}} \emph {et~al.},\ }\href {https://doi.org/10.3103/S1062873807040168}
  {\bibfield  {journal} {\bibinfo  {journal} {Bull. Russ. Acad. Sci. Phys.}\
  }\textbf {\bibinfo {volume} {71}},\ \bibinfo {pages} {494} (\bibinfo {year}
  {2007})},\ \Eprint {https://arxiv.org/abs/astro-ph/0612377}
  {arXiv:astro-ph/0612377} \BibitemShut {NoStop}%
\bibitem [{\citenamefont {Panov}\ \emph {et~al.}(2009)\citenamefont {Panov},
  \citenamefont {Adams}, \citenamefont {Ahn}, \citenamefont {Bashinzhagyan},
  \citenamefont {Watts}, \citenamefont {Wefel}, \citenamefont {Wu},
  \citenamefont {Ganel}, \citenamefont {Guzik}, \citenamefont {Zatsepin},
  \citenamefont {Isbert}, \citenamefont {Kim}, \citenamefont {Christl},
  \citenamefont {Kouznetsov}, \citenamefont {Panasyuk}, \citenamefont {Seo},
  \citenamefont {Sokolskaya}, \citenamefont {Chang}, \citenamefont {Schmidt},\
  and\ \citenamefont {Fazely}}]{Panov:2009iih}%
  \BibitemOpen
  \bibfield  {author} {\bibinfo {author} {\bibfnamefont {A.~D.}\ \bibnamefont
  {Panov}}, \bibinfo {author} {\bibfnamefont {J.~H.}\ \bibnamefont {Adams}},
  \bibinfo {author} {\bibfnamefont {H.~S.}\ \bibnamefont {Ahn}}, \bibinfo
  {author} {\bibfnamefont {G.~L.}\ \bibnamefont {Bashinzhagyan}}, \bibinfo
  {author} {\bibfnamefont {J.~W.}\ \bibnamefont {Watts}}, \bibinfo {author}
  {\bibfnamefont {J.~P.}\ \bibnamefont {Wefel}}, \bibinfo {author}
  {\bibfnamefont {J.}~\bibnamefont {Wu}}, \bibinfo {author} {\bibfnamefont
  {O.}~\bibnamefont {Ganel}}, \bibinfo {author} {\bibfnamefont {T.~G.}\
  \bibnamefont {Guzik}}, \bibinfo {author} {\bibfnamefont {V.~I.}\ \bibnamefont
  {Zatsepin}}, \bibinfo {author} {\bibfnamefont {I.}~\bibnamefont {Isbert}},
  \bibinfo {author} {\bibfnamefont {K.~C.}\ \bibnamefont {Kim}}, \bibinfo
  {author} {\bibfnamefont {M.}~\bibnamefont {Christl}}, \bibinfo {author}
  {\bibfnamefont {E.~N.}\ \bibnamefont {Kouznetsov}}, \bibinfo {author}
  {\bibfnamefont {M.~I.}\ \bibnamefont {Panasyuk}}, \bibinfo {author}
  {\bibfnamefont {E.~S.}\ \bibnamefont {Seo}}, \bibinfo {author} {\bibfnamefont
  {N.~V.}\ \bibnamefont {Sokolskaya}}, \bibinfo {author} {\bibfnamefont
  {J.}~\bibnamefont {Chang}}, \bibinfo {author} {\bibfnamefont {W.~K.~H.}\
  \bibnamefont {Schmidt}},\ and\ \bibinfo {author} {\bibfnamefont {A.~R.}\
  \bibnamefont {Fazely}},\ }\href {https://doi.org/10.3103/S1062873809050098}
  {\bibfield  {journal} {\bibinfo  {journal} {Bulletin of the Russian Academy
  of Sciences: Physics}\ }\textbf {\bibinfo {volume} {73}},\ \bibinfo {pages}
  {564} (\bibinfo {year} {2009})},\ \Eprint {https://arxiv.org/abs/1101.3246}
  {arxiv:1101.3246 [astro-ph.HE]} \BibitemShut {NoStop}%
\bibitem [{\citenamefont {Ahn}\ \emph {et~al.}(2010)\citenamefont {Ahn},
  \citenamefont {Allison}, \citenamefont {Bagliesi}, \citenamefont {Beatty},
  \citenamefont {Bigongiari}, \citenamefont {Childers}, \citenamefont
  {Conklin}, \citenamefont {Coutu}, \citenamefont {DuVernois}, \citenamefont
  {Ganel}, \citenamefont {Han}, \citenamefont {Jeon}, \citenamefont {Kim},
  \citenamefont {Lee}, \citenamefont {Lutz}, \citenamefont {Maestro},
  \citenamefont {Malinin}, \citenamefont {Marrocchesi}, \citenamefont
  {Minnick}, \citenamefont {Mognet}, \citenamefont {Nam}, \citenamefont {Nam},
  \citenamefont {Nutter}, \citenamefont {Park}, \citenamefont {Park},
  \citenamefont {Seo}, \citenamefont {Sina}, \citenamefont {Wu}, \citenamefont
  {Yang}, \citenamefont {Yoon}, \citenamefont {Zei},\ and\ \citenamefont
  {Zinn}}]{Ahn:2010gv}%
  \BibitemOpen
  \bibfield  {author} {\bibinfo {author} {\bibfnamefont {H.~S.}\ \bibnamefont
  {Ahn}}, \bibinfo {author} {\bibfnamefont {P.}~\bibnamefont {Allison}},
  \bibinfo {author} {\bibfnamefont {M.~G.}\ \bibnamefont {Bagliesi}}, \bibinfo
  {author} {\bibfnamefont {J.~J.}\ \bibnamefont {Beatty}}, \bibinfo {author}
  {\bibfnamefont {G.}~\bibnamefont {Bigongiari}}, \bibinfo {author}
  {\bibfnamefont {J.~T.}\ \bibnamefont {Childers}}, \bibinfo {author}
  {\bibfnamefont {N.~B.}\ \bibnamefont {Conklin}}, \bibinfo {author}
  {\bibfnamefont {S.}~\bibnamefont {Coutu}}, \bibinfo {author} {\bibfnamefont
  {M.~A.}\ \bibnamefont {DuVernois}}, \bibinfo {author} {\bibfnamefont
  {O.}~\bibnamefont {Ganel}}, \bibinfo {author} {\bibfnamefont {J.~H.}\
  \bibnamefont {Han}}, \bibinfo {author} {\bibfnamefont {J.~A.}\ \bibnamefont
  {Jeon}}, \bibinfo {author} {\bibfnamefont {K.~C.}\ \bibnamefont {Kim}},
  \bibinfo {author} {\bibfnamefont {M.~H.}\ \bibnamefont {Lee}}, \bibinfo
  {author} {\bibfnamefont {L.}~\bibnamefont {Lutz}}, \bibinfo {author}
  {\bibfnamefont {P.}~\bibnamefont {Maestro}}, \bibinfo {author} {\bibfnamefont
  {A.}~\bibnamefont {Malinin}}, \bibinfo {author} {\bibfnamefont {P.~S.}\
  \bibnamefont {Marrocchesi}}, \bibinfo {author} {\bibfnamefont
  {S.}~\bibnamefont {Minnick}}, \bibinfo {author} {\bibfnamefont {S.~I.}\
  \bibnamefont {Mognet}}, \bibinfo {author} {\bibfnamefont {J.}~\bibnamefont
  {Nam}}, \bibinfo {author} {\bibfnamefont {S.}~\bibnamefont {Nam}}, \bibinfo
  {author} {\bibfnamefont {S.~L.}\ \bibnamefont {Nutter}}, \bibinfo {author}
  {\bibfnamefont {I.~H.}\ \bibnamefont {Park}}, \bibinfo {author}
  {\bibfnamefont {N.~H.}\ \bibnamefont {Park}}, \bibinfo {author}
  {\bibfnamefont {E.~S.}\ \bibnamefont {Seo}}, \bibinfo {author} {\bibfnamefont
  {R.}~\bibnamefont {Sina}}, \bibinfo {author} {\bibfnamefont {J.}~\bibnamefont
  {Wu}}, \bibinfo {author} {\bibfnamefont {J.}~\bibnamefont {Yang}}, \bibinfo
  {author} {\bibfnamefont {Y.~S.}\ \bibnamefont {Yoon}}, \bibinfo {author}
  {\bibfnamefont {R.}~\bibnamefont {Zei}},\ and\ \bibinfo {author}
  {\bibfnamefont {S.~Y.}\ \bibnamefont {Zinn}},\ }\href
  {https://doi.org/10.1088/2041-8205/714/1/L89} {\bibfield  {journal} {\bibinfo
   {journal} {The Astrophysical Journal}\ }\textbf {\bibinfo {volume} {714}},\
  \bibinfo {pages} {L89} (\bibinfo {year} {2010})},\ \Eprint
  {https://arxiv.org/abs/1004.1123} {arxiv:1004.1123 [astro-ph.HE]}
  \BibitemShut {NoStop}%
\bibitem [{\citenamefont {Yoon}\ \emph {et~al.}(2017)\citenamefont {Yoon},
  \citenamefont {Anderson}, \citenamefont {Barrau}, \citenamefont {Conklin},
  \citenamefont {Coutu}, \citenamefont {Derome}, \citenamefont {Han},
  \citenamefont {Jeon}, \citenamefont {Kim}, \citenamefont {Kim}, \citenamefont
  {Lee}, \citenamefont {Lee}, \citenamefont {Lee}, \citenamefont {Lee},
  \citenamefont {Link}, \citenamefont {{Menchaca-Rocha}}, \citenamefont
  {Mitchell}, \citenamefont {Mognet}, \citenamefont {Nutter}, \citenamefont
  {Park}, \citenamefont {{Picot-Clemente}}, \citenamefont {Putze},
  \citenamefont {Seo}, \citenamefont {Smith},\ and\ \citenamefont
  {Wu}}]{Yoon:2017qjx}%
  \BibitemOpen
  \bibfield  {author} {\bibinfo {author} {\bibfnamefont {Y.~S.}\ \bibnamefont
  {Yoon}}, \bibinfo {author} {\bibfnamefont {T.}~\bibnamefont {Anderson}},
  \bibinfo {author} {\bibfnamefont {A.}~\bibnamefont {Barrau}}, \bibinfo
  {author} {\bibfnamefont {N.~B.}\ \bibnamefont {Conklin}}, \bibinfo {author}
  {\bibfnamefont {S.}~\bibnamefont {Coutu}}, \bibinfo {author} {\bibfnamefont
  {L.}~\bibnamefont {Derome}}, \bibinfo {author} {\bibfnamefont {J.~H.}\
  \bibnamefont {Han}}, \bibinfo {author} {\bibfnamefont {J.~A.}\ \bibnamefont
  {Jeon}}, \bibinfo {author} {\bibfnamefont {K.~C.}\ \bibnamefont {Kim}},
  \bibinfo {author} {\bibfnamefont {M.~H.}\ \bibnamefont {Kim}}, \bibinfo
  {author} {\bibfnamefont {H.~Y.}\ \bibnamefont {Lee}}, \bibinfo {author}
  {\bibfnamefont {J.}~\bibnamefont {Lee}}, \bibinfo {author} {\bibfnamefont
  {M.~H.}\ \bibnamefont {Lee}}, \bibinfo {author} {\bibfnamefont {S.~E.}\
  \bibnamefont {Lee}}, \bibinfo {author} {\bibfnamefont {J.~T.}\ \bibnamefont
  {Link}}, \bibinfo {author} {\bibfnamefont {A.}~\bibnamefont
  {{Menchaca-Rocha}}}, \bibinfo {author} {\bibfnamefont {J.~W.}\ \bibnamefont
  {Mitchell}}, \bibinfo {author} {\bibfnamefont {S.~I.}\ \bibnamefont
  {Mognet}}, \bibinfo {author} {\bibfnamefont {S.}~\bibnamefont {Nutter}},
  \bibinfo {author} {\bibfnamefont {I.~H.}\ \bibnamefont {Park}}, \bibinfo
  {author} {\bibfnamefont {N.}~\bibnamefont {{Picot-Clemente}}}, \bibinfo
  {author} {\bibfnamefont {A.}~\bibnamefont {Putze}}, \bibinfo {author}
  {\bibfnamefont {E.~S.}\ \bibnamefont {Seo}}, \bibinfo {author} {\bibfnamefont
  {J.}~\bibnamefont {Smith}},\ and\ \bibinfo {author} {\bibfnamefont
  {J.}~\bibnamefont {Wu}},\ }\href {https://doi.org/10.3847/1538-4357/aa68e4}
  {\bibfield  {journal} {\bibinfo  {journal} {The Astrophysical Journal}\
  }\textbf {\bibinfo {volume} {839}},\ \bibinfo {pages} {5} (\bibinfo {year}
  {2017})},\ \Eprint {https://arxiv.org/abs/1704.02512} {arxiv:1704.02512
  [astro-ph.HE]} \BibitemShut {NoStop}%
\bibitem [{\citenamefont {Adriani}\ \emph {et~al.}(2011)\citenamefont {Adriani}
  \emph {et~al.}}]{PAMELA:2011mvy}%
  \BibitemOpen
  \bibfield  {author} {\bibinfo {author} {\bibfnamefont {O.}~\bibnamefont
  {Adriani}} \emph {et~al.} (\bibinfo {collaboration} {PAMELA}),\ }\href
  {https://doi.org/10.1126/science.1199172} {\bibfield  {journal} {\bibinfo
  {journal} {Science}\ }\textbf {\bibinfo {volume} {332}},\ \bibinfo {pages}
  {69} (\bibinfo {year} {2011})},\ \Eprint {https://arxiv.org/abs/1103.4055}
  {arxiv:1103.4055 [astro-ph.HE]} \BibitemShut {NoStop}%
\bibitem [{\citenamefont {Aguilar}\ \emph
  {et~al.}(2015{\natexlab{a}})\citenamefont {Aguilar}, \citenamefont {Aisa},
  \citenamefont {Alpat} \emph {et~al.}}]{AMS:2015azc}%
  \BibitemOpen
  \bibfield  {author} {\bibinfo {author} {\bibfnamefont {M.}~\bibnamefont
  {Aguilar}}, \bibinfo {author} {\bibfnamefont {D.}~\bibnamefont {Aisa}},
  \bibinfo {author} {\bibfnamefont {B.}~\bibnamefont {Alpat}}, \emph {et~al.}
  (\bibinfo {collaboration} {AMS}),\ }\href
  {https://doi.org/10.1103/PhysRevLett.115.211101} {\bibfield  {journal}
  {\bibinfo  {journal} {Physical Review Letters}\ }\textbf {\bibinfo {volume}
  {115}},\ \bibinfo {pages} {211101} (\bibinfo {year}
  {2015}{\natexlab{a}})}\BibitemShut {NoStop}%
\bibitem [{\citenamefont {Aguilar}\ \emph
  {et~al.}(2015{\natexlab{b}})\citenamefont {Aguilar}, \citenamefont {Aisa},
  \citenamefont {Alpat} \emph {et~al.}}]{AMS:2015tnn}%
  \BibitemOpen
  \bibfield  {author} {\bibinfo {author} {\bibfnamefont {M.}~\bibnamefont
  {Aguilar}}, \bibinfo {author} {\bibfnamefont {D.}~\bibnamefont {Aisa}},
  \bibinfo {author} {\bibfnamefont {B.}~\bibnamefont {Alpat}}, \emph {et~al.}
  (\bibinfo {collaboration} {AMS}),\ }\href
  {https://doi.org/10.1103/PhysRevLett.114.171103} {\bibfield  {journal}
  {\bibinfo  {journal} {Physical Review Letters}\ }\textbf {\bibinfo {volume}
  {114}},\ \bibinfo {pages} {171103} (\bibinfo {year}
  {2015}{\natexlab{b}})}\BibitemShut {NoStop}%
\bibitem [{\citenamefont {Blasi}\ and\ \citenamefont
  {Amato}(2012)}]{Blasi:2011fm}%
  \BibitemOpen
  \bibfield  {author} {\bibinfo {author} {\bibfnamefont {P.}~\bibnamefont
  {Blasi}}\ and\ \bibinfo {author} {\bibfnamefont {E.}~\bibnamefont {Amato}},\
  }\href {https://doi.org/10.1088/1475-7516/2012/01/011} {\bibfield  {journal}
  {\bibinfo  {journal} {Journal of Cosmology and Astroparticle Physics}\
  }\textbf {\bibinfo {volume} {01}}\bibfield  {number} {\bibinfo  {number} {
  (01)},\ \bibinfo {pages} {011}},\ }\Eprint {https://arxiv.org/abs/1105.4529}
  {arxiv:1105.4529 [astro-ph.HE]} \BibitemShut {NoStop}%
\bibitem [{\citenamefont {Ahlers}\ and\ \citenamefont
  {Mertsch}(2017)}]{Ahlers:2016rox}%
  \BibitemOpen
  \bibfield  {author} {\bibinfo {author} {\bibfnamefont {M.}~\bibnamefont
  {Ahlers}}\ and\ \bibinfo {author} {\bibfnamefont {P.}~\bibnamefont
  {Mertsch}},\ }\href {https://doi.org/10.1016/j.ppnp.2017.01.004} {\bibfield
  {journal} {\bibinfo  {journal} {Progress in Particle and Nuclear Physics}\
  }\textbf {\bibinfo {volume} {94}},\ \bibinfo {pages} {184} (\bibinfo {year}
  {2017})},\ \Eprint {https://arxiv.org/abs/1612.01873} {arxiv:1612.01873
  [astro-ph.HE]} \BibitemShut {NoStop}%
\bibitem [{\citenamefont {Gamerman}(1997)}]{gamermanMarkovChainMonte1997}%
  \BibitemOpen
  \bibfield  {author} {\bibinfo {author} {\bibfnamefont {D.}~\bibnamefont
  {Gamerman}},\ }\href@noop {} {\emph {\bibinfo {title} {Markov Chain {{Monte
  Carlo}}: Stochastic Simulation for {{Bayesian}} Inference}}},\ Chapman and
  {{Hall}} Texts in Statistical Science Series\ (\bibinfo  {publisher}
  {{Chapman \& Hall}},\ \bibinfo {address} {{London}},\ \bibinfo {year}
  {1997})\BibitemShut {NoStop}%
\bibitem [{\citenamefont {Orusa}\ \emph {et~al.}(2022)\citenamefont {Orusa},
  \citenamefont {Di~Mauro}, \citenamefont {Donato},\ and\ \citenamefont
  {Korsmeier}}]{Orusa:2022pvp}%
  \BibitemOpen
  \bibfield  {author} {\bibinfo {author} {\bibfnamefont {L.}~\bibnamefont
  {Orusa}}, \bibinfo {author} {\bibfnamefont {M.}~\bibnamefont {Di~Mauro}},
  \bibinfo {author} {\bibfnamefont {F.}~\bibnamefont {Donato}},\ and\ \bibinfo
  {author} {\bibfnamefont {M.}~\bibnamefont {Korsmeier}},\ }\href
  {https://doi.org/10.1103/PhysRevD.105.123021} {\bibfield  {journal} {\bibinfo
   {journal} {Physical Review D}\ }\textbf {\bibinfo {volume} {105}},\ \bibinfo
  {pages} {123021} (\bibinfo {year} {2022})},\ \Eprint
  {https://arxiv.org/abs/2203.13143} {arxiv:2203.13143 [astro-ph.HE]}
  \BibitemShut {NoStop}%
\bibitem [{\citenamefont {Potgieter}(2013)}]{Potgieter:2013pdj}%
  \BibitemOpen
  \bibfield  {author} {\bibinfo {author} {\bibfnamefont {M.~S.}\ \bibnamefont
  {Potgieter}},\ }\href {https://doi.org/10.12942/lrsp-2013-3} {\bibfield
  {journal} {\bibinfo  {journal} {Living Reviews in Solar Physics}\ }\textbf
  {\bibinfo {volume} {10}},\ \bibinfo {pages} {3} (\bibinfo {year} {2013})},\
  \Eprint {https://arxiv.org/abs/1306.4421} {arxiv:1306.4421
  [physics.space-ph]} \BibitemShut {NoStop}%
\bibitem [{\citenamefont {Aguilar}\ \emph {et~al.}(2021)\citenamefont {Aguilar}
  \emph {et~al.}}]{AMS:2021nhj}%
  \BibitemOpen
  \bibfield  {author} {\bibinfo {author} {\bibfnamefont {M.}~\bibnamefont
  {Aguilar}} \emph {et~al.} (\bibinfo {collaboration} {AMS}),\ }\href
  {https://doi.org/10.1016/j.physrep.2020.09.003} {\bibfield  {journal}
  {\bibinfo  {journal} {Physics Reports}\ }\bibinfo {series} {The {{Alpha
  Magnetic Spectrometer}} ({{AMS}}) on the {{International Space Station}}:
  {{Part II}} - {{Results}} from the {{First Seven Years}}},\ \textbf {\bibinfo
  {volume} {894}},\ \bibinfo {pages} {1} (\bibinfo {year} {2021})}\BibitemShut
  {NoStop}%
\bibitem [{\citenamefont {Yuan}\ \emph {et~al.}(2014)\citenamefont {Yuan},
  \citenamefont {Bi}, \citenamefont {Chen}, \citenamefont {Guo}, \citenamefont
  {Lin},\ and\ \citenamefont {Zhang}}]{Yuan:2013eja}%
  \BibitemOpen
  \bibfield  {author} {\bibinfo {author} {\bibfnamefont {Q.}~\bibnamefont
  {Yuan}}, \bibinfo {author} {\bibfnamefont {X.-J.}\ \bibnamefont {Bi}},
  \bibinfo {author} {\bibfnamefont {G.-M.}\ \bibnamefont {Chen}}, \bibinfo
  {author} {\bibfnamefont {Y.-Q.}\ \bibnamefont {Guo}}, \bibinfo {author}
  {\bibfnamefont {S.-J.}\ \bibnamefont {Lin}},\ and\ \bibinfo {author}
  {\bibfnamefont {X.}~\bibnamefont {Zhang}},\ }\href
  {https://doi.org/10.1016/j.astropartphys.2014.05.005} {\bibfield  {journal}
  {\bibinfo  {journal} {Astroparticle Physics}\ }\textbf {\bibinfo {volume}
  {60}},\ \bibinfo {pages} {1} (\bibinfo {year} {2014})},\ \Eprint
  {https://arxiv.org/abs/1304.1482} {arxiv:1304.1482 [astro-ph.HE]}
  \BibitemShut {NoStop}%
\bibitem [{\citenamefont {Lewis}\ and\ \citenamefont
  {Bridle}(2002)}]{Lewis:2002ah}%
  \BibitemOpen
  \bibfield  {author} {\bibinfo {author} {\bibfnamefont {A.}~\bibnamefont
  {Lewis}}\ and\ \bibinfo {author} {\bibfnamefont {S.}~\bibnamefont {Bridle}},\
  }\href {https://doi.org/10.1103/PhysRevD.66.103511} {\bibfield  {journal}
  {\bibinfo  {journal} {Phys. Rev.}\ }\textbf {\bibinfo {volume} {D66}},\
  \bibinfo {pages} {103511} (\bibinfo {year} {2002})},\ \Eprint
  {https://arxiv.org/abs/astro-ph/0205436} {arXiv:astro-ph/0205436 [astro-ph]}
  \BibitemShut {NoStop}%
\bibitem [{\citenamefont {Lewis}(2013)}]{Lewis:2013hha}%
  \BibitemOpen
  \bibfield  {author} {\bibinfo {author} {\bibfnamefont {A.}~\bibnamefont
  {Lewis}},\ }\href {https://doi.org/10.1103/PhysRevD.87.103529} {\bibfield
  {journal} {\bibinfo  {journal} {Phys. Rev.}\ }\textbf {\bibinfo {volume}
  {D87}},\ \bibinfo {pages} {103529} (\bibinfo {year} {2013})},\ \Eprint
  {https://arxiv.org/abs/1304.4473} {arXiv:1304.4473 [astro-ph.CO]}
  \BibitemShut {NoStop}%
\bibitem [{\citenamefont {Torrado}\ and\ \citenamefont
  {Lewis}(2021)}]{Torrado:2020dgo}%
  \BibitemOpen
  \bibfield  {author} {\bibinfo {author} {\bibfnamefont {J.}~\bibnamefont
  {Torrado}}\ and\ \bibinfo {author} {\bibfnamefont {A.}~\bibnamefont
  {Lewis}},\ }\href {https://doi.org/10.1088/1475-7516/2021/05/057} {\bibfield
  {journal} {\bibinfo  {journal} {Journal of Cosmology and Astroparticle
  Physics}\ }\textbf {\bibinfo {volume} {2021}}\bibfield  {number} {\bibinfo
  {number} { (05)},\ \bibinfo {pages} {057}},\ }\Eprint
  {https://arxiv.org/abs/2005.05290} {arxiv:2005.05290 [astro-ph.IM]}
  \BibitemShut {NoStop}%
\bibitem [{\citenamefont {Strong}\ \emph {et~al.}(2007)\citenamefont {Strong},
  \citenamefont {Moskalenko},\ and\ \citenamefont {Ptuskin}}]{Strong:2007nh}%
  \BibitemOpen
  \bibfield  {author} {\bibinfo {author} {\bibfnamefont {A.~W.}\ \bibnamefont
  {Strong}}, \bibinfo {author} {\bibfnamefont {I.~V.}\ \bibnamefont
  {Moskalenko}},\ and\ \bibinfo {author} {\bibfnamefont {V.~S.}\ \bibnamefont
  {Ptuskin}},\ }\href {https://doi.org/10.1146/annurev.nucl.57.090506.123011}
  {\bibfield  {journal} {\bibinfo  {journal} {Annual Review of Nuclear and
  Particle Science}\ }\textbf {\bibinfo {volume} {57}},\ \bibinfo {pages} {285}
  (\bibinfo {year} {2007})},\ \Eprint {https://arxiv.org/abs/astro-ph/0701517}
  {arxiv:astro-ph/0701517} \BibitemShut {NoStop}%
\bibitem [{\citenamefont {Seo}\ and\ \citenamefont
  {Ptuskin}(1994)}]{seoStochasticReaccelerationCosmic1994}%
  \BibitemOpen
  \bibfield  {author} {\bibinfo {author} {\bibfnamefont {E.~S.}\ \bibnamefont
  {Seo}}\ and\ \bibinfo {author} {\bibfnamefont {V.~S.}\ \bibnamefont
  {Ptuskin}},\ }\href {https://doi.org/10.1086/174520} {\bibfield  {journal}
  {\bibinfo  {journal} {The Astrophysical Journal}\ }\textbf {\bibinfo {volume}
  {431}},\ \bibinfo {pages} {705} (\bibinfo {year} {1994})}\BibitemShut
  {NoStop}%
\bibitem [{\citenamefont {Berezinskii}\ \emph {et~al.}(1990)\citenamefont
  {Berezinskii}, \citenamefont {Bulanov}, \citenamefont {Dogiel},\ and\
  \citenamefont {Ptuskin}}]{berezinskiiAstrophysicsCosmicRays1990a}%
  \BibitemOpen
  \bibfield  {author} {\bibinfo {author} {\bibfnamefont {V.~S.}\ \bibnamefont
  {Berezinskii}}, \bibinfo {author} {\bibfnamefont {S.~V.}\ \bibnamefont
  {Bulanov}}, \bibinfo {author} {\bibfnamefont {V.~A.}\ \bibnamefont
  {Dogiel}},\ and\ \bibinfo {author} {\bibfnamefont {V.~S.}\ \bibnamefont
  {Ptuskin}},\ }\href@noop {} {\emph {\bibinfo {title} {Astrophysics of Cosmic
  Rays}}}\ (\bibinfo {year} {1990})\BibitemShut {NoStop}%
\bibitem [{\citenamefont {Yuan}\ \emph {et~al.}(2017)\citenamefont {Yuan},
  \citenamefont {Lin}, \citenamefont {Fang},\ and\ \citenamefont
  {Bi}}]{Yuan:2017ozr}%
  \BibitemOpen
  \bibfield  {author} {\bibinfo {author} {\bibfnamefont {Q.}~\bibnamefont
  {Yuan}}, \bibinfo {author} {\bibfnamefont {S.-J.}\ \bibnamefont {Lin}},
  \bibinfo {author} {\bibfnamefont {K.}~\bibnamefont {Fang}},\ and\ \bibinfo
  {author} {\bibfnamefont {X.-J.}\ \bibnamefont {Bi}},\ }\href
  {https://doi.org/10.1103/PhysRevD.95.083007} {\bibfield  {journal} {\bibinfo
  {journal} {Phys. Rev. D}\ }\textbf {\bibinfo {volume} {95}},\ \bibinfo
  {pages} {083007} (\bibinfo {year} {2017})},\ \Eprint
  {https://arxiv.org/abs/1701.06149} {arXiv:1701.06149 [astro-ph.HE]}
  \BibitemShut {NoStop}%
\bibitem [{\citenamefont {Weinrich}\ \emph {et~al.}(2020)\citenamefont
  {Weinrich}, \citenamefont {Boudaud}, \citenamefont {Derome}, \citenamefont
  {Genolini}, \citenamefont {Lavalle}, \citenamefont {Maurin}, \citenamefont
  {Salati}, \citenamefont {Serpico},\ and\ \citenamefont
  {Weymann-Despres}}]{Weinrich:2020ftb}%
  \BibitemOpen
  \bibfield  {author} {\bibinfo {author} {\bibfnamefont {N.}~\bibnamefont
  {Weinrich}}, \bibinfo {author} {\bibfnamefont {M.}~\bibnamefont {Boudaud}},
  \bibinfo {author} {\bibfnamefont {L.}~\bibnamefont {Derome}}, \bibinfo
  {author} {\bibfnamefont {Y.}~\bibnamefont {Genolini}}, \bibinfo {author}
  {\bibfnamefont {J.}~\bibnamefont {Lavalle}}, \bibinfo {author} {\bibfnamefont
  {D.}~\bibnamefont {Maurin}}, \bibinfo {author} {\bibfnamefont
  {P.}~\bibnamefont {Salati}}, \bibinfo {author} {\bibfnamefont
  {P.}~\bibnamefont {Serpico}},\ and\ \bibinfo {author} {\bibfnamefont
  {G.}~\bibnamefont {Weymann-Despres}},\ }\href
  {https://doi.org/10.1051/0004-6361/202038064} {\bibfield  {journal} {\bibinfo
   {journal} {Astron. Astrophys.}\ }\textbf {\bibinfo {volume} {639}},\
  \bibinfo {pages} {A74} (\bibinfo {year} {2020})},\ \Eprint
  {https://arxiv.org/abs/2004.00441} {arXiv:2004.00441 [astro-ph.HE]}
  \BibitemShut {NoStop}%
\bibitem [{\citenamefont {Genolini}\ \emph {et~al.}(2019)\citenamefont
  {Genolini}, \citenamefont {Boudaud}, \citenamefont {Batista}, \citenamefont
  {Caroff}, \citenamefont {Derome}, \citenamefont {Lavalle}, \citenamefont
  {Marcowith}, \citenamefont {Maurin}, \citenamefont {Poireau}, \citenamefont
  {Poulin}, \citenamefont {Rosier}, \citenamefont {Salati}, \citenamefont
  {Serpico},\ and\ \citenamefont {Vecchi}}]{Genolini:2019ewc}%
  \BibitemOpen
  \bibfield  {author} {\bibinfo {author} {\bibfnamefont {Y.}~\bibnamefont
  {Genolini}}, \bibinfo {author} {\bibfnamefont {M.}~\bibnamefont {Boudaud}},
  \bibinfo {author} {\bibfnamefont {P.~I.}\ \bibnamefont {Batista}}, \bibinfo
  {author} {\bibfnamefont {S.}~\bibnamefont {Caroff}}, \bibinfo {author}
  {\bibfnamefont {L.}~\bibnamefont {Derome}}, \bibinfo {author} {\bibfnamefont
  {J.}~\bibnamefont {Lavalle}}, \bibinfo {author} {\bibfnamefont
  {A.}~\bibnamefont {Marcowith}}, \bibinfo {author} {\bibfnamefont
  {D.}~\bibnamefont {Maurin}}, \bibinfo {author} {\bibfnamefont
  {V.}~\bibnamefont {Poireau}}, \bibinfo {author} {\bibfnamefont
  {V.}~\bibnamefont {Poulin}}, \bibinfo {author} {\bibfnamefont
  {S.}~\bibnamefont {Rosier}}, \bibinfo {author} {\bibfnamefont
  {P.}~\bibnamefont {Salati}}, \bibinfo {author} {\bibfnamefont {P.~D.}\
  \bibnamefont {Serpico}},\ and\ \bibinfo {author} {\bibfnamefont
  {M.}~\bibnamefont {Vecchi}},\ }\href
  {https://doi.org/10.1103/PhysRevD.99.123028} {\bibfield  {journal} {\bibinfo
  {journal} {Physical Review D}\ }\textbf {\bibinfo {volume} {99}},\ \bibinfo
  {pages} {123028} (\bibinfo {year} {2019})},\ \Eprint
  {https://arxiv.org/abs/1904.08917} {arxiv:1904.08917 [astro-ph.HE]}
  \BibitemShut {NoStop}%
\bibitem [{\citenamefont {Porter}\ \emph {et~al.}(2022)\citenamefont {Porter},
  \citenamefont {Johannesson},\ and\ \citenamefont
  {Moskalenko}}]{Porter:2021tlr}%
  \BibitemOpen
  \bibfield  {author} {\bibinfo {author} {\bibfnamefont {T.~A.}\ \bibnamefont
  {Porter}}, \bibinfo {author} {\bibfnamefont {G.}~\bibnamefont
  {Johannesson}},\ and\ \bibinfo {author} {\bibfnamefont {I.~V.}\ \bibnamefont
  {Moskalenko}},\ }\href {https://doi.org/10.3847/1538-4365/ac80f6} {\bibfield
  {journal} {\bibinfo  {journal} {The Astrophysical Journal Supplement Series}\
  }\textbf {\bibinfo {volume} {262}},\ \bibinfo {pages} {30} (\bibinfo {year}
  {2022})},\ \Eprint {https://arxiv.org/abs/2112.12745} {arxiv:2112.12745
  [astro-ph.HE]} \BibitemShut {NoStop}%
\bibitem [{\citenamefont {Porter}\ and\ \citenamefont
  {Strong}(2005)}]{Porter:2005qx}%
  \BibitemOpen
  \bibfield  {author} {\bibinfo {author} {\bibfnamefont {T.~A.}\ \bibnamefont
  {Porter}}\ and\ \bibinfo {author} {\bibfnamefont {A.~W.}\ \bibnamefont
  {Strong}},\ }\bibfield  {journal} {\bibinfo  {journal} {arXiv preprint
  arXiv:astro-ph/0507119}\ }\href
  {https://doi.org/10.48550/arXiv.astro-ph/0507119}
  {10.48550/arXiv.astro-ph/0507119} (\bibinfo {year} {2005}),\ \Eprint
  {https://arxiv.org/abs/astro-ph/0507119} {arxiv:astro-ph/0507119}
  \BibitemShut {NoStop}%
\bibitem [{\citenamefont {Pshirkov}\ \emph {et~al.}(2011)\citenamefont
  {Pshirkov}, \citenamefont {Tinyakov}, \citenamefont {Kronberg},\ and\
  \citenamefont {{Newton-McGee}}}]{Pshirkov:2011um}%
  \BibitemOpen
  \bibfield  {author} {\bibinfo {author} {\bibfnamefont {M.~S.}\ \bibnamefont
  {Pshirkov}}, \bibinfo {author} {\bibfnamefont {P.~G.}\ \bibnamefont
  {Tinyakov}}, \bibinfo {author} {\bibfnamefont {P.~P.}\ \bibnamefont
  {Kronberg}},\ and\ \bibinfo {author} {\bibfnamefont {K.~J.}\ \bibnamefont
  {{Newton-McGee}}},\ }\href {https://doi.org/10.1088/0004-637X/738/2/192}
  {\bibfield  {journal} {\bibinfo  {journal} {The Astrophysical Journal}\
  }\textbf {\bibinfo {volume} {738}},\ \bibinfo {pages} {192} (\bibinfo {year}
  {2011})},\ \Eprint {https://arxiv.org/abs/1103.0814} {arxiv:1103.0814
  [astro-ph.GA]} \BibitemShut {NoStop}%
\bibitem [{\citenamefont {Sun}\ \emph {et~al.}(2008)\citenamefont {Sun},
  \citenamefont {Reich}, \citenamefont {Waelkens},\ and\ \citenamefont
  {En{\ss}lin}}]{Sun:2007mx}%
  \BibitemOpen
  \bibfield  {author} {\bibinfo {author} {\bibfnamefont {X.~H.}\ \bibnamefont
  {Sun}}, \bibinfo {author} {\bibfnamefont {W.}~\bibnamefont {Reich}}, \bibinfo
  {author} {\bibfnamefont {A.}~\bibnamefont {Waelkens}},\ and\ \bibinfo
  {author} {\bibfnamefont {T.~A.}\ \bibnamefont {En{\ss}lin}},\ }\href
  {https://doi.org/10.1051/0004-6361:20078671} {\bibfield  {journal} {\bibinfo
  {journal} {Astronomy \& Astrophysics}\ }\textbf {\bibinfo {volume} {477}},\
  \bibinfo {pages} {573} (\bibinfo {year} {2008})},\ \Eprint
  {https://arxiv.org/abs/0711.1572} {arxiv:0711.1572 [astro-ph]} \BibitemShut
  {NoStop}%
\bibitem [{\citenamefont {Gleeson}\ and\ \citenamefont
  {Axford}(1968)}]{Gleeson:1968zza}%
  \BibitemOpen
  \bibfield  {author} {\bibinfo {author} {\bibfnamefont {L.~J.}\ \bibnamefont
  {Gleeson}}\ and\ \bibinfo {author} {\bibfnamefont {W.~I.}\ \bibnamefont
  {Axford}},\ }\href {https://doi.org/10.1086/149822} {\bibfield  {journal}
  {\bibinfo  {journal} {The Astrophysical Journal}\ }\textbf {\bibinfo {volume}
  {154}},\ \bibinfo {pages} {1011} (\bibinfo {year} {1968})}\BibitemShut
  {NoStop}%
\bibitem [{\citenamefont {Potgieter}(2014)}]{Potgieter:2014pka}%
  \BibitemOpen
  \bibfield  {author} {\bibinfo {author} {\bibfnamefont {M.~S.}\ \bibnamefont
  {Potgieter}},\ }\href {https://doi.org/10.1016/j.asr.2013.04.015} {\bibfield
  {journal} {\bibinfo  {journal} {Advances in Space Research}\ }\bibinfo
  {series} {Cosmic {{Ray Origins}}: {{Viktor Hess Centennial Anniversary}}},\
  \textbf {\bibinfo {volume} {53}},\ \bibinfo {pages} {1415} (\bibinfo {year}
  {2014})}\BibitemShut {NoStop}%
\bibitem [{\citenamefont {Cholis}\ \emph {et~al.}(2016)\citenamefont {Cholis},
  \citenamefont {Hooper},\ and\ \citenamefont {Linden}}]{Cholis:2015gna}%
  \BibitemOpen
  \bibfield  {author} {\bibinfo {author} {\bibfnamefont {I.}~\bibnamefont
  {Cholis}}, \bibinfo {author} {\bibfnamefont {D.}~\bibnamefont {Hooper}},\
  and\ \bibinfo {author} {\bibfnamefont {T.}~\bibnamefont {Linden}},\ }\href
  {https://doi.org/10.1103/PhysRevD.93.043016} {\bibfield  {journal} {\bibinfo
  {journal} {Physical Review D}\ }\textbf {\bibinfo {volume} {93}},\ \bibinfo
  {pages} {043016} (\bibinfo {year} {2016})},\ \Eprint
  {https://arxiv.org/abs/1511.01507} {arxiv:1511.01507 [astro-ph.SR]}
  \BibitemShut {NoStop}%
\bibitem [{\citenamefont {Trotta}\ \emph {et~al.}(2011)\citenamefont {Trotta},
  \citenamefont {Johannesson}, \citenamefont {Moskalenko}, \citenamefont
  {Porter}, \citenamefont {{de Austri}},\ and\ \citenamefont
  {Strong}}]{Trotta:2010mx}%
  \BibitemOpen
  \bibfield  {author} {\bibinfo {author} {\bibfnamefont {R.}~\bibnamefont
  {Trotta}}, \bibinfo {author} {\bibfnamefont {G.}~\bibnamefont {Johannesson}},
  \bibinfo {author} {\bibfnamefont {I.~V.}\ \bibnamefont {Moskalenko}},
  \bibinfo {author} {\bibfnamefont {T.~A.}\ \bibnamefont {Porter}}, \bibinfo
  {author} {\bibfnamefont {R.~R.}\ \bibnamefont {{de Austri}}},\ and\ \bibinfo
  {author} {\bibfnamefont {A.~W.}\ \bibnamefont {Strong}},\ }\href
  {https://doi.org/10.1088/0004-637X/729/2/106} {\bibfield  {journal} {\bibinfo
   {journal} {The Astrophysical Journal}\ }\textbf {\bibinfo {volume} {729}},\
  \bibinfo {pages} {106} (\bibinfo {year} {2011})},\ \Eprint
  {https://arxiv.org/abs/1011.0037} {arxiv:1011.0037 [astro-ph.HE]}
  \BibitemShut {NoStop}%
\bibitem [{\citenamefont {Achterberg}\ \emph {et~al.}(2001)\citenamefont
  {Achterberg}, \citenamefont {Gallant}, \citenamefont {Kirk},\ and\
  \citenamefont {Guthmann}}]{Achterberg:2001rx}%
  \BibitemOpen
  \bibfield  {author} {\bibinfo {author} {\bibfnamefont {A.}~\bibnamefont
  {Achterberg}}, \bibinfo {author} {\bibfnamefont {Y.~A.}\ \bibnamefont
  {Gallant}}, \bibinfo {author} {\bibfnamefont {J.~G.}\ \bibnamefont {Kirk}},\
  and\ \bibinfo {author} {\bibfnamefont {A.~W.}\ \bibnamefont {Guthmann}},\
  }\href {https://doi.org/10.1046/j.1365-8711.2001.04851.x} {\bibfield
  {journal} {\bibinfo  {journal} {Monthly Notices of the Royal Astronomical
  Society}\ }\textbf {\bibinfo {volume} {328}},\ \bibinfo {pages} {393}
  (\bibinfo {year} {2001})},\ \Eprint {https://arxiv.org/abs/astro-ph/0107530}
  {arxiv:astro-ph/0107530} \BibitemShut {NoStop}%
\bibitem [{\citenamefont {Stone}\ \emph {et~al.}(2013)\citenamefont {Stone},
  \citenamefont {Cummings}, \citenamefont {McDonald}, \citenamefont {Heikkila},
  \citenamefont {Lal},\ and\ \citenamefont
  {Webber}}]{stoneVoyagerObservesLowEnergy2013}%
  \BibitemOpen
  \bibfield  {author} {\bibinfo {author} {\bibfnamefont {E.~C.}\ \bibnamefont
  {Stone}}, \bibinfo {author} {\bibfnamefont {A.~C.}\ \bibnamefont {Cummings}},
  \bibinfo {author} {\bibfnamefont {F.~B.}\ \bibnamefont {McDonald}}, \bibinfo
  {author} {\bibfnamefont {B.~C.}\ \bibnamefont {Heikkila}}, \bibinfo {author}
  {\bibfnamefont {N.}~\bibnamefont {Lal}},\ and\ \bibinfo {author}
  {\bibfnamefont {W.~R.}\ \bibnamefont {Webber}},\ }\href
  {https://doi.org/10.1126/science.1236408} {\bibfield  {journal} {\bibinfo
  {journal} {Science}\ }\textbf {\bibinfo {volume} {341}},\ \bibinfo {pages}
  {150} (\bibinfo {year} {2013})}\BibitemShut {NoStop}%
\bibitem [{\citenamefont {Alt}\ \emph {et~al.}(2006)\citenamefont {Alt},
  \citenamefont {Anticic}, \citenamefont {Baatar} \emph
  {et~al.}}]{NA49:2005qor}%
  \BibitemOpen
  \bibfield  {author} {\bibinfo {author} {\bibfnamefont {C.}~\bibnamefont
  {Alt}}, \bibinfo {author} {\bibfnamefont {T.}~\bibnamefont {Anticic}},
  \bibinfo {author} {\bibfnamefont {B.}~\bibnamefont {Baatar}}, \emph {et~al.}
  (\bibinfo {collaboration} {NA49}),\ }\href
  {https://doi.org/10.1140/epjc/s2005-02391-9} {\bibfield  {journal} {\bibinfo
  {journal} {The European Physical Journal C - Particles and Fields}\ }\textbf
  {\bibinfo {volume} {45}},\ \bibinfo {pages} {343} (\bibinfo {year} {2006})},\
  \Eprint {https://arxiv.org/abs/hep-ex/0510009} {arxiv:hep-ex/0510009}
  \BibitemShut {NoStop}%
\bibitem [{\citenamefont {Anticic}\ \emph {et~al.}(2010)\citenamefont {Anticic}
  \emph {et~al.}}]{NA49:2009wth}%
  \BibitemOpen
  \bibfield  {author} {\bibinfo {author} {\bibfnamefont {T.}~\bibnamefont
  {Anticic}} \emph {et~al.} (\bibinfo {collaboration} {NA49}),\ }\href
  {https://doi.org/10.1140/epjc/s10052-010-1328-0} {\bibfield  {journal}
  {\bibinfo  {journal} {The European Physical Journal C}\ }\textbf {\bibinfo
  {volume} {68}},\ \bibinfo {pages} {1} (\bibinfo {year} {2010})},\ \Eprint
  {https://arxiv.org/abs/1004.1889} {arxiv:1004.1889 [hep-ex]} \BibitemShut
  {NoStop}%
\bibitem [{\citenamefont {Aduszkiewicz}\ \emph {et~al.}(2017)\citenamefont
  {Aduszkiewicz} \emph {et~al.}}]{NA61SHINE:2017fne}%
  \BibitemOpen
  \bibfield  {author} {\bibinfo {author} {\bibfnamefont {A.}~\bibnamefont
  {Aduszkiewicz}} \emph {et~al.} (\bibinfo {collaboration} {NA61/SHINE}),\
  }\href {https://doi.org/10.1140/epjc/s10052-017-5260-4} {\bibfield  {journal}
  {\bibinfo  {journal} {The European Physical Journal C}\ }\textbf {\bibinfo
  {volume} {77}},\ \bibinfo {pages} {671} (\bibinfo {year} {2017})},\ \Eprint
  {https://arxiv.org/abs/1705.02467} {arxiv:1705.02467 [nucl-ex]} \BibitemShut
  {NoStop}%
\bibitem [{\citenamefont {Strong}\ \emph {et~al.}(2011)\citenamefont {Strong},
  \citenamefont {Orlando},\ and\ \citenamefont {Jaffe}}]{Strong:2011wd}%
  \BibitemOpen
  \bibfield  {author} {\bibinfo {author} {\bibfnamefont {A.~W.}\ \bibnamefont
  {Strong}}, \bibinfo {author} {\bibfnamefont {E.}~\bibnamefont {Orlando}},\
  and\ \bibinfo {author} {\bibfnamefont {T.~R.}\ \bibnamefont {Jaffe}},\ }\href
  {https://doi.org/10.1051/0004-6361/201116828} {\bibfield  {journal} {\bibinfo
   {journal} {Astron. Astrophys.}\ }\textbf {\bibinfo {volume} {534}},\
  \bibinfo {pages} {A54} (\bibinfo {year} {2011})},\ \Eprint
  {https://arxiv.org/abs/1108.4822} {arxiv:1108.4822 [astro-ph.HE]}
  \BibitemShut {NoStop}%
\bibitem [{\citenamefont {Bringmann}\ \emph {et~al.}(2012)\citenamefont
  {Bringmann}, \citenamefont {Donato},\ and\ \citenamefont
  {Lineros}}]{Bringmann:2011py}%
  \BibitemOpen
  \bibfield  {author} {\bibinfo {author} {\bibfnamefont {T.}~\bibnamefont
  {Bringmann}}, \bibinfo {author} {\bibfnamefont {F.}~\bibnamefont {Donato}},\
  and\ \bibinfo {author} {\bibfnamefont {R.~A.}\ \bibnamefont {Lineros}},\
  }\href {https://doi.org/10.1088/1475-7516/2012/01/049} {\bibfield  {journal}
  {\bibinfo  {journal} {Journal of Cosmology and Astroparticle Physics}\
  }\textbf {\bibinfo {volume} {01}}\bibfield  {number} {\bibinfo  {number} {
  (01)},\ \bibinfo {pages} {049}},\ }\Eprint {https://arxiv.org/abs/1106.4821}
  {arxiv:1106.4821 [astro-ph.GA]} \BibitemShut {NoStop}%
\bibitem [{\citenamefont {Orlando}\ and\ \citenamefont
  {Strong}(2013)}]{Orlando:2013ysa}%
  \BibitemOpen
  \bibfield  {author} {\bibinfo {author} {\bibfnamefont {E.}~\bibnamefont
  {Orlando}}\ and\ \bibinfo {author} {\bibfnamefont {A.}~\bibnamefont
  {Strong}},\ }\href {https://doi.org/10.1093/mnras/stt1718} {\bibfield
  {journal} {\bibinfo  {journal} {Monthly Notices of the Royal Astronomical
  Society}\ }\textbf {\bibinfo {volume} {436}},\ \bibinfo {pages} {2127}
  (\bibinfo {year} {2013})},\ \Eprint {https://arxiv.org/abs/1309.2947}
  {arxiv:1309.2947 [astro-ph.GA]} \BibitemShut {NoStop}%
\bibitem [{\citenamefont
  {Profumo}(2017)}]{profumoIntroductionParticleDark2017}%
  \BibitemOpen
  \bibfield  {author} {\bibinfo {author} {\bibfnamefont {S.}~\bibnamefont
  {Profumo}},\ }\href@noop {} {\emph {\bibinfo {title} {An Introduction to
  Particle Dark Matter}}},\ Advanced Textbooks in Physics\ (\bibinfo
  {publisher} {{World Scientific}},\ \bibinfo {address} {{Hackensack, NJ}},\
  \bibinfo {year} {2017})\BibitemShut {NoStop}%
\bibitem [{\citenamefont {Cirelli}\ \emph {et~al.}(2011)\citenamefont
  {Cirelli}, \citenamefont {Corcella}, \citenamefont {Hektor}, \citenamefont
  {H{\"u}tsi}, \citenamefont {Kadastik}, \citenamefont {Panci}, \citenamefont
  {Raidal}, \citenamefont {Sala},\ and\ \citenamefont
  {Strumia}}]{Cirelli:2010xx}%
  \BibitemOpen
  \bibfield  {author} {\bibinfo {author} {\bibfnamefont {M.}~\bibnamefont
  {Cirelli}}, \bibinfo {author} {\bibfnamefont {G.}~\bibnamefont {Corcella}},
  \bibinfo {author} {\bibfnamefont {A.}~\bibnamefont {Hektor}}, \bibinfo
  {author} {\bibfnamefont {G.}~\bibnamefont {H{\"u}tsi}}, \bibinfo {author}
  {\bibfnamefont {M.}~\bibnamefont {Kadastik}}, \bibinfo {author}
  {\bibfnamefont {P.}~\bibnamefont {Panci}}, \bibinfo {author} {\bibfnamefont
  {M.}~\bibnamefont {Raidal}}, \bibinfo {author} {\bibfnamefont
  {F.}~\bibnamefont {Sala}},\ and\ \bibinfo {author} {\bibfnamefont
  {A.}~\bibnamefont {Strumia}},\ }\href
  {https://doi.org/10.1088/1475-7516/2012/10/E01} {\bibfield  {journal}
  {\bibinfo  {journal} {Journal of Cosmology and Astroparticle Physics}\
  }\textbf {\bibinfo {volume} {03}}\bibfield  {number} {\bibinfo  {number} {
  (03)},\ \bibinfo {pages} {051}},\ }\Eprint {https://arxiv.org/abs/1012.4515}
  {arxiv:1012.4515 [hep-ph]} \BibitemShut {NoStop}%
\bibitem [{\citenamefont {Navarro}\ \emph {et~al.}(1997)\citenamefont
  {Navarro}, \citenamefont {Frenk},\ and\ \citenamefont
  {White}}]{Navarro:1996gj}%
  \BibitemOpen
  \bibfield  {author} {\bibinfo {author} {\bibfnamefont {J.~F.}\ \bibnamefont
  {Navarro}}, \bibinfo {author} {\bibfnamefont {C.~S.}\ \bibnamefont {Frenk}},\
  and\ \bibinfo {author} {\bibfnamefont {S.~D.~M.}\ \bibnamefont {White}},\
  }\href {https://doi.org/10.1086/304888} {\bibfield  {journal} {\bibinfo
  {journal} {The Astrophysical Journal}\ }\textbf {\bibinfo {volume} {490}},\
  \bibinfo {pages} {493} (\bibinfo {year} {1997})},\ \Eprint
  {https://arxiv.org/abs/astro-ph/9611107} {arxiv:astro-ph/9611107}
  \BibitemShut {NoStop}%
\bibitem [{\citenamefont {Karukes}\ \emph {et~al.}(2019)\citenamefont
  {Karukes}, \citenamefont {Benito}, \citenamefont {Iocco}, \citenamefont
  {Trotta},\ and\ \citenamefont {{Geringer-Sameth}}}]{Karukes:2019jxv}%
  \BibitemOpen
  \bibfield  {author} {\bibinfo {author} {\bibfnamefont {E.~V.}\ \bibnamefont
  {Karukes}}, \bibinfo {author} {\bibfnamefont {M.}~\bibnamefont {Benito}},
  \bibinfo {author} {\bibfnamefont {F.}~\bibnamefont {Iocco}}, \bibinfo
  {author} {\bibfnamefont {R.}~\bibnamefont {Trotta}},\ and\ \bibinfo {author}
  {\bibfnamefont {A.}~\bibnamefont {{Geringer-Sameth}}},\ }\href
  {https://doi.org/10.1088/1475-7516/2019/09/046} {\bibfield  {journal}
  {\bibinfo  {journal} {Journal of Cosmology and Astroparticle Physics}\
  }\textbf {\bibinfo {volume} {09}}\bibfield  {number} {\bibinfo  {number} {
  (09)},\ \bibinfo {pages} {046}},\ }\Eprint {https://arxiv.org/abs/1901.02463}
  {arxiv:1901.02463 [astro-ph.GA]} \BibitemShut {NoStop}%
\bibitem [{\citenamefont {Benito}\ \emph {et~al.}(2019)\citenamefont {Benito},
  \citenamefont {Cuoco},\ and\ \citenamefont {Iocco}}]{Benito:2019ngh}%
  \BibitemOpen
  \bibfield  {author} {\bibinfo {author} {\bibfnamefont {M.}~\bibnamefont
  {Benito}}, \bibinfo {author} {\bibfnamefont {A.}~\bibnamefont {Cuoco}},\ and\
  \bibinfo {author} {\bibfnamefont {F.}~\bibnamefont {Iocco}},\ }\href
  {https://doi.org/10.1088/1475-7516/2019/03/033} {\bibfield  {journal}
  {\bibinfo  {journal} {Journal of Cosmology and Astroparticle Physics}\
  }\textbf {\bibinfo {volume} {03}}\bibfield  {number} {\bibinfo  {number} {
  (03)},\ \bibinfo {pages} {033}},\ }\Eprint {https://arxiv.org/abs/1901.02460}
  {arxiv:1901.02460 [astro-ph.GA]} \BibitemShut {NoStop}%
\bibitem [{\citenamefont {Merritt}\ \emph {et~al.}(2006)\citenamefont
  {Merritt}, \citenamefont {Graham}, \citenamefont {Moore}, \citenamefont
  {Diemand},\ and\ \citenamefont {Terzic}}]{Graham:2005xx}%
  \BibitemOpen
  \bibfield  {author} {\bibinfo {author} {\bibfnamefont {D.}~\bibnamefont
  {Merritt}}, \bibinfo {author} {\bibfnamefont {A.~W.}\ \bibnamefont {Graham}},
  \bibinfo {author} {\bibfnamefont {B.}~\bibnamefont {Moore}}, \bibinfo
  {author} {\bibfnamefont {J.}~\bibnamefont {Diemand}},\ and\ \bibinfo {author}
  {\bibfnamefont {B.}~\bibnamefont {Terzic}},\ }\href
  {https://doi.org/10.1086/508988} {\bibfield  {journal} {\bibinfo  {journal}
  {The Astronomical Journal}\ }\textbf {\bibinfo {volume} {132}},\ \bibinfo
  {pages} {2685} (\bibinfo {year} {2006})},\ \Eprint
  {https://arxiv.org/abs/astro-ph/0509417} {arxiv:astro-ph/0509417}
  \BibitemShut {NoStop}%
\bibitem [{\citenamefont {Navarro}\ \emph {et~al.}(2010)\citenamefont
  {Navarro}, \citenamefont {Ludlow}, \citenamefont {Springel}, \citenamefont
  {Wang}, \citenamefont {Vogelsberger}, \citenamefont {White}, \citenamefont
  {Jenkins}, \citenamefont {Frenk},\ and\ \citenamefont
  {Helmi}}]{Navarro:2008kc}%
  \BibitemOpen
  \bibfield  {author} {\bibinfo {author} {\bibfnamefont {J.~F.}\ \bibnamefont
  {Navarro}}, \bibinfo {author} {\bibfnamefont {A.}~\bibnamefont {Ludlow}},
  \bibinfo {author} {\bibfnamefont {V.}~\bibnamefont {Springel}}, \bibinfo
  {author} {\bibfnamefont {J.}~\bibnamefont {Wang}}, \bibinfo {author}
  {\bibfnamefont {M.}~\bibnamefont {Vogelsberger}}, \bibinfo {author}
  {\bibfnamefont {S.~D.~M.}\ \bibnamefont {White}}, \bibinfo {author}
  {\bibfnamefont {A.}~\bibnamefont {Jenkins}}, \bibinfo {author} {\bibfnamefont
  {C.~S.}\ \bibnamefont {Frenk}},\ and\ \bibinfo {author} {\bibfnamefont
  {A.}~\bibnamefont {Helmi}},\ }\href
  {https://doi.org/10.1111/j.1365-2966.2009.15878.x} {\bibfield  {journal}
  {\bibinfo  {journal} {Monthly Notices of the Royal Astronomical Society}\
  }\textbf {\bibinfo {volume} {402}},\ \bibinfo {pages} {21} (\bibinfo {year}
  {2010})},\ \Eprint {https://arxiv.org/abs/0810.1522} {arxiv:0810.1522
  [astro-ph]} \BibitemShut {NoStop}%
\bibitem [{\citenamefont {Burkert}(1996)}]{Burkert:1995yz}%
  \BibitemOpen
  \bibfield  {author} {\bibinfo {author} {\bibfnamefont {A.}~\bibnamefont
  {Burkert}},\ }\href {https://doi.org/10.1086/309560} {\bibfield  {journal}
  {\bibinfo  {journal} {The Astrophysical Journal}\ }\textbf {\bibinfo {volume}
  {171}},\ \bibinfo {pages} {175} (\bibinfo {year} {1996})},\ \Eprint
  {https://arxiv.org/abs/astro-ph/9504041} {arxiv:astro-ph/9504041}
  \BibitemShut {NoStop}%
\bibitem [{\citenamefont {Salucci}\ and\ \citenamefont
  {Burkert}(2000)}]{Salucci:2000ps}%
  \BibitemOpen
  \bibfield  {author} {\bibinfo {author} {\bibfnamefont {P.}~\bibnamefont
  {Salucci}}\ and\ \bibinfo {author} {\bibfnamefont {A.}~\bibnamefont
  {Burkert}},\ }\href {https://doi.org/10.1086/312747} {\bibfield  {journal}
  {\bibinfo  {journal} {The Astrophysical Journal}\ }\textbf {\bibinfo {volume}
  {537}},\ \bibinfo {pages} {L9} (\bibinfo {year} {2000})},\ \Eprint
  {https://arxiv.org/abs/astro-ph/0004397} {arxiv:astro-ph/0004397}
  \BibitemShut {NoStop}%
\bibitem [{\citenamefont {Derome}\ \emph {et~al.}(2019)\citenamefont {Derome},
  \citenamefont {Maurin}, \citenamefont {Salati}, \citenamefont {Boudaud},
  \citenamefont {G{\'e}nolini},\ and\ \citenamefont
  {Kunz{\'e}}}]{Derome:2019jfs}%
  \BibitemOpen
  \bibfield  {author} {\bibinfo {author} {\bibfnamefont {L.}~\bibnamefont
  {Derome}}, \bibinfo {author} {\bibfnamefont {D.}~\bibnamefont {Maurin}},
  \bibinfo {author} {\bibfnamefont {P.}~\bibnamefont {Salati}}, \bibinfo
  {author} {\bibfnamefont {M.}~\bibnamefont {Boudaud}}, \bibinfo {author}
  {\bibfnamefont {Y.}~\bibnamefont {G{\'e}nolini}},\ and\ \bibinfo {author}
  {\bibfnamefont {P.}~\bibnamefont {Kunz{\'e}}},\ }\href
  {https://doi.org/10.1051/0004-6361/201935717} {\bibfield  {journal} {\bibinfo
   {journal} {Astronomy \& Astrophysics}\ }\textbf {\bibinfo {volume} {627}},\
  \bibinfo {pages} {A158} (\bibinfo {year} {2019})},\ \Eprint
  {https://arxiv.org/abs/1904.08210} {arxiv:1904.08210 [astro-ph.HE]}
  \BibitemShut {NoStop}%
\bibitem [{\citenamefont {Cuoco}\ \emph {et~al.}(2019)\citenamefont {Cuoco},
  \citenamefont {Heisig}, \citenamefont {Klamt}, \citenamefont {Korsmeier},\
  and\ \citenamefont {Kr{\"a}mer}}]{Cuoco:2019kuu}%
  \BibitemOpen
  \bibfield  {author} {\bibinfo {author} {\bibfnamefont {A.}~\bibnamefont
  {Cuoco}}, \bibinfo {author} {\bibfnamefont {J.}~\bibnamefont {Heisig}},
  \bibinfo {author} {\bibfnamefont {L.}~\bibnamefont {Klamt}}, \bibinfo
  {author} {\bibfnamefont {M.}~\bibnamefont {Korsmeier}},\ and\ \bibinfo
  {author} {\bibfnamefont {M.}~\bibnamefont {Kr{\"a}mer}},\ }\href
  {https://doi.org/10.1103/PhysRevD.99.103014} {\bibfield  {journal} {\bibinfo
  {journal} {Physical Review D}\ }\textbf {\bibinfo {volume} {99}},\ \bibinfo
  {pages} {103014} (\bibinfo {year} {2019})},\ \Eprint
  {https://arxiv.org/abs/1903.01472} {arxiv:1903.01472 [astro-ph.HE]}
  \BibitemShut {NoStop}%
\bibitem [{\citenamefont {Heisig}\ \emph {et~al.}(2020)\citenamefont {Heisig},
  \citenamefont {Korsmeier},\ and\ \citenamefont {Winkler}}]{Heisig:2020nse}%
  \BibitemOpen
  \bibfield  {author} {\bibinfo {author} {\bibfnamefont {J.}~\bibnamefont
  {Heisig}}, \bibinfo {author} {\bibfnamefont {M.}~\bibnamefont {Korsmeier}},\
  and\ \bibinfo {author} {\bibfnamefont {M.~W.}\ \bibnamefont {Winkler}},\
  }\href {https://doi.org/10.1103/PhysRevResearch.2.043017} {\bibfield
  {journal} {\bibinfo  {journal} {Physical Review Research}\ }\textbf {\bibinfo
  {volume} {2}},\ \bibinfo {pages} {043017} (\bibinfo {year} {2020})},\ \Eprint
  {https://arxiv.org/abs/2005.04237} {arxiv:2005.04237 [astro-ph.HE]}
  \BibitemShut {NoStop}%
\bibitem [{\citenamefont {Fang}\ \emph {et~al.}(2017)\citenamefont {Fang},
  \citenamefont {Wang}, \citenamefont {Bi}, \citenamefont {Lin},\ and\
  \citenamefont {Yin}}]{Fang:2016wid}%
  \BibitemOpen
  \bibfield  {author} {\bibinfo {author} {\bibfnamefont {K.}~\bibnamefont
  {Fang}}, \bibinfo {author} {\bibfnamefont {B.-B.}\ \bibnamefont {Wang}},
  \bibinfo {author} {\bibfnamefont {X.-J.}\ \bibnamefont {Bi}}, \bibinfo
  {author} {\bibfnamefont {S.-J.}\ \bibnamefont {Lin}},\ and\ \bibinfo {author}
  {\bibfnamefont {P.-F.}\ \bibnamefont {Yin}},\ }\href
  {https://doi.org/10.3847/1538-4357/aa5b93} {\bibfield  {journal} {\bibinfo
  {journal} {The Astrophysical Journal}\ }\textbf {\bibinfo {volume} {836}},\
  \bibinfo {pages} {172} (\bibinfo {year} {2017})},\ \Eprint
  {https://arxiv.org/abs/1611.10292} {arxiv:1611.10292 [astro-ph.HE]}
  \BibitemShut {NoStop}%
\bibitem [{\citenamefont {Fang}\ \emph
  {et~al.}(2018{\natexlab{a}})\citenamefont {Fang}, \citenamefont {Bi},\ and\
  \citenamefont {Yin}}]{Fang:2017tvj}%
  \BibitemOpen
  \bibfield  {author} {\bibinfo {author} {\bibfnamefont {K.}~\bibnamefont
  {Fang}}, \bibinfo {author} {\bibfnamefont {X.-J.}\ \bibnamefont {Bi}},\ and\
  \bibinfo {author} {\bibfnamefont {P.-F.}\ \bibnamefont {Yin}},\ }\href
  {https://doi.org/10.3847/1538-4357/aaa710} {\bibfield  {journal} {\bibinfo
  {journal} {The Astrophysical Journal}\ }\textbf {\bibinfo {volume} {854}},\
  \bibinfo {pages} {57} (\bibinfo {year} {2018}{\natexlab{a}})},\ \Eprint
  {https://arxiv.org/abs/1711.10996} {arxiv:1711.10996 [astro-ph.HE]}
  \BibitemShut {NoStop}%
\bibitem [{\citenamefont {Accardo}\ \emph {et~al.}(2014)\citenamefont {Accardo}
  \emph {et~al.}}]{AMS:2014bun}%
  \BibitemOpen
  \bibfield  {author} {\bibinfo {author} {\bibfnamefont {L.}~\bibnamefont
  {Accardo}} \emph {et~al.} (\bibinfo {collaboration} {AMS}),\ }\href
  {https://doi.org/10.1103/PhysRevLett.113.121101} {\bibfield  {journal}
  {\bibinfo  {journal} {Physical Review Letters}\ }\textbf {\bibinfo {volume}
  {113}},\ \bibinfo {pages} {121101} (\bibinfo {year} {2014})}\BibitemShut
  {NoStop}%
\bibitem [{\citenamefont {Ghelfi}\ \emph {et~al.}(2017)\citenamefont {Ghelfi},
  \citenamefont {Maurin}, \citenamefont {Cheminet}, \citenamefont {Derome},
  \citenamefont {Hubert},\ and\ \citenamefont {Melot}}]{Ghelfi_2017}%
  \BibitemOpen
  \bibfield  {author} {\bibinfo {author} {\bibfnamefont {A.}~\bibnamefont
  {Ghelfi}}, \bibinfo {author} {\bibfnamefont {D.}~\bibnamefont {Maurin}},
  \bibinfo {author} {\bibfnamefont {A.}~\bibnamefont {Cheminet}}, \bibinfo
  {author} {\bibfnamefont {L.}~\bibnamefont {Derome}}, \bibinfo {author}
  {\bibfnamefont {G.}~\bibnamefont {Hubert}},\ and\ \bibinfo {author}
  {\bibfnamefont {F.}~\bibnamefont {Melot}},\ }\href
  {https://doi.org/10.1016/j.asr.2016.06.027} {\bibfield  {journal} {\bibinfo
  {journal} {Advances in Space Research}\ }\textbf {\bibinfo {volume} {60}},\
  \bibinfo {pages} {833} (\bibinfo {year} {2017})},\ \Eprint
  {https://arxiv.org/abs/1607.01976} {1607.01976} \BibitemShut {NoStop}%
\bibitem [{\citenamefont {Zhu}\ \emph {et~al.}(2018)\citenamefont {Zhu},
  \citenamefont {Yuan},\ and\ \citenamefont {Wei}}]{Zhu:2018jbk}%
  \BibitemOpen
  \bibfield  {author} {\bibinfo {author} {\bibfnamefont {C.-R.}\ \bibnamefont
  {Zhu}}, \bibinfo {author} {\bibfnamefont {Q.}~\bibnamefont {Yuan}},\ and\
  \bibinfo {author} {\bibfnamefont {D.-M.}\ \bibnamefont {Wei}},\ }\href
  {https://doi.org/10.3847/1538-4357/aacff9} {\bibfield  {journal} {\bibinfo
  {journal} {The Astrophysical Journal}\ }\textbf {\bibinfo {volume} {863}},\
  \bibinfo {pages} {119} (\bibinfo {year} {2018})},\ \Eprint
  {https://arxiv.org/abs/1807.09470} {arxiv:1807.09470 [astro-ph.HE]}
  \BibitemShut {NoStop}%
\bibitem [{\citenamefont {Koldobskiy}\ and\ \citenamefont
  {Usoskin}(2023)}]{Koldobskiy:2023prp}%
  \BibitemOpen
  \bibfield  {author} {\bibinfo {author} {\bibfnamefont {S.}~\bibnamefont
  {Koldobskiy}}\ and\ \bibinfo {author} {\bibfnamefont {I.}~\bibnamefont
  {Usoskin}},\ }\href {https://doi.org/10.22323/1.444.1325} {\bibfield
  {journal} {\bibinfo  {journal} {PoS}\ }\textbf {\bibinfo {volume}
  {ICRC2023}},\ \bibinfo {pages} {1325} (\bibinfo {year} {2023})}\BibitemShut
  {NoStop}%
\bibitem [{\citenamefont {Lin}\ \emph {et~al.}(2019)\citenamefont {Lin},
  \citenamefont {Bi},\ and\ \citenamefont {Yin}}]{Lin:2019ljc}%
  \BibitemOpen
  \bibfield  {author} {\bibinfo {author} {\bibfnamefont {S.-J.}\ \bibnamefont
  {Lin}}, \bibinfo {author} {\bibfnamefont {X.-J.}\ \bibnamefont {Bi}},\ and\
  \bibinfo {author} {\bibfnamefont {P.-F.}\ \bibnamefont {Yin}},\ }\href
  {https://doi.org/10.1103/PhysRevD.100.103014} {\bibfield  {journal} {\bibinfo
   {journal} {Physical Review D}\ }\textbf {\bibinfo {volume} {100}},\ \bibinfo
  {pages} {103014} (\bibinfo {year} {2019})},\ \Eprint
  {https://arxiv.org/abs/1903.09545} {arxiv:1903.09545 [astro-ph.HE]}
  \BibitemShut {NoStop}%
\bibitem [{\citenamefont {Steigman}\ \emph {et~al.}(2012)\citenamefont
  {Steigman}, \citenamefont {Dasgupta},\ and\ \citenamefont
  {Beacom}}]{Steigman:2012nb}%
  \BibitemOpen
  \bibfield  {author} {\bibinfo {author} {\bibfnamefont {G.}~\bibnamefont
  {Steigman}}, \bibinfo {author} {\bibfnamefont {B.}~\bibnamefont {Dasgupta}},\
  and\ \bibinfo {author} {\bibfnamefont {J.~F.}\ \bibnamefont {Beacom}},\
  }\href {https://doi.org/10.1103/PhysRevD.86.023506} {\bibfield  {journal}
  {\bibinfo  {journal} {Physical Review D}\ }\textbf {\bibinfo {volume} {86}},\
  \bibinfo {pages} {023506} (\bibinfo {year} {2012})},\ \Eprint
  {https://arxiv.org/abs/1204.3622} {arxiv:1204.3622 [hep-ph]} \BibitemShut
  {NoStop}%
\bibitem [{\citenamefont {Ibarra}\ \emph {et~al.}(2013)\citenamefont {Ibarra},
  \citenamefont {Tran},\ and\ \citenamefont {Weniger}}]{Ibarra:2013cra}%
  \BibitemOpen
  \bibfield  {author} {\bibinfo {author} {\bibfnamefont {A.}~\bibnamefont
  {Ibarra}}, \bibinfo {author} {\bibfnamefont {D.}~\bibnamefont {Tran}},\ and\
  \bibinfo {author} {\bibfnamefont {C.}~\bibnamefont {Weniger}},\ }\href
  {https://doi.org/10.1142/S0217751X13300408} {\bibfield  {journal} {\bibinfo
  {journal} {International Journal of Modern Physics A}\ }\textbf {\bibinfo
  {volume} {28}},\ \bibinfo {pages} {1330040} (\bibinfo {year} {2013})},\
  \Eprint {https://arxiv.org/abs/1307.6434} {arxiv:1307.6434 [hep-ph]}
  \BibitemShut {NoStop}%
\bibitem [{\citenamefont {Nardi}\ \emph {et~al.}(2009)\citenamefont {Nardi},
  \citenamefont {Sannino},\ and\ \citenamefont {Strumia}}]{Nardi:2008ix}%
  \BibitemOpen
  \bibfield  {author} {\bibinfo {author} {\bibfnamefont {E.}~\bibnamefont
  {Nardi}}, \bibinfo {author} {\bibfnamefont {F.}~\bibnamefont {Sannino}},\
  and\ \bibinfo {author} {\bibfnamefont {A.}~\bibnamefont {Strumia}},\ }\href
  {https://doi.org/10.1088/1475-7516/2009/01/043} {\bibfield  {journal}
  {\bibinfo  {journal} {Journal of Cosmology and Astroparticle Physics}\
  }\textbf {\bibinfo {volume} {01}}\bibfield  {number} {\bibinfo  {number} {
  (01)},\ \bibinfo {pages} {043}},\ }\Eprint {https://arxiv.org/abs/0811.4153}
  {arxiv:0811.4153 [hep-ph]} \BibitemShut {NoStop}%
\bibitem [{\citenamefont {Meade}\ \emph {et~al.}(2010)\citenamefont {Meade},
  \citenamefont {Papucci}, \citenamefont {Strumia},\ and\ \citenamefont
  {Volansky}}]{Meade:2009iu}%
  \BibitemOpen
  \bibfield  {author} {\bibinfo {author} {\bibfnamefont {P.}~\bibnamefont
  {Meade}}, \bibinfo {author} {\bibfnamefont {M.}~\bibnamefont {Papucci}},
  \bibinfo {author} {\bibfnamefont {A.}~\bibnamefont {Strumia}},\ and\ \bibinfo
  {author} {\bibfnamefont {T.}~\bibnamefont {Volansky}},\ }\href
  {https://doi.org/10.1016/j.nuclphysb.2010.01.012} {\bibfield  {journal}
  {\bibinfo  {journal} {Nuclear Physics B}\ }\textbf {\bibinfo {volume}
  {831}},\ \bibinfo {pages} {178} (\bibinfo {year} {2010})},\ \Eprint
  {https://arxiv.org/abs/0905.0480} {arxiv:0905.0480 [hep-ph]} \BibitemShut
  {NoStop}%
\bibitem [{\citenamefont {Fang}\ \emph
  {et~al.}(2018{\natexlab{b}})\citenamefont {Fang}, \citenamefont {Bi},\ and\
  \citenamefont {Yin}}]{Fang:2017nww}%
  \BibitemOpen
  \bibfield  {author} {\bibinfo {author} {\bibfnamefont {K.}~\bibnamefont
  {Fang}}, \bibinfo {author} {\bibfnamefont {X.-J.}\ \bibnamefont {Bi}},\ and\
  \bibinfo {author} {\bibfnamefont {P.-f.}\ \bibnamefont {Yin}},\ }\href
  {https://doi.org/10.1093/mnras/sty1463} {\bibfield  {journal} {\bibinfo
  {journal} {Monthly Notices of the Royal Astronomical Society}\ }\textbf
  {\bibinfo {volume} {478}},\ \bibinfo {pages} {5660} (\bibinfo {year}
  {2018}{\natexlab{b}})},\ \Eprint {https://arxiv.org/abs/1706.03745}
  {arxiv:1706.03745 [astro-ph.HE]} \BibitemShut {NoStop}%
\end{thebibliography}%

%
\end{document}